%% file: paper.tex
\documentclass[doctor,english,final]{kaist-ucs}
\usepackage{amssymb}
\usepackage{amsmath} 
\usepackage{rotating}

\title[korean] {음향 이벤트 검출을 위한 주파수 동적 컨볼루션}
\title[english]{Frequency Dynamic Convolutions for Sound Event Detection}

\author[korean]{남}{현 욱}
\author[korean2]{남}{현욱}    
\author[chinese]{南}{炫 旭}
\author[english]{Nam}{Hyeonuk}

\advisor[major]{박 용 화}{Yong-Hwa Park}{signed}
\advisor[major2]{박용화}{Yong-Hwa Park}{signed}    
\advisorinfo{Professor of Mechanical Engineering} 
\department{ME}{engineering}{c}

\referee[1]{박 용 화}
\referee[2]{윤 용 진}
\referee[3]{이 승 철}
\referee[4]{최 정 우}
\referee[5]{김 성 영}

\approvaldate{2025}{6}{10}

\refereedate{2025}{6}{10}

\gradyear{2025}

\begin{document}


   \thesisinfo
    \begin{summary}      
    \input{sections/0_1_abstract-korean}
    \end{summary}
   
    \begin{Korkeyword}
    음향 이벤트 검출, 이동 등변화성, 주파수 동적 컨볼루션, 시간 집중 풀링
    \end{Korkeyword}

    \begin{abstract}
    \input{sections/0_2_abstract-english}
    \end{abstract} 
     
    \begin{Engkeyword}
    sound event detection, translational equivariacne, frequency dynamic convolution, temproal attention pooling
    \end{Engkeyword}

    \addtocounter{pagemarker}{1}                 
    \newpage

    \tableofcontents

    \listoftables

    \listoffigures



\chapter{Introduction}
\input{sections/1_Introduction}

\chapter{Related Works}
\input{sections/2_Relatedworks}

\chapter{Proposed Methods: Frequency Dynamic Convolutions}
\input{sections/3_methodologies}

\chapter{Experiments \& Results}
\input{sections/4_experiments}

\chapter{Case Studies}
\input{sections/5_casestudy}

\chapter{Conclusion}
\input{sections/6_Conclusionandfutureworks}



\acknowledgment[4]
\input{sections/7_acknowledgement}

\curriculumvitae[4]

    \begin{personaldata}
        \name       {남 현 욱}
        \dateofbirth{1993}{10}{09}
        \email      {frednam@kaist.ac.kr}
    \end{personaldata}

    \begin{education}
        \item[2009. 1.\ --\ 2012. 12.] NUS High School of Math and Science, Singapore (NUS High diploma)
        \item[2013. 9.\ --\ 2018. 2.] 한국과학기술원 기계공학과 (학사)
        \item[2018. 3.\ --\ 2020. 2.] 한국과학기술원 기계공학과 (석사)
        \item[2020. 3.\ --\ 2025. 8.] 한국과학기술원 기계공학과 (박사)
    \end{education}

    \begin{career}
        \item[2017. 3.\ --\ 2024. 2.] 한국과학기술원 기계공학과 조교
    \end{career}


    \begin{publication}
        \item {\bf H. Nam}, B. Y. Ko, G. T. Lee, S. -H. Kim, W. H. Jung, S. M. Choi and Y. -H. Park,
``Heavily augmented sound event detection utilizing weak predictions,"
\textit{DCASE Challenge Tech. rep.}, 2021.
        \item {\bf H. Nam}, S. -H. Kim and Y. -H. Park,
``FilterAugment: An Acoustic Environmental Data Augmentation Method,"
\textit{International Conference on Acoustics, Speech and Signal Processing (ICASSP)}, 2022.
        \item S. -H. Kim, {\bf H. Nam} and Y. -H. Park,
``Temporal dynamic convolutional neural network for text-independent speaker verification and phonemic analysis,"
\textit{International Conference on Acoustics, Speech and Signal Processing (ICASSP)}, 2022.
        \item {\bf H. Nam}, S. -H. Kim, D. Min, B. Y. Ko, S. D. Choi and Y. -H. Park,
``Frequency dependent sound event detection for DCASE 2022 Challenge Task 4,"
\textit{DCASE Challenge Tech. rep.}, 2022.
        \item B. Y. Ko, {\bf H. Nam}, S. -H. Kim, D. Min, S. Choi and Y. -H. Park,
``Data augmentation and squeeze-and-excitation network on multiple dimension for sound event localization and detection in real scenes,"
\textit{DCASE Challenge Tech. rep.}, 2022.
        \item {\bf H. Nam}, S. -H. Kim, B. Y. Ko and Y. -H. Park,
``Frequency Dynamic Convolution: Frequency-Adaptive Pattern Recognition for Sound Event Detection,"
\textit{Interspeech}, 2022.
        \item G. T. Lee, {\bf H. Nam}, S. -H. Kim, S. M. Choi and Y. -H. Park,
``Deep learning based cough detection camera using enhanced features,"
\textit{Expert Systems with Applications}, 2022.
        \item J. Lee, {\bf H. Nam} and Y. -H. Park,
``VIFS: An end-to-end variational inference for foley sound synthesis,"
\textit{DCASE Challenge Tech. rep.}, 2023.
        \item D. Min, {\bf H. Nam} and Y. -H. Park,
``Application of spectro-temporal receptive field on soft labeled sound event detection,"
\textit{DCASE Challenge Tech. rep.}, 2023.
        \item S. -H. Kim, {\bf H. Nam} and Y. -H. Park,
``Analysis-based optimization of temporal dynamic convolutional neural network for text-independent speaker verification,"
\textit{IEEE Access}, 2023.
        \item {\bf H. Nam}, S. -H. Kim, D. Min and Y. -H. Park,
``Frequency \& Channel Attention for Computationally Efficient Sound Event Detection,"
\textit{DCASE Workshop}, 2023.
        \item D. Min, {\bf H. Nam} and Y. -H. Park,
``Auditory Neural Response Inspired Sound Event Detection Based on Spectro-temporal Receptive Field,"
\textit{DCASE Workshop}, 2023.
        \item B. Y. Ko, G. T. Lee, {\bf H. Nam} and Y. -H. Park,
``PRTFNet: HRTF Individualization for Accurate Spectral Cues Using a Compact PRTF,"
\textit{IEEE Access}, 2023.
        \item S. -H. Kim, {\bf H. Nam}, S. M. Choi and Y. -H. Park,
``Real-Time Sound Recognition System for Human Care Robot Considering Custom Sound Events,"
\textit{IEEE Access}, 2024.
        \item I. Choi, {\bf H. Nam}, D. Min, S. Choi and Y. -H. Park,
``ChatGPT Caption Paraphrasing and FENSE-based Caption Filtering for Automated Audio Captioning,"
\textit{DCASE Challenge Tech. rep.}, 2024.
        \item {\bf H. Nam}, S. -H. Kim, D. Min, J. Lee and Y. -H. Park,
``Diversifying and Expanding Frequency-Adaptive Convolution Kernels for Sound Event Detection,"
\textit{Interspeech}, 2024.
        \item {\bf H. Nam}, D. Min, S. Choi, I. Choi and Y. -H. Park,
``Self Training and Ensembling Frequency Dependent Networks with Coarse Prediction Pooling and Sound Event Bounding Boxes,"
\textit{DCASE Workshop}, 2024.
        \item {\bf H. Nam} and Y. -H. Park,
``Coherence-based Phonemic Analysis on the Effect of Reverberation to Practical Automatic Speech Recognition,"
\textit{Applied Acoustics}, 2025.
        \item {\bf H. Nam}, S. -H. Kim, D. Min, B. Y. Ko and Y. -H. Park,
``Towards Understanding of Frequency Dependence on Sound Event Detection,"
\textit{arXiv}, 2025.
        \item {\bf H. Nam} and Y. -H. Park,
``Pushing the limit of sound event detection with multi-dilated frequency dynamic convolution,"
\textit{arXiv}, 2025.
        \item {\bf H. Nam} and Y. -H. Park,
``JiTTER: Jigsaw Temporal Transformer for Event Reconstruction for Self-Supervised Sound Event Detection,"
\textit{arXiv}, 2025.
        \item {\bf H. Nam} and Y. -H. Park,
``Temporal Attention Pooling for Frequency Dynamic Convolution in Sound Event Detection,"
\textit{arXiv}, 2025.
        \item B. Y. Ko, D. Min, {\bf H. Nam} and Y. -H. Park,
``DNN based HRIRs Identification with a Continuously Rotating Speaker Array,"
\textit{arXiv}, 2025.
    \end{publication}

  \label{paperlastpagelabel}     
\end{document}

%% file: sections/0_1_abstract-korean.tex
딥러닝 기반의 음향 이벤트 검출(Sound Event Detection, SED) 연구는 최근 합성곱 순환 신경망(Convolutional Recurrent Neural Network, CRNN) 및 트랜스포머(Transformer) 모델을 중심으로 발전해왔다. 그러나 기존 2D convolution 기반 모델은 시간 및 주파수 축에서 동일한 이동 불변성(shift invariance)을 가정하며, 이는 주파수 종속적 특성을 갖는 음향 신호와의 불일치를 초래한다. 이를 해결하기 위해, 본 연구에서는 주파수 동적 컨볼루션(Frequency Dynamic Convolution, FDY conv)을 제안하며, 입력 신호의 주파수 구성에 따라 커널을 동적으로 조정하여 SED 성능을 향상시킨다. FDY conv는 주파수별로 학습된 attention weight를 기반으로 여러 개의 기반 커널(basis kernel)을 가중합하여 최적의 주파수 응답을 형성하는 구조를 가진다. 실험 결과, FDY conv를 CRNN에 적용 시 DESED 데이터셋에서 베이스라인 CRNN 대비 성능이 7.56\% 향상되었다. 그러나 FDY conv는 모든 주파수에서 동일한 형태의 기반 커널을 조합하여 사용하므로, 주파수별 특성을 더욱 다양하게 반영하는 데 한계가 존재한다. 또한, 기반 커널의 사이즈가 3×3으로 넓은 주파수 범위를 고려하기에 부족하다. 이를 개선하기 위해, 본 연구에서는 FDY conv의 확장 모델군을 제안한다. Dilated FDY conv (DFD conv)는 다양한 dilation 크기를 적용한 커널을 활용하여 주파수 적응형 컨볼루션의 수용 영역을 확장하고 주파수별 정보 표현력을 강화하였다. 실험 결과, 베이스라인 대비 성능이 9.27\% 향상되었다. Partial FDY conv (PFD conv)는 FDY conv에서 모든 합성곱 연산을 동적 커널로 수행할 경우 복수의 기반 커널로 인해 연산량이 크며, 정적인 음향 이벤트에 대해 불필요한 적응성을 가질 수 있으므로, 정적 2D 합성곱과 주파수 적응형 커널을 결합하여 연산량을 줄이면서도 성능을 유지할 수 있도록 설계하였다. 실험 결과, FDY conv와 마찬가지로 베이스라인 대비 성능이 7.80\% 향상되었으며, FDY conv 대비 파라미터 수가 54.4\% 감소하였다. Multi-Dilated FDY conv (MDFD conv)는 DFD conv가 dilation을 적용하여 주파수 축의 수용 영역을 확장하지만, 모든 주파수에서 동일한 dilation을 적용하는 구조적 한계를 가지는 문제를 해결하기 위해, 여러 dilation 크기를 동시에 활용하는 다중 커널 구조를 적용하여 다양한 주파수 의존적 패턴을 효과적으로 학습하도록 설계되었다. 실험 결과, MDFD conv는 베이스라인 CRNN 대비 성능이 10.98\% 향상되는 높은 성능을 기록하였다. 또한 기존 FDY conv는 시간 축에서 모든 프레임을 동일한 가중치로 처리하는 Temporal Average Pooling을 사용하므로, 순간적인 이벤트를 효과적으로 반영하지 못하는 문제가 있다. 이를 해결하기 위해, 본 연구에서는 두드러지는 특성에 집중하는 시간적 집중 풀링(Time Attention Pooling, TA)과 일시적인 특성에 집중하는 속도 주의 풀링(Velocity Attention Pooling, VA), 그리고 정적인 특성에 집중하는 AP(Average Pooling)를 함께 적용한 TAP-FDY (TFD) conv를 제안하여 순간적인 이벤트의 검출 성능을 향상시켰다. TFD conv는 MDFD conv와 동일한 성능을 기록했지만, 파라미터 수가 12.703M으로 MDFD conv(18.157M) 대비 약 30.01\% 감소하여 동일한 성능을 더 적은 연산량으로 달성하였다. 클래스별 성능 분석 결과, FDY conv는 비정상(non-stationary) 이벤트에서, DFD conv는 넓은 주파수 범위를 가지는 이벤트에서, PFD conv는 정적인(quasi-stationary) 이벤트에서 각각 성능이 향상되었다. 또한 TFD conv(TFD-CRNN)는 순간적인(transient) 이벤트 탐지에서 강점을 보였다. 케이스 스터디 분석 결과, 전차 동력장치 결함 인식에서는 PFD conv가 정적인 신호 패턴을 효과적으로 학습하며 높은 검출 성능을 기록하였고, 가변속도 모터 고장진단에서는 넓은 주파수에 걸친 하모닉 성분을 인식하는 DFD conv가 최적의 검출 성능을 보였으며, 해상 환경의 아크 검출에서는 TAP-FDY conv가 순간적인 신호 특성을 반영하여 우수한 성능을 보였다. 본 연구 결과는 주파수 적응형 컨볼루션과 그 확장 모델군들이 오디오 딥러닝 분야에서 기존 2D convolution을 대체할 강력한 대안이 될 수 있음을 시사한다.

%% file: sections/0_2_abstract-english.tex
Recent research in deep learning-based Sound Event Detection (SED) has primarily focused on Convolutional Recurrent Neural Networks (CRNNs) and Transformer models. However, conventional 2D convolution-based models assume shift invariance along both the temporal and frequency axes, leading to inconsistencies when dealing with frequency-dependent characteristics of acoustic signals. To address this issue, this study proposes Frequency Dynamic Convolution (FDY conv), which dynamically adjusts convolutional kernels based on the frequency composition of the input signal to enhance SED performance. FDY conv constructs an optimal frequency response by adaptively weighting multiple basis kernels based on frequency-specific attention weights. Experimental results show that applying FDY conv to CRNNs improves performance on the DESED dataset by 7.56\% compared to the baseline CRNN. However, FDY conv has limitations in that it combines basis kernels of the same shape across all frequencies, restricting its ability to capture diverse frequency-specific characteristics. Additionally, the 3×3 basis kernel size is insufficient to capture a broader frequency range. To overcome these limitations, this study introduces an extended family of FDY conv models. Dilated FDY conv (DFD conv) applies convolutional kernels with various dilation rates to expand the receptive field along the frequency axis and enhance frequency-specific feature representation. Experimental results show that DFD conv improves performance by 9.27\% over the baseline. Partial FDY conv (PFD conv) addresses the high computational cost of FDY conv, which results from performing all convolution operations with dynamic kernels. Since FDY conv may introduce unnecessary adaptivity for quasi-stationary sound events, PFD conv integrates standard 2D convolutions with frequency-adaptive kernels to reduce computational complexity while maintaining performance. Experimental results demonstrate that PFD conv improves performance by 7.80\% over the baseline while reducing the number of parameters by 54.4\% compared to FDY conv. Multi-Dilated FDY conv (MDFD conv) extends DFD conv by addressing its structural limitation of applying the same dilation across all frequencies. By utilizing multiple convolutional kernels with different dilation rates, MDFD conv effectively captures diverse frequency-dependent patterns. Experimental results indicate that MDFD conv achieves the highest performance, improving the baseline CRNN performance by 10.98\%. Furthermore, standard FDY conv employs Temporal Average Pooling, which assigns equal weight to all frames along the time axis, limiting its ability to effectively capture transient events. To overcome this, this study proposes TAP-FDY conv (TFD conv), which integrates Temporal Attention Pooling (TA) that focuses on salient features, Velocity Attention Pooling (VA) that emphasizes transient characteristics, and Average Pooling (AP) that captures stationary properties. TAP-FDY conv achieves the same performance as MDFD conv but reduces the number of parameters by approximately 30.01\% (12.703M vs. 18.157M), achieving equivalent accuracy with lower computational complexity. Class-wise performance analysis reveals that FDY conv improves detection of non-stationary events, DFD conv is particularly effective for events with broad spectral features, and PFD conv enhances the detection of quasi-stationary events. Additionally, TFD conv (TFD-CRNN) demonstrates strong performance in detecting transient events. In the case studies, PFD conv effectively captures stable signal patterns in tank powertrain fault recognition, DFD conv recognizes wide harmonic spectral patterns on speed-varying motor fault recognition, while TFD conv outperforms other models in detecting transient signals in offshore arc detection. These results suggest that frequency-adaptive convolutions and their extended variants provide a robust alternative to conventional 2D convolutions in deep learning-based audio processing.

%% file: sections/1_Introduction.tex
\section{Background and Motivation}

Auditory intelligence, which aims to replicate human auditory perception, is an essential area of research with wide-ranging applications, including human-machine interaction, robotics, automation, surveillance, and multimedia information retrieval \cite{CASSE}. However, it has received significantly less attention than visual intelligence, which mimics human vision. Most deep learning research has centered around computer vision tasks such as image classification and object detection, while auditory perception has often been sidelined. Even within the auditory domain, speech recognition has dominated the field, despite being only a narrow subset of auditory intelligence. This imbalance largely reflects human reliance on vision over hearing in daily life. Nevertheless, auditory perception plays a critical and often complementary role, especially in situations where visual information is unavailable or obstructed—such as dark environments or occluded scenes. In such contexts, sound provides indispensable cues that enable effective perception and decision-making. Therefore, for machines and robots to attain human-like perceptual capabilities, auditory intelligence must be developed in parallel with visual intelligence.

Sound event detection (SED) is a fundamental task in machine listening that plays a vital role in auditory intelligence, enabling applications in AI-driven perception, smart environments, and bioacoustic monitoring \cite{CASSE, DCASEtask4, crnn, sedmetrics, PSDS, scaper, freqdepinternoise}. Beyond SED, extensive research has been conducted in related auditory tasks, including speech and speaker recognition \cite{specaug, conformer, mpc, wav2vec2.0, hubert, ASP, SAP, tdyaccess, freqse, c2datt}, sound event recognition \cite{PANN, coughcam, etri, AST, beats}, and sound event localization and detection (SELD) \cite{seld2019, starss22, 2022t3report}. Emerging research areas such as automated audio captioning \cite{dcaseaac, clotho, chatgptaugaac}, few-shot bioacoustic detection \cite{dcasebed2024, bioacousticstrf}, and human auditory perception modeling \cite{prtfnet, brainstem, contHRTF} further highlight the broad impact of auditory intelligence across various real-world applications. Additionally, advancements in sound synthesis \cite{audioldm, audiogen, vifs} have explored generative modeling of sound events from textual descriptions, providing new perspectives in sound representation learning. Within this context, SED has emerged as a key technology for identifying and temporally localizing sound events in audio signals. It has been applied in diverse domains such as environmental monitoring, smart home automation, and industrial anomaly detection. Recent advances in deep learning have greatly enhanced SED performance, with convolutional recurrent neural networks (CRNNs) and transformer-based models becoming dominant approaches. However, despite their success, these models still face fundamental challenges in effectively capturing the frequency-dependent characteristics of sound events.

Traditional two-dimensional convolution-based models assume shift invariance along both the time and frequency axes. However, this assumption is inconsistent with the nature of audio signals, where shifting a pattern along the frequency axis alters its perceptual characteristics. As a result, standard two-dimensional convolutions struggle to perform frequency-dependent pattern recognition, leading to suboptimal performance in SED tasks. To overcome these limitations, frequency-dependent convolutional methods have been proposed to adapt convolutional kernels to the spectral characteristics of input signals. Among them, frequency dynamic convolution (FDY conv) has demonstrated significant improvements in SED performance by introducing frequency-adaptive kernels that enhance frequency-dependent feature extraction \cite{FDY}. However, FDY conv also has inherent limitations, including restricted spectral receptive fields, high computational costs, and temporal average pooling performed to obtain frequency-adpative attention weight. To further refine frequency-adaptive modeling, several enhanced architectures have been proposed, including dilated frequency dynamic convolution (DFD conv) \cite{DFD}, partial frequency dynamic convolution (PFD conv) \cite{PFD}, multi-dilated frequency dynamic convolution (MDFD conv) \cite{PFD}, and temporal attention pooling frequency dynamic convolution (TFD conv) \cite{TFD}. These models incrementally address the weaknesses of FDY conv, improving the ability to model diverse spectral-temporal characteristics in sound events.

This dissertation systematically investigates the evolution of frequency-adaptive convolution methods for SED. By analyzing the effectiveness of different convolutional architectures, we aim to provide insights into optimizing frequency-dependent modeling while balancing computational efficiency and detection accuracy.

\section{Challenges in Existing Methods}

SED has significantly advanced with the development of deep learning-based models, particularly convolutional neural networks (CNNs) and transformer-based architectures \cite{crnn, dcase2023a_1st, dcase2024mytechrep}. These models have demonstrated impressive performance improvements in recognizing sound events. Despite these advancements, existing SED methods still face fundamental challenges in effectively capturing the frequency-dependent characteristics of sound events. Traditional SED systems are predominantly based on 2D CNNs, which assume translation equivariance across both the time and frequency dimensions. This assumption is suitable for image processing tasks, where spatial features remain consistent regardless of positional shifts. However, in audio signals, shifting a pattern along the frequency axis alters its perceptual characteristics, making the direct application of 2D CNNs suboptimal for SED. Consequently, CNN-based SED models struggle to learn frequency-dependent representations, leading to degraded performance in detecting spectrally diverse sound events.

To further improve SED performance, pre-trained transformer-based models have been introduced. Transformers leverage self-attention mechanisms to capture long-range dependencies in the time-frequency domain, offering improved contextual understanding. While their positional encoding schemes may allow some degree of frequency-related pattern learning, they do not explicitly address frequency-dependent transformations. Moreover, most transformer-based SED systems still incorporate convolutional front-ends for local feature extraction, which inherit the same limitations of conventional 2D CNNs. In addition, transformer models typically require large-scale datasets and high computational resources, making them less accessible for real-time or resource-constrained applications.

Attention-based methods have been widely explored to enhance CNN architectures by introducing adaptive mechanisms into convolutional operations. The squeeze-and-excitation (SE) module is a lightweight attention mechanism that recalibrates channel-wise feature responses by modeling inter-channel dependencies through global average pooling and two fully connected layers \cite{senet}. The selective kernel (SK) module extends this idea by dynamically selecting among multiple convolutional kernels with different receptive field sizes, allowing the model to adjust its spatial sensitivity according to the input context \cite{SKnet}. Similarly, dynamic convolution (DY conv) generates input-adaptive kernels by computing attention weights over a set of basis kernels and combining them dynamically during inference \cite{dyconv}. This approach was further extended along the temporal axis in speaker verification tasks, resulting in temporal dynamic convolution (TDY conv), which applies time-dependent kernel adaptation to better capture rapidly changing audio patterns \cite{tdycnn}. However, despite their adaptive nature, these methods do not explicitly address the issue of frequency shift-variance in audio signals, and their effectiveness in modeling frequency-dependent patterns remains limited in the context of sound event detection.

Finally, existing models show class-dependent performance variation due to differences in temporal and spectral structure across sound events. For example, quasi-stationary sounds, such as vacuum cleaners and electric shavers, exhibit relatively stable spectral profiles and are well recognized by conventional CNN-based models. In contrast, transient-heavy sounds—such as alarms, door knocks, and speech plosives—require precise temporal modeling, which average pooling mechanisms and coarse temporal resolutions often fail to provide. Although transformer-based approaches have improved temporal modeling capabilities, their advantage across specific sound event types remains inconclusive and inconsistent across datasets.

To address these challenges, it is essential to develop frequency-adaptive SED models that dynamically adjust their feature extraction mechanisms based on spectral content. This dissertation introduces a series of frequency-adaptive convolutional models that systematically overcome the limitations of existing methods. The proposed models aim to balance computational efficiency with robust frequency-dependent modeling, ultimately enhancing the overall performance of SED systems in diverse real-world applications.

\section{Contributions}

This dissertation proposes a series of advancements in frequency-adaptive convolutional methods for SED, ultimately leading to the development of an efficient and high-performing model. The key contributions of this work are as follows.

First, we introduce frequency dynamic Convolution (FDY conv), a novel approach that applies frequency-adaptive kernels to overcome the shift-invariance limitation of traditional 2D convolutions. Unlike conventional convolutions, which enforce translation equivariance across both time and frequency dimensions, FDY conv dynamically adjusts its kernels based on frequency-specific characteristics, significantly improving the recognition of frequency-dependent sound events.

Building on FDY conv, we propose dilated frequency dynamic convolution (DFD conv) to enhance spectral feature extraction by expanding the receptive field along the frequency axis. This allows the model to capture broader spectral patterns, which is particularly beneficial for sound events with rich spectral structures. Additionally, we introduce partial frequency dynamic convolution (PFD conv) and multi-dilated frequency dynamic convolution (MDFD conv), which improve computational efficiency by balancing dynamic and static convolutions. PFD conv selectively applies frequency-adaptive convolutions to only a portion of the feature maps, reducing computational overhead while maintaining strong performance for quasi-stationary sound events. MDFD conv further extends this idea by integrating multiple dilation rates within a single model, enabling robust feature extraction across diverse spectral characteristics.

A major contribution of this work is the introduction of Temporal Attention Pooling (TAP) to address the limitations of previous FDY-based models in detecting transient sound events. Traditional SED models rely on temporal average pooling, which inadequately represents short-duration, high-intensity acoustic events. TAP incorporates multiple attention-based pooling mechanisms, including Time Attention Pooling (TA) which enhances salient temporal regions, Velocity Attention Pooling (VA) which emphasizes transient features by considering temporal differences, and average pooling which captures stationary features. By integrating TAP into FDY conv, we develop the final proposed model, TFD conv, which significantly improves transient sound event detection while maintaining strong performance across other sound categories.

To better understand the impact of these advancements, we conduct a comprehensive class-wise performance analysis. Our results demonstrate that FDY conv is particularly effective for non-stationary sound events with intricate spectral structures, while DFD conv enhances the recognition of broad spectral events. PFD conv optimizes computational efficiency while maintaining strong performance for quasi-stationary sound events, and TFD conv achieves superior results for transient-heavy sound events, such as alarms, speech plosives, and impulsive noises.

Beyond standard SED benchmarks, we validate the practical applicability of our models through two real-world case studies. The first focuses on tank powertrain fault recognition, where FDY conv and varaints are used to detect mechanical failures in tank powertrains based on acoustic signals. The second case study applies our methods to arc detection in offshore environments, identifying abnormal electrical discharge events to prevent fire hazards in marine vessels. These studies highlight the robustness and versatility of proposed frequency-adaptive convolutional models in industrial sound event recognition.

By developing a series of frequency-adaptive convolutional architectures and integrating temporal attention mechanisms, this dissertation makes significant contributions to the field of auditory intelligence. The proposed models enhance frequency-aware feature extraction, improve transient event detection, and offer practical benefits for real-world applications, including industrial fault detection and intelligent audio monitoring systems.

\section{Organization}

This dissertation is organized into six chapters. 

Chapter 2 provides a review of related works in sound event detection (SED), focusing on three key categories: convolution-based methods, frequency-aware approaches, and attention-based techniques. It highlights the limitations of existing models, particularly their inability to capture frequency-dependent variations, which motivates the development of the proposed methods.

Chapter 3 introduces the proposed frequency-adaptive convolutional models. It begins with FDY conv, which applies frequency-adaptive kernels to address the shift-invariance limitation of traditional two-dimensional CNNs. The chapter then presents its extended versions: dilated FDY conv (DFD conv), which expands the receptive field along the frequency axis; partial FDY conv (PFD conv) and multi-dilated FDY conv (MDFD conv), which enhance computational efficiency and model robustness; and finally, TFD conv, which integrates Temporal Attention Pooling (TAP) to improve transient sound event detection. 

Chapter 4 describes the experimental setup and evaluation results. It includes a comprehensive performance comparison of the proposed models against baseline architectures, along with an analysis of frequency-adaptive feature extraction through Principal Component Analysis (PCA) and class-wise performance comparisons. Additionally, computational cost analysis is conducted to assess the efficiency of frequency-adaptive convolutions.

Chapter 5 presents two case studies demonstrating the real-world applicability of the proposed models. The first case study focuses on tank powertrain fault recognition, where the proposed methods are used to detect mechanical failures in powertrains based on acoustic signals. The second case study addresses arc detection in offshore environments, applying frequency-adaptive SED models to identify abnormal electrical discharge events for fire prevention in marine vessels.

Finally, Chapter 6 concludes the dissertation by summarizing the key contributions and discussing potential directions for future research. This chapter outlines opportunities to further enhance frequency-adaptive SED models and expand their applications beyond SED to broader auditory intelligence tasks.

Through this structured approach, this dissertation aims to advance the development of frequency-adaptive modeling in SED, providing a comprehensive framework for improving sound event detection across diverse real-world scenarios.

%% file: sections/2_Relatedworks.tex
\section{Sound Event Detection and Deep Learning}

SED has undergone substantial advancements over the past two decades, particularly with the integration of deep learning techniques. The goal of SED is to detect and temporally localize various sound events occurring in an audio stream, which may be spontaneous, overlapping, and context-dependent. Initially dominated by traditional machine learning pipelines that relied on manually crafted acoustic features and statistical models, the field has now transitioned into a data-driven paradigm powered by neural networks.

While SED shares foundational elements with related speech tasks—such as automatic speech recognition (ASR), speaker verification, and music genre classification—it poses a distinct set of challenges. Unlike ASR, which deals with structured linguistic signals, or speaker verification, which focuses on speaker identity, SED must handle a wide variety of heterogeneous, often overlapping, environmental sound events. These events can be highly non-stationary, vary significantly in duration and spectral content, and may occur simultaneously in noisy or reverberant environments. Therefore, successful SED systems must possess strong generalization capabilities, robust spectral-temporal modeling, and efficient computational strategies.

\subsection{Early Handcrafted Feature-based SED Methods}

Before the emergence of deep learning, traditional SED approaches predominantly relied on hand-engineered features. Popular representations included Mel-Frequency Cepstral Coefficients (MFCCs), log-mel spectrograms, zero-crossing rates, and spectral centroid. These features were designed to capture relevant auditory cues such as pitch, timbre, and formant structure.

Once extracted, features were typically modeled using statistical classifiers. Gaussian Mixture Models (GMMs) and Hidden Markov Models (HMMs) were widely used due to their probabilistic nature and ability to model temporal sequences. For instance, an HMM-based approach would model each sound event as a sequence of acoustic states, using Viterbi decoding to infer event boundaries. These models, however, often suffered from poor generalization when applied to real-world data. Their performance degraded in the presence of overlapping sound events, varying acoustic conditions, and non-stationary noise, primarily because the hand-crafted features could not fully capture the complex spectral-temporal patterns inherent in environmental audio.

\subsection{Evolution of Deep Learning in SED}

The introduction of deep learning revolutionized SED by enabling end-to-end learning of both feature representations and classifiers. CNNs became a foundational tool due to their ability to extract local spectral and temporal features from 2D audio representations such as log-mel spectrograms. CNN-based models were particularly effective in modeling stationary and quasi-stationary sound events, such as engine hums, vacuum cleaners, and alarms \cite{FDY}.

To better handle temporal dynamics, Convolutional Recurrent Neural Networks (CRNNs) were introduced, combining CNN layers for feature extraction with Recurrent Neural Networks (RNNs) for temporal sequence modeling. Common choices for RNN units included Long Short-Term Memory (LSTM) and Gated Recurrent Units (GRU), which are well-suited for learning long-range temporal dependencies in sequential data. CRNNs became the de facto standard for many SED tasks, balancing robustness and efficiency \cite{crnn}.

More recently, Transformer-based models have been applied to SED, following their success in natural language processing (NLP) computer vision and ASR domains. Models such as AST \cite{AST, astsed} and BEATs \cite{beats, dcase2023a_1st, dcase2023a_2nd} have demonstrated strong performance by leveraging self-attention mechanisms to capture global temporal and spectral dependencies. However, these models are computationally intensive and require large-scale pre-training, which may not be feasible for all SED applications, particularly in edge computing or real-time deployment scenarios. Hybrid models that integrate CNN front-ends with Transformer encoders have shown promising results in recent DCASE challenges \cite{dcase2023a_1st, DCASEtask4, dcase2024mytechrep, dcase2024_1st}.

\subsection{Challenges in SED: Overlapping and Diverse Sound Events}

SED poses a series of challenges that differentiate it from other audio tasks. Chief among these is the frequent occurrence of overlapping sound events, where multiple sources may emit sounds concurrently. For example, in a domestic scene, a baby crying may coincide with a microwave beeping and people talking in the background. Standard classification approaches often struggle with such polyphonic events, necessitating multi-label output frameworks and robust time-frequency representation learning.

In addition, sound events vary drastically in duration and spectral content. Some events, such as door knocks or speech plosives, are highly transient, while others like running water or vacuum cleaners are more sustained and stationary. These diverse temporal profiles require models capable of capturing both short-term transients and long-term patterns.

Furthermore, environmental factors—such as background noise, reverberation, microphone placement, and recording conditions—introduce variability that can hinder generalization. This variability becomes particularly problematic in real-world deployment scenarios, where test conditions often differ significantly from training datasets. Addressing this issue requires not only robust modeling architectures but also effective training strategies.

To mitigate these challenges, recent research has explored various techniques:
\begin{itemize}
    \item \textbf{Data Augmentation:} Methods such as mixup \cite{mixup}, SpecAugment \cite{specaug} and FilterAugment \cite{filtaug} have been employed to improve generalization.
    \item \textbf{Semi-Supervised and Self-Supervised Learning:} Leveraging unlabeled audio data through mean-teacher training \cite{meanteacher, DCASEtask4}, self-training \cite{dcase2023a_1st, mytechreport}, or self-supervised learning \cite{matsed, jitter} has shown strong performance gains.
    \item \textbf{Weakly-Supervised Learning:} Using clip-wise weak labels (e.g., presence/absence annotations) to train localization-capable models through multiple instance learning frameworks \cite{DCASEtask4}.
\end{itemize}

Despite these advances, SED remains a challenging task. Researchers continue to explore novel neural architectures, training paradigms, and frequency-aware modeling techniques to achieve a better balance between detection accuracy, model complexity, and real-world robustness. The next sections delve into these innovations in detail, starting with CNN-based methods (Section 2.2), followed by Transformer-based models (Section 2.3) and frequency-adaptive methods (Section 2.4).

\section{Convolution-based Approaches for SED}

CNNs have been widely used in deep learning-based SED due to their effectiveness in learning local spectral-temporal patterns. CNNs apply convolutional filters to 2D time-frequency representations such as spectrograms or log-mel spectrograms, enabling the automatic extraction of hierarchical features across different scales. Their inherent inductive bias towards locality, translation equivariance, and weight sharing makes them particularly effective for detecting structured acoustic patterns such as harmonics or noise bursts.

Since SED requires frame-wise prediction of sound event activity, CNNs are especially well-suited for extracting low-level and mid-level features that are informative for event classification. In recent years, CNN-based architectures have formed the foundation for most state-of-the-art (SOTA) SED systems, often in combination with temporal modeling layers such as RNNs or transformers.

\subsection{Standard CNNs for Spectral Feature Extraction}

The earliest CNN-based approaches to SED treated it as a clip-level or frame-level classification task. These models typically consisted of a series of 2D convolutional layers, followed by fully connected layers for final prediction \cite{PANN}. The input was segmented into fixed-length patches, each labeled according to the presence of sound events. Although these models showed promising results, they had several limitations: they struggled with handling long-duration sound events and failed to capture temporal dynamics adequately, especially in polyphonic or highly variable acoustic environments.

\subsection{Convolutional Recurrent Neural Networks (CRNNs)}

CRNNs were proposed to address the inability of CNNs to model long-range temporal dependencies \cite{crnn}. A typical CRNN consists of a CNN front-end that extracts spatial-spectral features, followed by one or more recurrent layers—commonly Long Short-Term Memory (LSTM) or Gated Recurrent Unit (GRU) layers—that capture sequential patterns over time.

This architecture became the de facto standard in SED, with several DCASE challenge-winning models employing CRNNs as their backbone \cite{dcase2021_1st, dcase2022_1st}. The strength of CRNNs lies in their ability to learn both local time-frequency features (via CNNs) and global temporal context (via RNNs). They are particularly effective for modeling quasi-stationary events (e.g., engine hums) as well as dynamic temporal behaviors (e.g., speech, alarms).

However, CRNNs still have limitations. RNNs suffer from vanishing gradient issues and are inherently sequential, making them computationally expensive and less parallelizable than CNNs or Transformer-based models. Moreover, their fixed memory capacity limits the extent of temporal context that can be captured, especially in long audio clips with complex event compositions.

\subsection{Advanced Convolutional Techniques}

To further enhance the capacity of CNNs for SED, researchers have proposed several architectural modifications. One prominent technique is the selective kernel (SK) convolution, which allows the model to dynamically select among multiple kernel sizes (e.g., 3$\times$3, 5$\times$5) based on the input context \cite{SKnet, dcase2021_1st}. This adaptivity enables the network to adjust its receptive field on the fly, capturing both fine-grained and coarse-grained spectral patterns. SK convolution has shown benefits in capturing spectral variations that differ across sound event classes.

Another enhancement involves incorporating attention mechanisms directly into the convolutional layers. Channel-frequency attention modules learn to assign dynamic weights across frequency bins and channels, enabling the model to emphasize salient frequency regions associated with specific sound events \cite{senet, freqatt}. These modules often use global average pooling, 1D convolution, and gating mechanisms to generate attention maps.

\subsection{Pre-trained CNNs for SED}

Another major trend in recent SED research is the use of pre-trained audio models. Pre-trained Audio Neural Networks (PANNs) \cite{PANN}, trained on large-scale audio datasets like AudioSet, have become a standard backbone for many SED systems. These models are typically fine-tuned on downstream tasks with smaller labeled datasets. The benefit of using PANNs lies in their ability to generalize well across domains and event classes, thanks to their exposure to a broad range of acoustic patterns during pre-training.

In addition to feature reuse, knowledge distillation techniques have been explored to transfer information from larger, transformer-based teacher models to smaller, efficient CNN-based student models \cite{audiodynet}. These distilled models achieve a favorable trade-off between accuracy and computational efficiency, making them suitable for real-time and edge-device applications.

\section{Transformer-based Approaches for SED}

Transformer-based architectures have demonstrated remarkable success in various sequential modeling tasks, particularly in NLP and ASR \cite{transformer, conformer, bert, hubert, wav2vec2.0}. Their ability to capture long-range dependencies through self-attention mechanisms has motivated their application in SED, especially for modeling complex temporal dynamics and polyphonic audio scenes.

Unlike speech, which follows relatively structured linguistic patterns, sound events in SED are highly diverse, non-stationary, and often overlapping. This makes temporal context modeling essential for accurate detection and localization. While RNNs such as LSTMs and GRUs have traditionally been used for this purpose, transformers offer an alternative that allows for parallel computation and greater flexibility in modeling global dependencies. Furthermore, success of transformers in computer vision such as vision transformer (ViT) enabled replacement of CNN module in CRNN structure by pretrained transformer as well \cite{vit, AST}.

Despite their potential, Transformer-based SED models have not always shown consistent superiority over conventional CRNN-based models. Several challenges, such as computational cost, data efficiency, and sensitivity to short-duration events, continue to limit their standalone adoption. However, in recent years, a growing number of studies have proposed hybrid architectures that combine the strengths of CNNs and transformers to achieve state-of-the-art performance in SED.

\subsection{Early Transformer-based SED Models}

One of the first successful applications of Transformer models in SED was the CNN-transformer and CNN-conformer \cite{transformer, conformer, dcase2020_1st}. By replacing RNN module in CRNN structure, it has won 1st place in DCASE 2020 challenge. However, the same structure failed to outperform CRNN in terms of polyphonic sound detection score (PSDS) in the folliwng challenges \cite{PSDS}.

On the other hand, an effort to replace CNN module in CRNN was made using Audio Spectrogram Transformer (AST) \cite{AST}, which adopted the Vision Transformer (ViT) framework for audio classification. AST replaces convolutional front-ends with Transformer encoders, processing log-mel spectrograms as input tokens. While AST demonstrated promising performance on large-scale classification tasks such as AudioSet, its performance on fine-grained, time-resolved SED tasks was less impressive. AST models lacked the temporal resolution and inductive biases necessary for local pattern recognition in complex auditory scenes.

Bidirectional Encoder Representations from Audio Transformers (BEATs) \cite{beats} built upon AST by introducing self-supervised learning and audio-language pretraining strategies. These models improved feature representation quality and generalization. Nevertheless, the direct application of BEATs to SED still encountered difficulties in modeling transient and overlapping sound events, as their temporal modeling was coarse-grained and optimized for tagging tasks rather than fine-scale event localization. These limitations highlighted the need for architectural adaptations that could better handle the specific demands of SED.

\subsection{Hybrid CNN-Transformer Approaches}

To address the shortcomings of pure Transformer models, hybrid architectures that combine CNNs and transformers have gained widespread adoption in recent SED research. These models leverage CNNs to extract local time-frequency patterns and reduce the dimensionality of spectrogram inputs before passing the encoded features to Transformer layers, which model long-range temporal dependencies \cite{dcase2023a_1st, dcase2023b_1st, dcase2024_1st, dcase2024mytechrep}. The division of labor between local feature extraction and global context modeling enables a balance between resolution, context, and efficiency.

One of the earliest hybrid structures was the CNN-Transformer and CNN-Conformer model \cite{transformer, conformer, dcase2020_1st}, which replaced the RNN component of CRNNs with Transformer-based modules. This approach led to first-place results in the DCASE 2020 Challenge. However, in subsequent years, the same architecture failed to outperform strong CRNN baselines, particularly in terms of polyphonic sound detection score (PSDS) \cite{PSDS}, revealing limitations in capturing fine-grained acoustic event boundaries.

Meanwhile, efforts to replace the CNN front-end led to models like the Audio Spectrogram Transformer (AST) \cite{AST}, which adopted the Vision Transformer (ViT) architecture for audio by tokenizing log-mel spectrograms into 2D patches. While AST performed well on large-scale tagging tasks such as AudioSet, its coarse input patching and lack of local inductive bias limited its effectiveness on SED tasks that require fine temporal localization.

To mitigate these issues, BEATs (Bidirectional Encoder Representations from Audio Transformers) \cite{beats} introduced a self-supervised audio-language pretraining framework that improved feature richness and generalization across domains. However, BEATs still relied on relatively large patch sizes (e.g., 160ms), and its performance in event boundary detection remained suboptimal due to insufficient temporal granularity. Nonetheless, hybrid CNN-BEATs structure resulted in outstanding performance in SED by wining 1st rank in DCASE 2023 Challenge \cite{dcase2023a_1st, dcase2023a_2nd}

Building on this, ATST (Audio Teacher-Student Transformer) \cite{ATST, ATSTclipframe, atstsed} proposed a frame-level variant called ATST-Frame, which generates 40ms-resolution embeddings explicitly designed for frame-wise SED. To further enhance performance, ATST introduced a two-stage fine-tuning strategy: (1) stabilizing pseudo labels using a downstream CRNN with frozen ATST-Frame encoder, and (2) jointly fine-tuning both networks using strong regularization techniques such as MeanTeacher and Interpolation Consistency Training (ICT), along with data augmentation methods like frequency warping and mixup. This strategy led to state-of-the-art results on SED \cite{atstsed}.

These developments illustrate the growing importance of hybrid CNN-Transformer architectures in SED. By combining CNN-based inductive bias with the contextual power of transformers—and in some cases, integrating RNNs for sequential refinement—recent systems have achieved significant gains in both accuracy and generalization. In particular, the progression from AST to BEATs to ATST highlights how architectural design and fine-tuning strategies must be aligned with the temporal resolution demands of SED tasks.

\subsection{Full Transformer Approaches}

Recent studies have explored fully Transformer-based architectures for SED that replace CRNN entirely and rely solely on self-attention mechanisms to capture both local and global temporal dependencies. These models are typically trained with self-supervised pretext tasks and fine-tuned on small labeled datasets, making them attractive in low-resource settings. Representative models include MAT-SED, JiTTER, and PMAM, each proposing a distinct pretraining strategy designed to enhance event-aware representation learning.

MAT-SED (Masked Audio Transformer for SED) \cite{matsed} employs a masked block prediction task inspired by masked language modeling. It utilizes PaSST \cite{passt} as an encoder and a transformer with relative positional encoding (RPE) as the context network. The model is trained to reconstruct randomly masked time blocks in spectrogram inputs, which encourages the learning of temporal dependencies and robust representations. While effective, MAT-SED faces two key limitations: (1) the removal of audio segments can disrupt short transient events, and (2) the reconstruction task does not explicitly enforce the temporal order of events.

To address these limitations, JiTTER (Jigsaw Temporal Transformer for Event Reconstruction) \cite{jitter} introduces a hierarchical temporal shuffle reconstruction framework. Instead of masking, JiTTER perturbs the input by shuffling audio segments at two levels: block-level and frame-level. The model is then trained to restore the correct temporal order. Block-level shuffling disrupts long-range event dependencies while preserving local coherence, whereas frame-level shuffling introduces subtle local perturbations to refine transient event modeling. In addition, noise injection is applied to maintain weak structural cues and enhance robustness. JiTTER achieves significant improvements in PSDS over MAT-SED, particularly in detecting events with sharp onset-offset boundaries. The multitask configuration, which combines both shuffle types, was found to yield the best performance, indicating the benefit of modeling temporal structure across multiple scales.

More recently, PMAM (Prototype-based Masked Audio Model) \cite{pmam} proposes a novel self-supervised framework using semantically rich pseudo labels derived from prototypical distributions. While PMAM adopts CNN module with PaSST encoder, it still heavily relies on transformer structures. Unlike masked reconstruction or jigsaw-based ordering, PMAM performs masked prediction of latent prototypes using a Gaussian Mixture Model (GMM). It models frame-level embeddings with soft assignment to multiple prototypes, which better captures the polyphonic nature of SED where multiple sound events can co-occur. A key innovation is the use of prototype-wise binary cross-entropy (BCE) loss instead of InfoNCE, allowing the model to independently learn from multiple prototype associations. PMAM further applies iterative Expectation-Maximization (EM) updates to refine the pseudo labels during training.

\subsection{Challenges of Transformers in SED}

While Transformer-based and hybrid architectures have advanced the state of the art in SED, they also introduce a new set of challenges that distinguish them from convolutional or recurrent models. These limitations must be carefully considered when designing Transformer-based SED systems, particularly for real-world or low-resource deployment scenarios.

\begin{itemize}
    \item \textbf{Computational Complexity:} The self-attention mechanism used in transformers has quadratic complexity with respect to sequence length, leading to substantial memory and computation costs when applied to long-duration audio or high-resolution spectrograms. This presents a barrier for edge deployment and real-time inference, where latency and efficiency are critical.

    \item \textbf{Temporal Resolution and Local Structure:} transformers excel at modeling global context but lack inherent mechanisms for capturing fine-grained local features such as abrupt transients or high-frequency textures. Without a convolutional front-end, purely Transformer-based models like MAT-SED or PMAM may struggle to accurately localize short-duration events (e.g., gunshots, knocks), especially when the input patching is coarse or the masking strategy removes key segments.

    \item \textbf{Pretext Task Sensitivity:} In self-supervised frameworks, the performance of full Transformer models is heavily dependent on the quality of the pretext task. For instance, the masked reconstruction used in MAT-SED may oversmooth representations, while JiTTER's effectiveness hinges on how well temporal ordering is restored. PMAM mitigates this via semantically meaningful prototypes, but at the cost of increased modeling complexity (e.g., EM refinement of GMMs).

    \item \textbf{Data Requirements and Generalization:} Transformer-based models typically require large-scale training data to learn robust representations. When labeled data is limited, self-supervised learning (e.g., JiTTER, PMAM) helps, but fine-tuning still benefits from task-specific annotations. In many practical SED scenarios, domain mismatch or lack of diverse audio samples can lead to overfitting or degraded generalization performance.

    \item \textbf{Design and Integration Complexity:} Hybrid models that combine CNNs, transformers, and RNNs may offer the best of all worlds, but at the cost of architectural complexity. Balancing component depths, attention heads, embedding dimensions, and integration strategies (e.g., skip connections, positional encodings) introduces substantial design overhead. This can increase tuning time and limit model reproducibility.
\end{itemize}

Despite these limitations, Transformer-based approaches remain among the most promising directions for SED research due to their scalability, flexibility, and ability to integrate multimodal information. Ongoing developments in efficient attention mechanisms, local-global hybridization, and task-specific self-supervised learning are expected to further address these challenges and expand the applicability of transformers in event-level audio understanding.

\section{Frequency-dependent Processing in Speaker Verification}

In speaker verification (SV), frequency-adaptive processing has become a key strategy to improve robustness and capture fine-grained speaker-specific information. Below, we describe three representative approaches that have demonstrated strong performance in recent studies.

\subsection{Learnable Frequency Filters}

Conventional speaker verification systems typically use Mel-filterbank features, which apply fixed triangular filters based on perceptual scales. However, these handcrafted filters are not necessarily optimal for SV tasks, especially in terms of capturing narrow-band speaker characteristics. To address this, Learnable Frequency Filters (LFF) were proposed as a trainable front-end module directly applied to the STFT magnitude spectrum \cite{freqfilter}.

Each filter in LFF is defined by learnable parameters representing center frequency and bandwidth, which are optimized end-to-end with the downstream embedding network. Two parameterizations are explored: triangular filters (LFF-T) and bell-shaped filters (LFF-B), both of which allow flexible adjustment in spectral focus. Experiments on VoxCeleb and CN-Celeb datasets demonstrate that LFF consistently outperforms MFBANK in equal error rate (EER), particularly when narrow bandwidths are learned in high-frequency regions. Moreover, LFF incurs much lower computational cost than waveform-based CNN front-ends such as SincNet and Gabor filters, as it operates on precomputed STFTs and does not require small convolution strides.

\subsection{Frequency-wise Squeeze-and-Excitation with Positional Encoding}

Squeeze-and-Excitation (SE) modules are widely used for channel-wise feature recalibration in speaker models. However, the original SE applies the same scalar across all frequencies, neglecting the importance of frequency-local context. To overcome this, frequency-wise SE (fwSE) modules were introduced, which compute attention weights along the frequency axis \cite{freqse}.

Further, frequency position information is injected into the model via learnable frequency positional encodings. These encodings are added to the intermediate feature maps within ResNet blocks, allowing the model to maintain frequency awareness even when using shift-invariant 2D convolutions. This hybrid design (fwSE + positional encoding) was shown to enhance performance on short-duration and cross-lingual SV tasks in SdSVC-21, with notable improvements in both EER and minDCF.

\subsection{Convolutional Channel-Frequency Attention (C2D-Att)}

To achieve fine-grained and efficient frequency adaptation, the C2D-Att module proposes a 2D convolution-based attention mechanism across both channel and frequency dimensions \cite{c2datt}. This approach contrasts with MLP-based attention mechanisms, which are often parameter-heavy. C2D-Att uses temporal pooling to form a channel-frequency matrix, and applies a stack of 2D convolutions to produce attention weights specific to each channel-frequency bin.

This module is integrated into a modified ResNet34 backbone for speaker embedding extraction, replacing standard SE blocks. Extensive experiments on VoxCeleb1 and VoxCeleb2 show that C2D-Att consistently outperforms SE and fwSE in both EER and computational efficiency, with only a small number of additional parameters. Visualization of attention maps confirms that C2D-Att can capture nuanced spectral-temporal patterns associated with speaker identity.

\section{Frequency-dependent Processing in SED}

CNNs have become the backbone of many SED systems due to their strong performance in learning hierarchical patterns from spectrograms. However, conventional CNNs make a critical assumption: convolutional kernels are applied uniformly across both the time and frequency dimensions. While temporal shift-invariance is often desirable and reasonable in audio modeling, frequency shift-invariance is a more problematic assumption. In reality, shifting a sound event in frequency can lead to entirely different perceptual meanings—for instance, moving a dog bark an octave higher results in an unnatural and unrealistic sound.

This discrepancy has prompted researchers to develop frequency-dependent modeling techniques. The central motivation is that sound events exhibit frequency-localized structures, and these structures are non-stationary across the frequency axis. In particular, environmental sounds often concentrate energy in specific spectral bands, and event categories differ markedly in their frequency distribution. As such, frequency-invariant filters may obscure meaningful patterns or introduce spurious correlations. Frequency-adaptive methods aim to overcome this by allowing convolutional kernels or attention mechanisms to vary their behavior depending on the frequency location or content of the input.

\subsection{FilterAugment: Frequency-domain Data Augmentation}

While frequency masking has become a popular data augmentation technique in audio learning due to its simplicity and effectiveness \cite{specaug}, it introduces information loss by entirely zeroing out random frequency bands. To address this limitation and better reflect realistic acoustic environments, FilterAugment \cite{filtaug} was proposed as a frequency-domain augmentation method that simulates acoustic filtering behaviors without requiring detailed signal processing.

FilterAugment is motivated by the observation that in real-world environments, sound signals are often altered by their surrounding physical context. For example, high-frequency components may be attenuated due to air absorption, object occlusion, or wall reflections. These effects can be approximated by applying acoustic filters such as high-pass, low-pass, or band-stop filters. However, designing and applying such filters in training pipelines can be computationally expensive and acoustically inconsistent.

FilterAugment simplifies this idea by randomly amplifying or attenuating energy across random frequency bands in the log-mel spectrogram. This method simulates diverse acoustic filtering conditions while preserving the integrity of the original sound. It allows the model to learn to extract informative cues from broader and potentially less discriminative frequency regions, improving robustness to spectral variation across environments.

Three types of FilterAugment are proposed:
\begin{itemize}
    \item \textbf{Step-type:} Random frequency bands are selected, and each band is assigned a constant weight drawn from a predefined dB range. This leads to abrupt energy transitions at band boundaries, simulating crude filtering conditions.
    
    \item \textbf{Linear-type:} Interpolated weights are applied across band boundaries, generating smoother spectral modulations. This produces more natural distortion and was shown to yield slightly better SED performance than the step type.
    
    \item \textbf{Mixed-type:} During training, each batch is augmented using either the step or linear method with a fixed probability (e.g., 0.7/0.3). However, experiments found that mixed-type augmentation may cause inconsistent regularization, often underperforming both individual methods.
\end{itemize}

Experiments on the DCASE 2021 Task 4 dataset demonstrated that FilterAugment outperforms both the baseline and frequency masking methods. Specifically, linear FilterAugment improved polyphonic sound detection score (PSDS) by 6.5\% over the baseline, whereas frequency masking yielded only a 2.13\% gain. These results validate the effectiveness of filter-based augmentation in training acoustic models to generalize across frequency-varying environments.

Compared to frequency masking, which completely removes information from selected frequency regions, FilterAugment allows partial suppression and even enhancement, more closely mimicking realistic audio variations. By exposing the model to repeated presentations of the same sample with different emphasized bands, FilterAugment encourages robust and redundant encoding of sound events.

Beyond SED, FilterAugment has also been successfully applied to speaker verification tasks, where it achieved lower equal error rate (EER) than models trained with frequency masking. This suggests its potential as a general-purpose frequency-domain augmentation technique across various audio modeling domains.

\subsection{Frequency-wise Squeeze-and-Excitation (Freq-SE)}

One of the earliest frequency-adaptive mechanisms proposed for SED was Frequency-wise Squeeze-and-Excitation (fwSE), a variant of the original SE module \cite{senet, freqse, freqatt}. The original SE module globally pools across spatial dimensions and applies a channel-wise attention mechanism. fwSE adapts this idea by applying global pooling only across the time axis and applying excitation (i.e., weighting) across the frequency axis. This enables the model to selectively enhance or suppress specific frequency bands depending on the context.

In tfwSE \cite{freqatt}, this idea was further extended by conditioning the frequency attention on each time frame independently, allowing the model to capture temporally dynamic frequency saliency. This is particularly helpful for transient sound events or those whose spectral content evolves rapidly over time (e.g., speech, footsteps, alarms).

These methods offer a lightweight and effective way to add frequency sensitivity to standard CNNs, but they operate in a post-activation manner—i.e., they reweight features after convolution. As such, they do not adapt the convolutional kernels themselves, which limits their flexibility in capturing structural variations in spectral input.

\subsection{Transformer-based Frequency-aware Encoding}

Another direction for incorporating frequency sensitivity is through Transformer architectures. The AST-SED framework \cite{astsed} modifies the standard AST by introducing frequency-wise Transformer encoder (FTE) blocks, which apply self-attention along the frequency axis rather than time. This allows the model to explicitly capture harmonic relationships and spectral dependencies that are crucial for separating overlapping events. The FTE module is often combined with standard temporal Transformer layers to form a two-branch encoder structure, achieving improved performance over AST baselines.

\subsection{Multi-dimensional Frequency Dynamic Convolution}

Beyond post-convolutional attention, Multi-Dimensional Frequency Dynamic Convolution (MDFD-Conv) \cite{mdfdy} dynamically generates frequency-aware convolution kernels for each frequency bin, and modulates both the input and output features using attention weights. This allows fine-grained spectral adaptation not only at the kernel level but also across the feature propagation process. When combined with semi-supervised training (e.g., mean teacher), MDFD-Conv demonstrated robust generalization in low-resource settings such as the DESED dataset.

These approaches reflect an ongoing effort to unify attention-based and convolution-based frequency modeling, pushing the boundary of spectral awareness in both architecture and learning strategy.

%% file: sections/3_methodologies.tex
\section{Overview}

SED models rely on time-frequency representations such as Mel spectrograms to extract spectral and temporal features from audio signals. Traditional CNNs have been widely used in SED due to their ability to efficiently learn hierarchical spectral representations. However, conventional 2D convolution modules enforce translation equivariance along both time and frequency axes. This means the output of a convolution with a translated input remains the same as with the original non-translated input, as shown in the following equation:

\begin{equation}
T(F(x)) = F(T(x))
\end{equation}
where $T(x)$ refers to translation, and $F(x)$ refers to any translation-equivariant function. This assumption is suboptimal for SED since frequency shifts carry distinct perceptual and semantic meanings in audio signals. The frequency axis is shift-variant, unlike the time or spatial dimensions, as frequency shifts alter the pitch of sounds, which can drastically change how a sound is perceived.

\begin{figure}[t]
\centerline{\includegraphics[width=10cm]{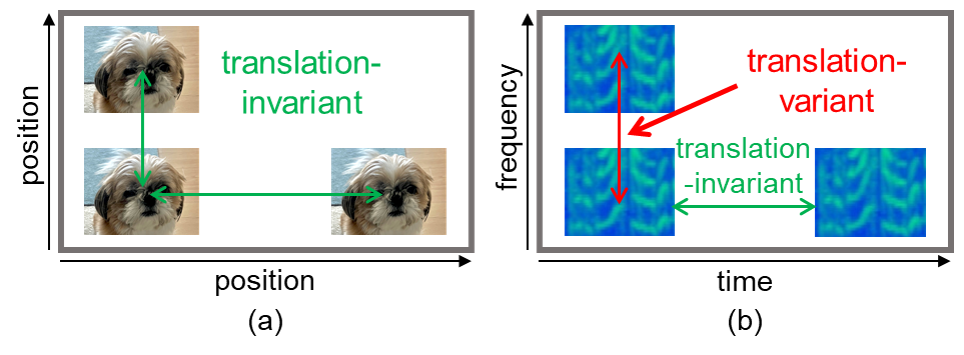}}
\caption{An illustration of shift invariance of (a) image and (b) 2D audio data.}
\label{fig:trsinv}
\end{figure}

As illustrated in Figure \ref{fig:trsinv}, images are shift-invariant along both the position and frequency axes. The same characteristics are preserved when the image is shifted. This property of images makes 2D convolution's translation equivariance adequate. However, for 2D audio data, translation equivariance along the frequency axis is inadequate because shifting a sound event along the frequency axis results in different auditory characteristics, particularly altering its pitch or timbre. Thus, translation equivariance is not suitable for the frequency axis in audio data. To overcome this limitation, frequency-dependent convolutional methods have been developed. These methods introduce spectral adaptability into convolutional architectures, allowing them to better model frequency-dependent features.

Simple frequency-dependent approaches such as frequency-kernel convolution (FK conv) and frequency-weighted convolution (FW conv) aimed to enhance SED models by applying predefined frequency-dependent transformations to convolutional kernels. While FW conv improved feature extraction over traditional 2D convolution, it lacked dynamic kernel modulation, limiting its ability to generalize across diverse sound events. To address these limitations, we propose frequency dynamic convolution (FDY conv), a fully dynamic, frequency-adaptive convolutional framework that allows convolutional kernels to reconstruct spectral feature representations dynamically based on the input frequency characteristics.

Furthermore, we propose three extensions of FDY conv to address its limitations:
\begin{itemize}
\item Dilated frequency dynamic convolution (DFD conv): Expands the receptive field in the frequency domain using multiple dilation sizes, capturing broader spectral dependencies while maintaining parameter efficiency.
\item Partial frequency dynamic convolution (PFD conv): Reduces computational overhead by introducing a hybrid convolutional structure that combines static and dynamic branches, preserving frequency adaptability while lowering complexity.
\item Multi-dimensional frequency dynamic convolution (MDFD conv): Enhances FDY conv by incorporating multiple dilated dynamic kernel branches to improve dynamic feature extraction.
\end{itemize}

To complement frequency-adaptive processing, we enhance temporal modeling using Temporal Attention Pooling (TAP). TAP integrates time attention pooling (TA), velocity attention pooling (VA), and average pooling (AP) to improve the detection of transient and stationary events. This enables SED models to better distinguish between dynamic sound events (e.g., speech, sirens) and steady-state events (e.g., engine noise, background hum).

The methodologies we propose are integrated into a unified framework, combining frequency-adaptive convolutions with enhanced temporal modeling. We validate our approach through extensive experiments on the DESED dataset, showing improvements in overall SED performance and class-wise detection accuracy. The following sections provide a detailed description of each component, beginning with a review of frequency-dependent convolution methods.

\section{Frequency Dependent Convolutions}

Traditional CNNs assume shift invariance along both the time and frequency axes, applying the same convolutional kernel uniformly across all frequency bins. However, in SED, frequency components carry distinct semantic meanings, making this assumption suboptimal. A frequency shift in an audio signal does not have the same perceptual effect as a shift in time, unlike in image processing. This has motivated the development of frequency-dependent convolutional methods, which introduce spectral adaptability into convolutional architectures.

\begin{figure}[t]
\centerline{\includegraphics[width=17cm]{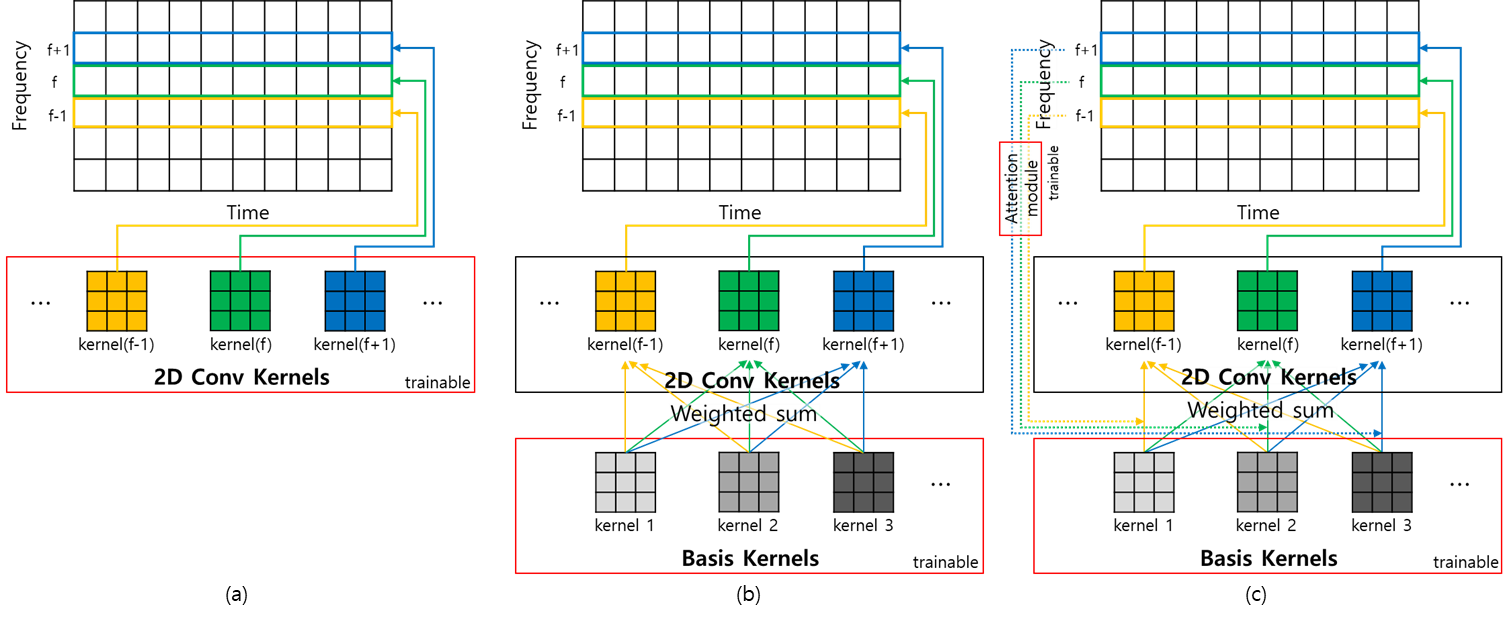}}
\caption{Comparative illustration of frequency dependent convolution methods: (a) frequency-wise kernel convolution, (b) frequency-wise weighted convolution and (c) frequency dynamic convolution.}
\label{fig:fdconvs}
\end{figure}

Several methods have been proposed to make CNN-based feature extraction more frequency-aware, allowing convolutional kernels to dynamically or statically adjust based on frequency characteristics. Frequency-kernel convolution (FK conv) and frequency-weighted convolution (FW conv) aimed to enhance frequency adaptability by applying predefined frequency-dependent transformations. Frequency dynamic convolution (FDY conv) introduced a fully dynamic approach, allowing the convolutional kernels to be adaptively reconstructed based on input frequency characteristics.

\subsection{Frequency-Kernel Convolution}

One of the simplest frequency-dependent CNN methods, frequency-kernel convolution (FK conv), modifies standard convolutional operations by assigning different kernel weights to each frequency bin as shown in figure \ref{fig:fdconvs} (a) \cite{freqdeptalsp}. Instead of applying a single convolutional kernel across all frequency bins, FK conv allocates a unique kernel per frequency bin, improving spectral feature extraction compared to conventional CNNs. This method allows the model to account for frequency-specific variations, making it better suited for SED than standard CNNs.

However, FK conv has a major drawback: the large number of parameters that need to be trained. Since FK conv assigns a unique convolution kernel to each frequency bin, the number of kernels scales linearly with the number of frequency bins. This results in a significant increase in the parameter count compared to standard convolutional models. As the number of kernels increases, so does the complexity of the model, which not only requires more computational resources but also increases the risk of overfitting.

For models trained on relatively small datasets, this excessive parameterization can be problematic. Insufficient data may not be able to effectively train all the convolution kernels assigned to each frequency bin, leading to underutilization of the model's capacity. In particular, for tasks such as SED, where the diversity of sound events can vary greatly, it becomes difficult to generalize across different sound patterns if the training data is not sufficiently representative of all possible frequency characteristics. Consequently, the model might struggle to learn the optimal weights for each kernel, potentially limiting the performance and robustness of the learned features.

Furthermore, as the model complexity increases, training time and memory consumption also grow. This becomes a practical challenge when working with real-time or resource-constrained environments, where the need for both high performance and efficiency is critical. Therefore, while FK conv improves the frequency-adaptive feature extraction, its scalability remains a key limitation, particularly in the context of large-scale and diverse sound event datasets.

\subsection{Frequency-Weighted Convolution}

To improve upon the limitations of FK conv, frequency-weighted convolution (FW conv) was introduced. FW conv incorporates trainable frequency-dependent weighting functions, which apply dynamic scaling factors to the convolution kernels based on the frequency content of the input as shown in figure \ref{fig:fdconvs} (b) \cite{freqdeptalsp}. Unlike FK conv, which assigns a distinct kernel for each frequency bin, FW conv uses a weighted sum of basis kernels. This weighted sum is determined by the learned frequency-specific weights, allowing the model to dynamically adjust the convolutional response to match the spectral characteristics of the input data.

One of the primary advantages of FW conv is its ability to reduce the number of parameters compared to FK conv, as it avoids the need to learn a separate kernel for each frequency bin. By instead learning a smaller set of frequency-dependent weights, FW conv significantly decreases the overall model complexity while still maintaining the flexibility of frequency-dependent processing. This reduction in parameters is particularly beneficial when training on limited datasets, as it reduces the risk of overfitting and helps to generalize better to unseen data.

However, despite these improvements, FW conv still relies on fixed kernel structures. The key limitation of FW conv is that the frequency-dependent weights are static, meaning that they are fixed once learned and do not change during inference based on the input frequency content. While the weights are learned during training, they do not adapt dynamically to the varying spectral content of different sound events during inference. This static nature of the frequency weights restricts FW conv's ability to fully adapt to the complexity of frequency-dependent sound patterns, especially in environments where the spectral characteristics of sound events vary significantly. As a result, the model may struggle to handle more diverse or overlapping sound events, limiting its ability to generalize to a wide range of acoustic environments.

\subsection{Towards fully Dynamic Frequency-Dependent Convolution}

Fully dynamic frequency-adaptive convolutional methods, where kernels are dynamically generated based on the spectral content of the input, offer a significant improvement over traditional approaches. Frequency dynamic convolution (FDY conv) represents a key advancement in this direction, enabling convolutional kernels to dynamically adapt the weighting for basis kernels depending on the frequency contents, as shown in figure \ref{fig:fdconvs} (c) \cite{FDY, freqdeptalsp}.

Unlike FK conv and FW conv, FDY conv introduces a frequency adaptive attention weight to modulate the basis kernels, allowing for dynamic adjustment based on the input's frequency spectrum. In FW conv, the approach of combining basis kernels remains similar, but the key difference lies in the use of frequency-specific attention weights. While FW conv uses static frequency-dependent weights applied across frequency bins, FDY conv uses attention weights that allow the model to adaptively combine the basis kernels according to the input's frequency content. This enables FDY conv to capture more complex spectral features and better model frequency-dependent characteristics, making it more effective for non-stationary and overlapping sound events.

\subsection{Results on Frequency Dependent Convolutions}
\input{figs/table_freqdepconvs}
We assessed and compared the efficacy of three convolution methods employing frequency-dependent kernels on the baseline SED model, presenting the performance of each method in Table \ref{tab:freqdepconvs}. The results indicate a decrease in performance with FK conv (-14.7\%), a slight enhancement with FW conv (+3.72\%), and a significant improvement with FDY conv (+8.95\%). Comparing FK conv and FW conv performances suggests that employing separate kernels for each frequency bin is not effective. While time-frequency patterns depend on frequency, they also exhibit similarities. Adjacent frequency bins share similarities, allowing for recognition of sound information even with slight pitch shifts. In this context, FW conv outperforms the baseline by employing frequency-dependent kernels with shared basis kernels. Each frequency bin utilizes kernels that are linear combinations of basis kernels, enabling learning of shared sound patterns across the frequency dimension.

In FDY conv, the attention module infers frequency contents from the convolution input and applies an appropriate frequency-dependent kernel to extract sound information effectively, outperforming the baseline by 8.95\%. This underscores the effectiveness of the attention mechanism, which adapts to the convolution input contents. Additionally, considering the principle of superposition in acoustic waves, where simultaneous sound events and scenes can be described as a superposition of waveforms, using a weighted sum of basis kernels can be seen as recognizing various superposed sound event patterns. This elucidates the synergy between the attention mechanism and FW conv: the attention mechanism identifies which patterns should be emphasized given the data, and FW conv extracts corresponding sound information and superposes it accordingly.

\begin{figure}[t]
\centerline{\includegraphics[width=13cm]{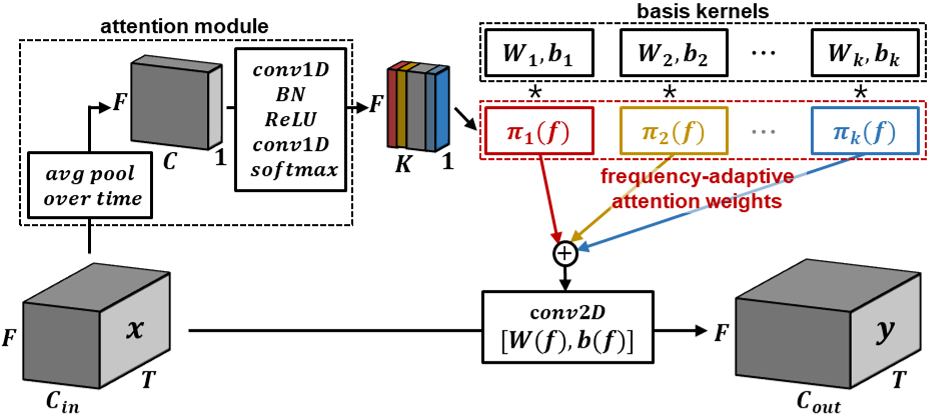}}
\caption{An illustration of frequency dynamic convolution operation. $x$ and $y$ are input and output of frequency dynamic convolution layer. $T$, $F$ and $C$ are dimension size of time, frequency and channel. $K$ is number of basis kernels, $W_i$ and $b_i$ are weight and bias of $i$th basis kernel and $\pi_i(f)$ is frequency-adaptive attention weight for $i$th basis kernel.}
\label{fig:fdyconv}
\end{figure}

\section{Frequency Dynamic Convolution (FDY conv)}

Traditional convolution-based SED models apply fixed convolutional kernels across all frequency bins, assuming shift invariance along both the time and frequency axes. This assumption implies that the convolutional operation generates the same output when the input is translated along the time or frequency dimension. However, in the context of audio signals, frequency components carry distinct perceptual meanings, and shifting these components along the frequency axis can result in significant changes in the sound’s characteristics, such as pitch shift. As a result, the assumption of shift invariance is suboptimal for modeling audio signals, especially in SED tasks where frequency-dependent characteristics play a critical role.

To address this issue, we propose frequency dynamic convolution (FDY conv), a fully dynamic frequency-adaptive convolutional framework. Unlike traditional convolutional methods, FDY conv allows convolutional kernels to adaptively reconstruct spectral feature representations based on the input frequency characteristics as illustrated in figure \ref{fig:fdyconv}. This enables the model to apply frequency-specific kernel weights that adjust to the spectral content of the input, improving the model’s ability to capture complex frequency-dependent patterns.

In FDY conv, the convolutional operation is performed by learning a set of basis kernels, which are then combined in a frequency-specific manner based on attention weights. These attention weights dynamically adjust the combination of the basis kernels depending on the input's frequency content. This process allows FDY conv to adaptively select and combine filters that best match the spectral features of the input, making it more effective at capturing non-stationary and overlapping sound events.

\subsection{Concept and Mechanism}

Unlike previous approaches such as frequency-kernel convolution (FK conv) and frequency-weighted convolution (FW conv), which rely on predefined or static transformations applied to frequency bins, FDY conv introduces a fully dynamic mechanism to adaptively construct frequency-dependent kernels. While FK conv assigns fixed kernels to individual frequency bins and FW conv applies static frequency-specific scaling factors to convolution kernels, FDY conv enables the model to learn the optimal frequency-adaptive kernels on-the-fly during training. This dynamic learning mechanism makes FDY conv highly adaptable to the varying spectral content of input audio, offering superior flexibility compared to static methods.

In FDY conv, the process starts with learning a set of basis kernels that serve as fundamental components for convolution. The model then computes frequency-specific attention weights, which determine how much each basis kernel should contribute to the final convolutional filter. This weight assignment is computed using a learnable attention mechanism that considers the spectral characteristics of the input data, allowing the model to assign greater importance to the relevant basis kernels for each frequency range. Once the attention weights are calculated, FDY conv generates the final convolutional filter by applying a weighted sum of the basis kernels. Each kernel is weighted based on its relevance to the specific frequency content of the input. This means that the convolutional filter is not static but dynamically reconstructed based on the spectral features of the input, making the convolution process more adaptive to different sound events. The dynamic nature of FDY conv allows it to adjust the receptive field across different frequency regions, enabling the model to focus more on the frequency bands that carry significant information for the detection task while ignoring irrelevant frequencies.

This ability to dynamically adjust the receptive field and frequency-specific weighting enables FDY conv to be particularly robust in detecting non-stationary and overlapping sound events, where the frequency content can change over time or different sound sources overlap. By learning frequency-specific attention weights, FDY conv can more effectively capture the complex spectral dependencies in audio signals, which are crucial for accurate sound event detection.

Thus, FDY conv represents a significant step forward in frequency-dependent convolutional methods, as it allows the convolutional kernels to adaptively respond to the changing frequency patterns in input signals, enabling better performance in real-world, complex acoustic environments.

\subsection{Mathematical Formulation}

Given an input feature map $\mathbf{x} \in \mathbb{R}^{B \times C \times F \times T}$, where $B$ is the batch size, $C$ is the number of input channels, $F$ is the number of frequency bins, and $T$ is the number of time frames, the frequency dynamic convolution (FDY conv) operation is defined as:

\begin{equation}
    \mathbf{y(f)} = \left(\sum_{i=1}^{K} \mathbf{W}_i \odot \mathbf{\pi(f, x)}_i\right) * \mathbf{x(f)} + \sum_{i=1}^{K} \mathbf{b}_i
\end{equation}

\noindent where:
\begin{itemize}
    \item $\mathbf{W}_i \in \mathbb{R}^{C_{in} \times C_{out} \times h \times w}$ and $\mathbf{b}_i \in \mathbb{R}^{\times C_{out}}$ represents the $i$-th basis kernel.
    \item $\mathbf{\pi(x)}_i \in \mathbb{R}^{B \times 1 \times F \times 1}$ represents the frequency-wise attention weight for kernel $i$, which modulates the contribution of each basis kernel.
    \item $\odot$ denotes element-wise multiplication, ensuring frequency-specific modulation.
    \item $*$ denotes the convolution operation, which applies the frequency-modulated kernel to the input feature map.
    \item $K$ is the total number of basis kernels.
\end{itemize}

\noindent This formulation describes how the FDY conv layer uses dynamically learned attention weights to adjust the contribution of each basis kernel, based on the frequency content of the input. The learned basis kernels are weighted according to the frequency-specific attention weights, allowing the model to focus on the most relevant frequency regions for each sound event.

\noindent \textbf{Efficient Implementation:}  
The procedure described above illustrates the concept of frequency dynamic convolution in a simple way, but the official implementation of FDY conv uses a slightly different procedure to reduce computational cost, as described in previous works \cite{tdycnn, FDY}. While the algorithm still extracts frequency-adaptive attention weights in the same way, the application of convolution kernels is optimized for computational efficiency.

In the actual algorithm, outputs from each basis kernel are obtained first as in Equation (\ref{eqn:fdy_eff_1}), and then a weighted sum is applied as shown in Equation (\ref{eqn:fdy_eff_2}) to calculate the final output:

\begin{equation}
    y_i(f) = W_i * x(f) + b_i
\label{eqn:fdy_eff_1}
\end{equation}
\begin{equation}
    y(f,x) = \sum_{i=1}^{K}\pi_i(f,x) \odot y_i(f)
\label{eqn:fdy_eff_2}
\end{equation}

\noindent where:
\begin{itemize}
    \item $t$ is the time index, $f$ is the frequency index, and $x$ and $y$ are the input and output of a single frequency dynamic convolution layer.
    \item $W_i$ and $b_i$ are the weight and bias for the $i$-th basis kernel.
    \item $y_i$ represents the output from the $i$-th basis kernel.
    \item $\pi_i(f,x)$ is the frequency-adaptive attention weight for the $i$-th basis kernel.
    \item $K$ is the total number of basis kernels.
\end{itemize}

\noindent This procedure is equivalent to the one described earlier, but with optimized computational efficiency, reducing the complexity by applying dynamic kernel weights after calculating outputs from individual basis kernels. This allows FDY conv to maintain high adaptability while being computationally more efficient than directly applying frequency-dependent convolutions in a fully dynamic manner across the entire input.

\subsection{Performance Analysis}
Experimental results on the DESED dataset show that FDY conv outperforms standard CNN-based models by 8.95\% in polyphonic sound detection score (PSDS), proving its effectiveness in frequency-adaptive learning. Table \ref{tab:freqdepconvs} demonstrates the improvement in performance when applying FDY conv, highlighting the advantages of dynamically adjusting the convolutional kernels according to the frequency content of the input audio. FDY conv is superior in ability to learn frequency-dependent features compared to conventional CNN-based methods. This improvement underscores the importance of incorporating frequency-adaptive convolutional mechanisms for sound event detection tasks, where frequency-dependent characteristics are crucial for accurate classification.

\subsection{PCA Analysis on Attention Weights}

\begin{figure}[t]
\centerline{\includegraphics[width=\linewidth]{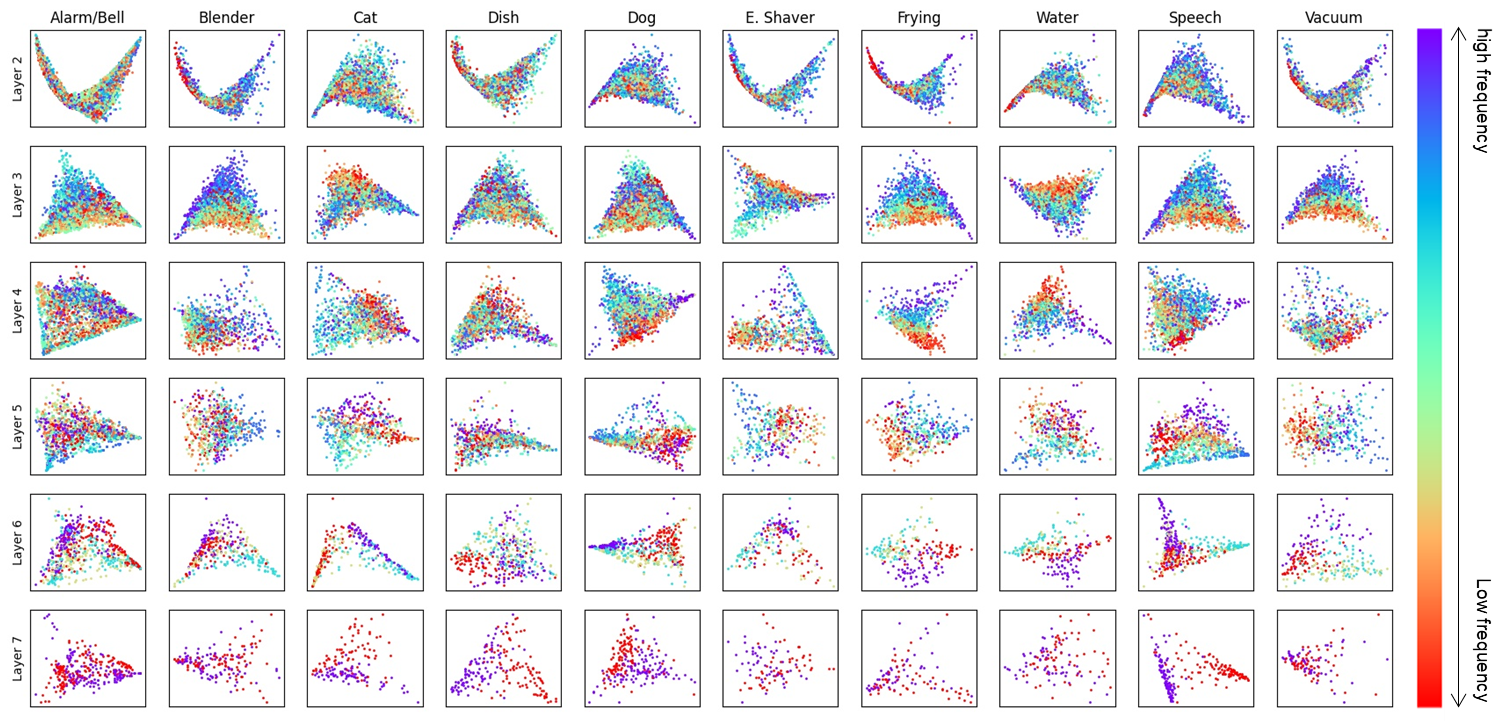}}
\caption{Examples of PCA analysis on the attention weights used in frequency dynamic convolution.}
\label{fig:pca}
\end{figure}

Figure \ref{fig:pca} displays PCA analysis plots of the frequency-adaptive attention weights used in FDY conv. Each plot shows the first two principal components of the attention weights on the x and y axes. Each column corresponds to a specific sound event class, and each row represents a specific FDY conv layer. The frequency bins are represented by colored dots, with spectrum colors ranging from red to purple representing low to high frequencies, respectively.

From the plots, it is observed that dots with similar colors tend to cluster together, suggesting that FDY conv applies similar kernels to adjacent frequency bins for the same sound event class. This indicates that FDY conv learns frequency-dependent attention weights that adaptively assign kernels based on the spectral characteristics of each sound event class. If the attention weights were independent of frequency, the PCA plots would appear more random, making it difficult to observe structured spectral patterns within the data.

More specifically, we observe a clearer distinction between the colors in the later layers of FDY conv. This suggests that the frequency-dependent attention weights become more distinctive across frequency bins as the layers progress. In the early layers (Layer 2), the plots show dots of various colors spread out, but there is some indication of color patterns emerging. However, from Layer 3 onward, the similar colors are grouped together, demonstrating the increasing specialization of frequency-specific features. This progression indicates that FDY conv is effectively learning how to prioritize certain frequency bands, adapting dynamically as the depth of the network increases.

The results from the PCA analysis highlight the importance of frequency-dependent attention weights in improving the model’s ability to identify and classify different sound event classes. The spectral flow patterns observed in the plots suggest that FDY conv produces frequency-adaptive convolution kernels that are optimized for the unique frequency characteristics of each sound event class, significantly enhancing the model's ability to handle complex acoustic scenes.

\begin{figure}[t]
\centerline{\includegraphics[width=13cm]{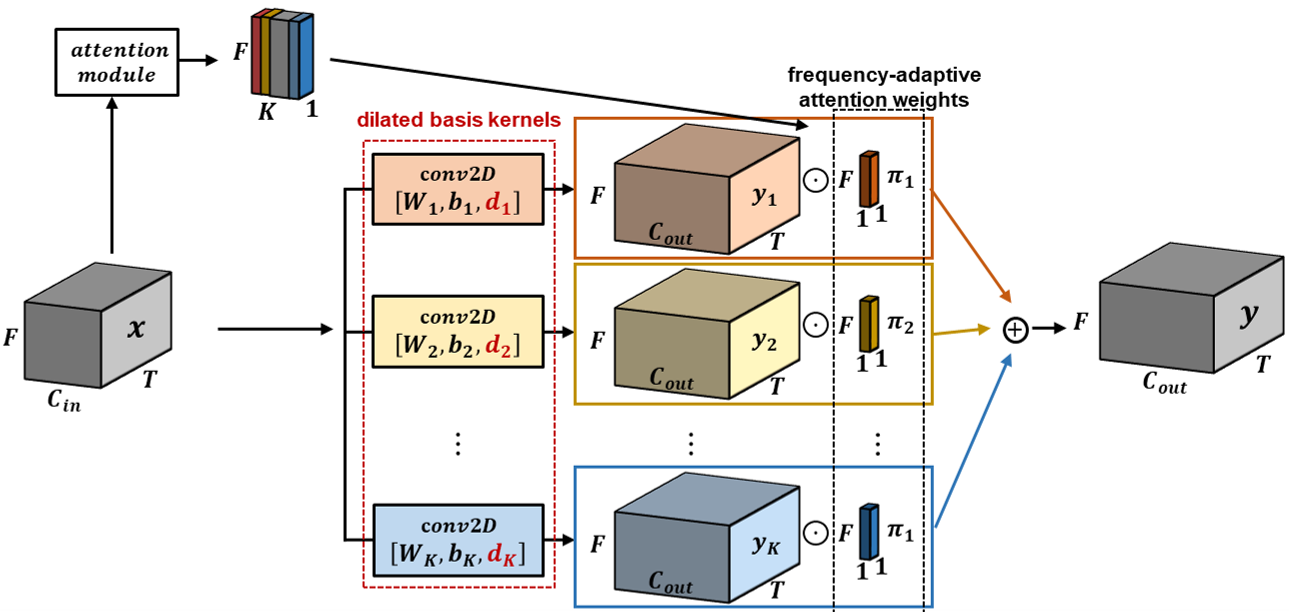}}
\caption{An illustration of dilated frequency dynamic convolution operation. $x$ and $y$ are input and output of DFD conv. $K$ is number of basis kernels, $W_i$ and $b_i$, $d_i$ and $\pi_i$ are weight, bias, dilation size and frequency-adaptive attention weight for $i$-th basis kernel.}
\label{fig:DFDconv}
\end{figure}

\section{Dilated Frequency Dynamic Convolution (DFD conv)}
While FDY conv effectively adapts convolutional kernels to input frequency characteristics, it relies on a fixed receptive field size (e.g., $3 \times 3$ kernels), which may limit its ability to capture diverse frequency-dependent patterns. Additionally, since all $K$ basis kernels are structurally the same, their roles are not explicitly different. This lack of specialization between kernels might hinder FDY conv's performance in capturing more complex spectral variations. To address these limitations, we propose dilated frequency dynamic convolution (DFD conv), an extension of FDY conv that introduces dilation to its basis kernels, thereby expanding the receptive field without increasing the number of trainable parameters as shown in figure \ref{fig:DFDconv} \cite{DFD}. This approach allows the model to capture broader spectral dependencies, improving performance across a wider range of sound events.

\subsection{Concept and Mechanism}
DFD conv applies dilated convolutions to the basis kernels of FDY conv, enabling the model to increase the receptive field without the need for additional parameters. Dilated convolutions work by spreading the kernel over a larger region of the input data, allowing the model to capture larger-scale spectral patterns. This expansion of the receptive field helps the model learn longer-range dependencies without increasing the computational burden, making the approach more efficient than using larger kernels.

Unlike selective kernel (SK) convolution, which selects a kernel from a fixed set (e.g., $3 \times 3$ or $5 \times 5$), DFD conv generates a diverse range of frequency-adaptive kernels dynamically \cite{SKnet}. The dynamic nature of DFD conv enables the model to adaptively choose the most suitable kernel for each frequency range. This capability allows better feature extraction for frequency-dependent patterns, particularly in complex acoustic environments.

The main distinction of DFD conv lies in the application of varying dilation sizes to different basis kernels. In traditional dilated convolutions, a single dilation size is applied uniformly across all frequency bins. In contrast, DFD conv uses multiple dilation rates for different kernels, allowing each kernel to specialize in different frequency ranges. For example:
\begin{itemize}
    \item Smaller dilation sizes capture fine-grained frequency details, which are important for detecting subtle changes in sound characteristics.
    \item Larger dilation sizes capture broader spectral information, useful for detecting more stable or global patterns in sound events.
\end{itemize}
This variability in dilation sizes improves the model's ability to extract both local and global spectral features, making it highly effective for polyphonic sound events (where multiple sound sources overlap) and non-stationary events (where the spectral content evolves over time).

\subsection{Mathematical Formulation}
In the DFD conv process illustrated in Figure \ref{fig:DFDconv}, $\mathbf{x}$ and $\mathbf{y}$ represent the input and output of DFD conv, respectively. The dimensions of the input are denoted as $T$, $F$, $C_{\text{in}}$, and $C_{\text{out}}$, representing time, frequency, and input/output channel sizes. It is important to note that both FDY conv and DFD conv do not alter the input size, maintaining the same dimensions for the output as the input.

The DFD conv operation begins by convolving the input $\mathbf{x}$ with $K$ basis kernels, each with its corresponding dilation size $d_i$, producing $K$ intermediate outputs $y_i$ for $i = 1, 2, ..., K$. The operation is described as:

\begin{equation}
    y_i = \mathbf{W}_i * \mathbf{x} + \mathbf{b}_i
\end{equation}

where $\mathbf{W}_i$ represents the $i$-th basis kernel, $\mathbf{b}_i$ is the corresponding bias, and $*$ denotes the convolution operation. The dilation size $d_i$ is used to expand the receptive field of each basis kernel, allowing it to capture larger-scale spectral dependencies without increasing the number of trainable parameters.

After obtaining the intermediate outputs $y_i$, each output is modulated by frequency-adaptive attention weights $\pi_i(f, x)$, which are learned during training. These attention weights allow the model to dynamically adjust the contribution of each basis kernel based on the frequency content of the input. The output after applying the attention weights is given by:

\begin{equation}
    y_i^{\text{mod}} = \pi_i(f, x) \odot y_i
\end{equation}

where $\odot$ denotes element-wise multiplication. The attention weights $\pi_i(f, x)$ enable the model to prioritize certain frequency bands over others, making the convolution operation frequency-adaptive.

Finally, the model computes the final output $\mathbf{y}$ by applying a weighted sum of the intermediate outputs $y_i^{\text{mod}}$:

\begin{equation}
    \mathbf{y}(f, x) = \sum_{i=1}^{K} y_i^{\text{mod}}
\end{equation}

This operation combines the contributions of all basis kernels, each modulated by its corresponding attention weight, to produce the final frequency-adaptive representation.

In summary, the DFD conv process involves convolving the input with multiple basis kernels, adjusting these kernels using frequency-dependent attention weights, and then combining the weighted outputs to generate a dynamic frequency response. This process allows the model to dynamically adjust its receptive field based on the input’s frequency characteristics, providing flexibility and improving the detection of frequency-dependent patterns in sound events.

The dilation size $d_i$ allows the model to expand the receptive field of each basis kernel, capturing larger-scale spectral patterns while maintaining the number of trainable parameters. This enables DFD conv to capture both fine-grained and broad spectral features, making it highly effective for detecting non-stationary sound events, where spectral content evolves over time.

\input{figs/table_kernelsize}
\subsection{Preliminary Experiments on Receptive Field Sizes}
To study the effect of receptive field size in convolution modules, we conducted preliminary experiments using a CRNN model with conventional 2D convolution modules, testing larger kernel sizes and dilation sizes along the frequency dimension. The results of these experiments, presented in Table \ref{tab:size_study}, show that increasing the kernel size along the frequency dimension leads to improved performance, as a larger receptive field allows the model to capture more spectral information. Specifically, dilated kernels with a dilation size of 2 slightly improved performance, while dilated kernels with a size of 3 actually worsened the performance. This suggests that although larger kernels can enhance performance by capturing broader spectral patterns, they also lead to a significant increase in the model size, making the model less efficient.

On the other hand, applying dilation to the kernels only marginally improved performance, yet did not involve an increase in the number of parameters. This indicates that dilated convolutions offer a computationally efficient way to increase the receptive field without adding to the model’s parameter complexity. Based on these findings, we hypothesize that by combining FDY conv with varying dilation sizes applied to different basis kernels, we could achieve the benefits of both larger receptive fields and parameter efficiency. The frequency-adaptive kernels in FDY conv, formed by the weighted sum of basis kernels with different dilation sizes, would provide a receptive field as large as a $3 \times 5$ or $3 \times 7$ kernel without increasing the model size. This would allow the model to capture broader spectral features while maintaining computational efficiency, combining the advantages of both larger kernel sizes and dilated convolutions.

\input{figs/table_dfdtimefreq}
\subsection{Experiments on Dilation Dimensions}
To further investigate the effect of dilation on the frequency and time dimensions, we performed experiments comparing the PSDS of DFD-CRNNs, where one basis kernel was dilated with size two along either the time or frequency dimension. In these experiments, we note that a dilation size of one implies no dilation. Additionally, we experimented with a DFD-CRNN model that applied dilation on both time and frequency dimensions, by using one time-dilated basis kernel and one frequency-dilated basis kernel, both with a dilation size of two. The results are shown in Table \ref{tab:tf_results}, where we observed that applying the frequency-dilated basis kernel improved the PSDS by 1\%, whereas the time-dilated and both-dilated basis kernels adversely affected the performance. This finding aligns with the previous conclusion from section 4.1, which indicated that dilation is only advantageous when applied to the frequency dimension, and time dilation does not significantly improve performance.

Furthermore, we conducted additional experiments with DFD-CRNN using five basis kernels to examine whether it is more beneficial to dilate new basis kernels or to dilate the existing ones. The results, shown in Table \ref{tab:tf_results}, are consistent with our earlier experiments: adding a frequency-dilated basis kernel improved performance, while adding a time-dilated basis kernel did not make a significant difference. However, the performance gain from adding a dilated basis kernel was not substantial when compared to dilating existing basis kernels. Moreover, introducing new basis kernels increases the number of parameters, which can lead to additional computational cost without yielding significant performance improvements. Therefore, we concluded that dilation improves SED performance when applied to the frequency dimension rather than the time dimension, and it is more efficient to dilate existing basis kernels instead of introducing new dilated kernels.

\input{figs/table_dfdsize}
\subsection{Experiments on Varying Dilation Sizes}
Experiments were conducted on DFD-CRNN with various dilation sizes along the frequency dimension to evaluate their impact on SED performance. The results are summarized in Table \ref{tab:size_results}, with parameter counts omitted, as all models in the table share the same parameter counts. In the first set of experiments, we tested the effect of using one, two, three, and four basis kernels, each with a dilation size of two, to explore how the number of dilated basis kernels influences performance. The results show that using two basis kernels with dilation size of two along the frequency dimension yields the best performance. However, applying dilation to all basis kernels resulted in a performance decrease, indicating that the model benefits from diversifying the dilation sizes across kernels rather than applying the same dilation size to all of them.

Next, we tested the effect of applying dilation sizes of three and four to one or two basis kernels. As shown in Table \ref{tab:size_results}, the results demonstrate that, when applied to one or two basis kernels, a dilation size of three performs similarly to a dilation size of two. However, applying a dilation size of four resulted in limited performance gain, primarily due to the sparsity of the convolution kernel. The larger dilation leads to broader receptive fields, but this can also dilute the model's ability to capture fine-grained frequency details, reducing its effectiveness in detecting smaller-scale features in the audio data.

Furthermore, we experimented with applying varied dilation sizes across different basis kernels, as shown in the lower part of Table \ref{tab:size_results}. The best results were achieved when applying one basis kernel with no dilation, one basis kernel with dilation size of two, and two basis kernels with dilation size of three. This combination outperformed the baseline FDY-CRNN by 2.43\% in PSDS, demonstrating the effectiveness of applying different dilation sizes to various kernels. We had initially expected that using four basis kernels with dilation sizes ranging from one to four would produce a more diverse set of frequency-adaptive kernels, but it turned out to be less effective. This result aligns with the previous experiment where DFD-CRNN with a dilation size of four showed only insignificant improvement, indicating that excessive dilation can be counterproductive in certain cases.

In conclusion, our experiments show that varying the dilation sizes across basis kernels improves SED performance by diversifying the frequency-adaptive kernels. While larger dilation sizes can capture broader spectral features, smaller dilation sizes are essential for capturing fine-grained frequency details, and applying too many large dilations can lead to diminishing returns. The optimal combination of dilation sizes, therefore, strikes a balance between capturing both local and global frequency features, ensuring better performance across a wide range of sound events.

\begin{figure}[t]
\centerline{\includegraphics[width=12cm]{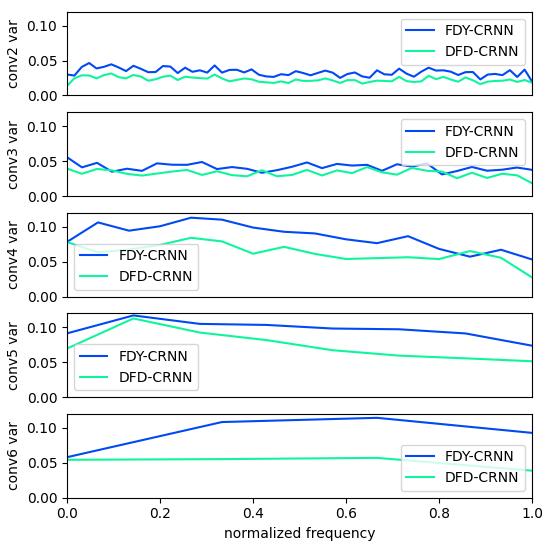}}
\caption{Plots comparing variance of attention weights on 2nd - 6th convolution layers in FDY-CRNN and DFD-CRNN.}
\label{fig:attvar}
\end{figure}

\subsection{Attention Weight Variance Comparison}
To further analyze the effect of diversifying basis kernels with dilation, we compared the variance of attention weight vectors across the 2nd to 6th dynamic convolution layers of FDY-CRNN and DFD-CRNN with dilation sizes of 1, 2, 3, and 3 along the frequency dimension, as illustrated in Fig. \ref{fig:attvar}. The purpose of this analysis is to evaluate how the diversification of basis kernels through dilation affects the variability of the learned attention weights, which is crucial for understanding the model’s frequency-adaptive behavior.

The variance of the attention weight vectors was computed by feeding the foreground train soundbank from the DESED dataset into both FDY-CRNN and DFD-CRNN. The attention weight vectors, denoted as $\mathbf{w} \in \mathbb{R}^K$, are produced by the attention module in Fig. \ref{fig:DFDconv} along a single frequency bin. The distance variance of these vectors was calculated for each frequency bin in each convolution layer using the following equation:

\begin{equation}
\mathrm{var}_{lf} = \frac{1}{N}\sum_{i=1}^{N} \left\Vert \frac{1}{N}\sum_{j=1}^{N} \mathbf{w_{jlf}} - \mathbf{w_{ilf}} \right\Vert _2 ^2
\end{equation}
where:
\begin{itemize}
    \item $l$ is the index of the convolution layer,
    \item $f$ is the frequency index,
    \item $N$ is the total number of samples in the foreground train soundbank.
\end{itemize}

We observed that, except for a few exceptions in the 3rd and 4th convolution layers, DFD-CRNN consistently shows smaller variance in attention weights compared to FDY-CRNN. This suggests that DFD-CRNN's frequency-adaptive kernels are more stable in their attention, allowing them to learn more consistent representations across different frequency bins. The fact that DFD-CRNN outperforms FDY-CRNN in terms of SED performance further supports the idea that the basis kernels in DFD conv are effectively diversified, leading to more stable attention weights.

By reducing the variance in attention weights, DFD conv is able to perform SED more effectively, as the frequency-adaptive kernels learn more robust and consistent frequency-specific features. This behavior is especially beneficial in polyphonic and overlapping sound events, where multiple sound sources must be identified with varying frequency characteristics. In conclusion, we can infer that the diversification of basis kernels through dilation in DFD conv enables the model to learn better frequency-adaptive representations, ultimately leading to improved SED performance with more stable attention weights.

\begin{figure}[t]
\centerline{\includegraphics[width=10cm]{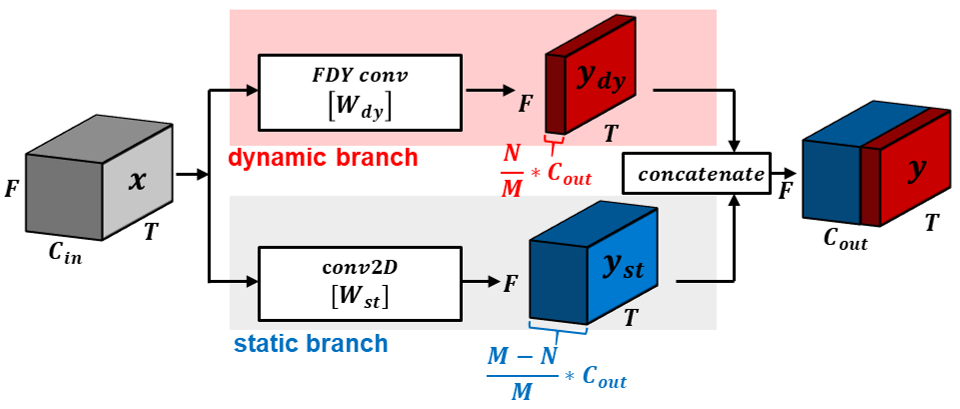}}
\caption{An illustration of partial frequency dynamic convolution operation.}
\label{fig:PFDconv}
\end{figure}

\section{Partial Frequency Dynamic Convolution (PFD conv)}
While FDY conv successfully adapts convolutional kernels to the frequency characteristics of the input, it introduces a significant computational cost. This cost arises because FDY conv requires dynamic kernel generation for each input, which leads to an increase in the number of parameters, especially as the number of frequency bins grows or the complexity of the input signal increases. While FDY conv excels at capturing frequency-dependent features, the computational overhead can become prohibitive, particularly in real-time or resource-constrained environments.

To address this issue, we propose partial frequency dynamic convolution (PFD conv), a hybrid convolutional approach that combines the benefits of frequency-dependent convolution with computational efficiency. PFD conv introduces a dual-branch structure, consisting of static and dynamic branches, which allows the model to maintain frequency adaptability while reducing the number of parameters and improving computational efficiency.

The static branch applies standard 2D convolution, which does not adapt to the frequency content of the input. This branch is particularly effective for stationary sound events, such as background noise or continuous mechanical sounds, where the frequency characteristics do not change drastically over time. The dynamic branch, on the other hand, employs FDY conv to capture frequency-adaptive features for more dynamic and non-stationary sound events, such as speech, alarms, and environmental sounds with rapidly changing spectral characteristics.

By combining both branches, PFD conv effectively balances the trade-off between model complexity and performance. The static branch reduces the computational burden, while the dynamic branch ensures that the model can still capture the rich, frequency-dependent features needed for accurate SED. This hybrid structure allows PFD conv to be both computationally efficient and capable of learning complex frequency-adaptive features, making it an ideal solution for SED tasks that require real-time processing.

\subsection{Concept and Mechanism}
The key idea behind PFD conv is to combine static 2D convolution with dynamic frequency-adaptive convolution, reducing the number of dynamically computed kernels while maintaining the adaptability of the model. This hybrid structure allows PFD conv to capture the benefits of both approaches without the computational overhead typically associated with fully dynamic methods like FDY conv.

PFD conv introduces two branches:

\begin{itemize}
    \item Static branch: This branch applies standard 2D convolutions with fixed filters across all frequency bins. Since the filters do not change during inference, this branch is computationally efficient and works well for stationary sound events, such as background noise, mechanical hums, or any sound that has a relatively consistent frequency spectrum over time. By using fixed filters, the model can perform faster while maintaining performance for such stable events.
    \item Dynamic branch: This branch utilizes FDY conv to apply frequency-adaptive kernels. These kernels are learned during training and dynamically adjust based on the input's frequency characteristics. The dynamic branch is crucial for detecting non-stationary and overlapping sound events, such as speech, alarms, and other sounds that change their frequency content over time. By learning frequency-specific adaptations, the dynamic branch improves the model's ability to handle more complex audio scenarios.
\end{itemize}

These two branches are then concatenated along the channel dimension, combining the benefits of both static and dynamic convolutions. The static branch provides computational efficiency while the dynamic branch captures the frequency-dependent variations in sound events. This combination allows PFD conv to perform accurate sound event detection while maintaining efficiency, ensuring that both stable and dynamic sound events are adequately modeled.

By controlling the proportion of the dynamic convolution applied (i.e., adjusting how much of the model uses the dynamic branch), PFD conv can adjust the balance between efficiency and adaptability based on the complexity of the sound event being detected. For simpler, stationary sounds, more weight is given to the static branch, while for complex, non-stationary sounds, the dynamic branch takes over. This flexibility makes PFD conv a highly efficient and scalable solution for real-world sound event detection tasks.

\subsection{Mathematical Formulation}
Given an input feature map $\mathbf{X} \in \mathbb{R}^{B \times C \times F \times T}$, where $B$ is the batch size, $C$ is the number of input channels, $F$ is the number of frequency bins, and $T$ is the number of time frames, the PFD conv operation is defined as follows:

\begin{equation}
    \mathbf{y}_{\text{PFD}} = \text{Concat}(\mathbf{y}_{\text{static}}, \mathbf{y}_{\text{dynamic}}, \text{dim}=C)
\end{equation}

where:
\begin{itemize}
    \item $\mathbf{y}_{\text{static}} = \mathbf{W}_{\text{static}} * \mathbf{x} + \mathbf{b}_{\text{static}}$ represents the output of the static 2D convolution branch, where $\mathbf{W}_{\text{static}}$ is the weight and $\mathbf{b}_{\text{static}}$ is the bias for the static convolution kernel. This branch uses fixed filters to process the input $\mathbf{x}$ across all frequency bins and is particularly efficient for stationary sound events.
    \item $\mathbf{y}_{\text{dynamic}} = \left(\sum_{i=1}^{K} \mathbf{W}_i \odot \mathbf{\pi(f, x)}_i\right) * \mathbf{x(f)} + \sum_{i=1}^{K} \mathbf{b}_i$ represents the output of the FDY conv branch. In this equation, $\mathbf{W}_i$ represents the $i$-th basis kernel, and $\mathbf{\pi(f, x)}_i$ is the frequency-adaptive attention weight for the $i$-th kernel. The operation $\odot$ denotes element-wise multiplication, and $\mathbf{b}_i$ is the bias for each basis kernel. The summation of the weighted kernels is applied to the frequency-modulated input $\mathbf{x(f)}$ to produce the dynamic output $\mathbf{y}_{\text{dynamic}}$.
    \item $K$ is the total number of basis kernels in the dynamic branch, which are learned to adapt to different frequency content based on the attention weights.
\end{itemize}

The static-to-dynamic processing ratio is controlled by adjusting the number of output channels allocated to each branch. By varying this ratio, PFD conv provides a flexible trade-off between computational efficiency and adaptability. The static branch ensures that simple, stationary sound events are processed efficiently, while the dynamic branch captures the frequency-dependent variations in more complex events. This structure allows the model to adaptively select the most appropriate convolution method for each sound event type, making it highly effective for real-world sound event detection.

\input{figs/table_PFD}
\subsection{Experimental Results and Analysis}
The results of PFD conv using various proportions of dynamic and static branches are shown in Table \ref{fig:PFDconv}. In the table, PFD-CRNN ($1/N$) denotes that $1/N$ of the output channels are obtained from the dynamic branch, while the remaining channels are obtained from the static branch. We experimented with different proportions of dynamic and static channels, specifically $1/32$, $1/16$, and $n/8$ where $n = 1, 2, ..., 7$, as the number of channels in the CNN module of CRNN is a multiple of 32 (i.e., 32, 64, 128, 256, 256, 256, 256 from the 1st to the 7th convolution layers). 

It is important to note that FDY-CRNN is equivalent to PFD-CRNN with a proportion of $8/8$, where all channels come from the dynamic branch. The results indicate that when the proportion is above $1/8$, PFD-CRNN's PSDS is either slightly worse or similar to that of FDY-CRNN, except for the model with a proportion of $5/8$, where PFD-CRNN achieves better performance. However, PFD-CRNN models with proportions of $1/16$ and $1/32$ perform worse than FDY-CRNN, which suggests that a very low proportion of dynamic convolution does not capture enough frequency-adaptive information, leading to a performance degradation.

Among all the tested models, the most efficient configuration is PFD-CRNN with a proportion of $1/8$, which introduces only 22.0\% of additional parameters compared to CRNN and reduces the number of parameters by 51.9\% when compared to FDY-CRNN. Despite the reduction in model size, this configuration retains the performance level of FDY-CRNN, demonstrating the effectiveness of the hybrid structure in balancing computational efficiency and adaptability. The results suggest that PFD conv can achieve a good trade-off between parameter efficiency and model performance. In particular, the $1/8$ proportion strikes a balance that allows the model to retain most of the performance benefits of FDY-CRNN while significantly reducing the number of parameters, making it an ideal choice for real-time and resource-constrained environments.

\begin{figure}[t]
\centerline{\includegraphics[width=10cm]{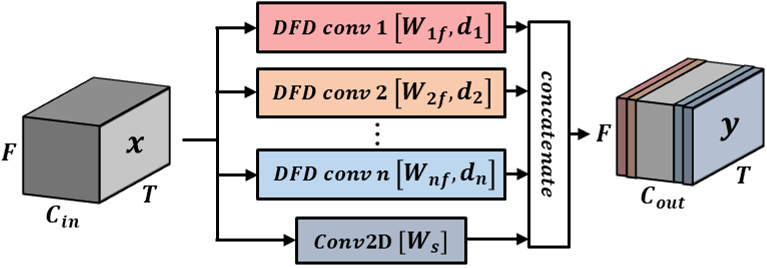}}
\caption{An illustration of multi-dilated frequency dynamic convolution operation.}
\label{fig:MDFDconv}
\end{figure}

\section{Multi-Dimensional frequency dynamic convolution (MDFD conv)}
While DFD conv enhances the receptive field of FDY conv by applying dilation to its basis kernels, it still faces limitations in handling multi-resolution spectral dependencies. As sound events can exhibit varying spectral characteristics across different frequency ranges, it is essential to capture these variations at multiple resolutions. DFD conv with a single dilation rate can only capture a limited range of frequency patterns, making it difficult to model sound events with both fine-grained and coarse spectral features.

To address this limitation, we introduce MDFD conv, an extension of DFD conv that incorporates multiple dilated dynamic branches. This architecture enables the model to capture broad spectral patterns at multiple resolutions, allowing for a more comprehensive representation of the frequency spectrum. By introducing multiple branches with different dilation rates, MDFD conv increases the diversity of frequency-adaptive kernels, improving the model's ability to adapt to varying spectral content.

In MDFD conv, each dilated branch operates with a different dilation size, allowing the model to specialize in capturing different levels of frequency details. Smaller dilation sizes are effective for capturing fine-grained frequency details, while larger dilation sizes are better suited for capturing broader spectral features. By combining these multiple branches, MDFD conv can learn a more comprehensive set of frequency-dependent features, making it more effective for sound event detection tasks that require the recognition of both local and global frequency patterns.

The flexibility of this approach allows MDFD conv to adapt to complex, overlapping sound events, where different frequency components need to be captured at multiple resolutions. This makes MDFD conv particularly suitable for detecting sound events that span a wide range of frequencies, such as speech with varying pitch, environmental sounds with multiple sources, and polyphonic music.

\subsection{Concept and Mechanism}
MDFD conv is designed to combine the strengths of both dilated convolution and multi-branch dynamic kernel processing. By introducing multiple dynamic branches with varying dilation sizes, the model can adaptively learn fine-grained frequency patterns while also capturing broader spectral dependencies. The ability to apply different dilation rates to different frequency bins enables MDFD conv to handle a wider range of frequency-dependent sound events, improving its performance on both local and global spectral features.

The model architecture consists of:
\begin{itemize}
    \item Multiple dilated dynamic branches, each with different dilation sizes, which allow the model to learn a diverse range of frequency-adaptive kernels. These branches capture varying scales of frequency information, with smaller dilation sizes focusing on fine-grained frequency details, and larger dilation sizes capturing broader spectral patterns.
    \item A static 2D convolution branch, which provides computational efficiency by applying fixed kernels across the frequency bins. This branch is particularly effective for modeling more consistent and stationary frequency patterns, such as background noise or constant sounds, and reduces the computational load compared to the fully dynamic branches.
\end{itemize}

The outputs from both the dynamic and static branches are then concatenated along the channel dimension, similar to PFD conv. This allows the model to combine the benefits of both convolution types: the efficiency of static convolutions and the adaptability of dynamic convolutions. The dynamic branches provide the flexibility to capture frequency-specific characteristics, while the static branch maintains computational efficiency and focuses on stable features that do not require dynamic adjustment.

By combining these two types of convolutions, MDFD conv enables the model to adapt to complex sound events that require both local and global frequency feature extraction, improving overall performance without compromising computational efficiency. This hybrid structure provides greater diversity in the learned features, allowing MDFD conv to handle non-stationary and overlapping sound events more effectively than previous models.

\subsection{Mathematical Formulation}
Given an input feature map $\mathbf{X} \in \mathbb{R}^{B \times C \times F \times T}$, where $B$ is the batch size, $C$ is the number of input channels, $F$ is the number of frequency bins, and $T$ is the number of time frames, the MDFD conv operation is defined as:

\begin{equation}
    \mathbf{y}_{\text{MDFD}} = \text{Concat}(\mathbf{y}_{\text{static}}, \mathbf{y}_{\text{dynamic-1}}, \mathbf{y}_{\text{dynamic-2}}, ..., \mathbf{y}_{\text{dynamic-N}}, \text{dim}=C)
    \label{eqn:MDFD}
\end{equation}

where:
\begin{itemize}
    \item $\mathbf{y}_{\text{static}} = \mathbf{W}_{\text{static}} * \mathbf{x} + \mathbf{b}_{\text{static}}$ represents the output of the static 2D convolution branch, where $\mathbf{W}_{\text{static}}$ is the weight and $\mathbf{b}_{\text{static}}$ is the bias for the static convolution kernel. The static branch uses fixed filters applied across all frequency bins. This approach is particularly efficient for stationary sound events that do not require dynamic frequency adaptation, such as constant background noise or mechanical hums.
    
    \item $\mathbf{y}_{\text{dynamic-n}} = \left(\sum_{i=1}^{K} \mathbf{W}_{in} \odot \mathbf{\pi(f, x)}_{in}\right) * \mathbf{x(f)} + \sum_{i=1}^{K} \mathbf{b}_{in}$ represents the output of the FDY conv branch for the $n$-th dynamic branch, where $\mathbf{W}_{in}$ is the $i$-th basis kernel in the $n$-th dynamic branch, and $\mathbf{\pi(f, x)}_{in}$ is the frequency-adaptive attention weight for the $i$-th kernel in the $n$-th branch. The element-wise multiplication $\odot$ applies the attention-modulated kernels to the frequency-modulated input $\mathbf{x(f)}$. The summation of weighted kernels produces the output $\mathbf{y}_{\text{dynamic-n}}$, where $n = 1, 2, ..., N$ denotes the number of dynamic branches.
    
    \item $K$ is the total number of basis kernels in each dynamic branch, which are learned to adapt to different frequency content based on the attention weights $\mathbf{\pi(f, x)}_{in}$. The dynamic branches allow the model to capture frequency-dependent patterns by modulating the contribution of each kernel based on the input signal's frequency content.
\end{itemize}

\noindent The final output $\mathbf{y}_{\text{MDFD}}$ is generated by concatenating the results of both the static and dynamic branches along the channel dimension, as shown in Equation (\ref{eqn:MDFD}). This structure enables MDFD conv to combine the computational efficiency of the static branch with the adaptive frequency-specific processing of the dynamic branches. By doing so, it captures a wide range of spectral features, from fine-grained details to broader spectral patterns, making the model well-suited for handling complex and overlapping sound events. This approach provides flexibility in learning both local and global spectral features, which is essential for accurate sound event detection across diverse acoustic environments.

\input{figs/table_MFD_one8th}
\subsection{MDFD-CRNNs with Non-Dilated Dynamic Branches}
To test the effect of multiple FDY conv modules, we experimented with MDFD conv without dilation in this subsection. This variant, which we call multi-frequency dynamic convolution (MFD conv), explores the impact of multiple dynamic branches in the model without introducing dilation to the basis kernels. The results are shown in Table \ref{tab:MFD_one-eighth}, where \#DYbr denotes the number of dynamic branches used in the model. We experimented with four dynamic branches, each with a proportion of $1/32$, and two dynamic branches with a proportion of $1/16$ to test whether multiple dynamic branches, summing up to $1/8$ of the output channels, would be beneficial.

The results indicate that increasing the number of dynamic branches beyond a proportion of $1/8$ leads to a performance drop. This suggests that dynamic branches with a proportion of $1/8$ represent the minimum necessary amount for maintaining performance, while further increasing the proportion does not provide additional benefits. These results align with the previous finding that FDY conv with a proportion of $1/8$ yields the best trade-off between performance and computational efficiency.

Furthermore, we conducted additional experiments with multiple $1/8$-sized FDY modules to test if further performance enhancement could be achieved by stacking multiple dynamic branches. The results show that using five and six dynamic branches provides the most significant improvement in SED performance, while still being lighter than the FDY conv model. This indicates that increasing the number of dynamic branches can enhance performance by enabling the model to learn different dynamic patterns from each branch. 

Additionally, we observed that the static branch also plays a crucial role in performance. The proportion of the static branch in the total number of output channels must be between $1/4$ and $3/8$ for the model to perform optimally. A lower proportion of the static branch leads to performance degradation, suggesting that the static branch is essential for efficiently modeling more stable, stationary sound events. The balance between dynamic and static branches is critical for achieving the best SED performance.

\input{figs/table_MDFD_5}
\subsection{MDFD-CRNNs with Five Dynamic Branches}
In this experiment, we introduced dilation to the MDFD-CRNN model with five dynamic branches, which had previously shown a good balance between model size and performance. The goal was to diversify the roles of the dynamic branches by applying dilation to some of the five dynamic branches, as proposed in previous work \cite{DFD}. This approach aimed to enhance the model's ability to capture diverse frequency-dependent patterns through dilated basis kernels in certain branches, while keeping others non-dilated for comparison.

The results of the experiments are shown in Table \ref{tab:MDFD_5}, where the dilation sizes are notated to show only frequency-wise dilation. Each set of parentheses describes the dilation size configuration for a single dynamic branch. For instance, $(1)\times3+(2,3)\times2$ means that three dynamic branches are non-dilated, while two dynamic branches have two non-dilated basis kernels, one kernel with a dilation size of 2, and the last kernel with a dilation size of 3. In total, all dynamic branches consist of four basis kernels.

The results demonstrate that dilating the basis kernels in the dynamic branches leads to either similar or slightly better performance compared to the baseline configuration without dilation \cite{DFD}. This indicates that applying dilated kernels to the dynamic branches can enhance the model’s ability to capture broader spectral patterns without significantly increasing the model's complexity. 

Furthermore, we interpret the results as showing that, just as the DFD conv branch benefits SED, the FDY conv branch also contributes well. This suggests that while DFD conv excels in capturing large-scale frequency dependencies through dilation, FDY conv continues to play an important role in modeling fine-grained frequency features that are crucial for accurate sound event detection.

In conclusion, the introduction of dilation to the MDFD-CRNN model with five dynamic branches allows the model to capture a wider range of spectral features, improving performance for complex sound event detection tasks. The combination of both dilated and non-dilated dynamic branches provides a flexible framework that balances the need for fine-grained detail and the ability to capture broader spectral patterns.

\input{figs/table_preMDFD}
\subsection{Multi-Dilated Frequency Dynamic Convolution with Pre-convolution and Varying Channel Sizes}
We conducted experiments to investigate the effect of adding dynamic branches with expanded channel sizes in MDFD conv. The results of these experiments are presented in Table \ref{tab:preMDFD}. In addition to the expansion of the dynamic branches, we also introduced a pre-convolution module with an output channel size of 16 before the CNN module. This pre-convolution module is used to inject MDFD conv into the 1st convolution layer. Without the pre-convolution, the input channel size of the first convolution layer would be just one, making it impossible to extract $K$ frequency-adaptive attention weights from a single channel, which would significantly limit the model's ability to learn frequency-specific patterns.

The results show that adding three dilated dynamic branches with varying dilation sizes improved performance significantly, achieving PSDS1 of 0.455, which outperforms FDY-CRNN by 3.17\%. This demonstrates the effectiveness of introducing multiple dilated branches with different dilation sizes to capture a wider range of spectral features, improving the model’s ability to adapt to complex sound events. However, we also found that introducing more dilated dynamic branches beyond a certain point resulted in worse performance. This suggests that too many dynamic branches can be counterproductive, as they may introduce excessive complexity and disrupt the model’s ability to learn efficiently. 

Interestingly, the experiments show that while adding multiple dilated dynamic branches leads to an increase in the number of parameters, several dynamic branches without dilation still help improve performance. This emphasizes that the right balance between dilated and non-dilated branches is crucial. The model benefits from a moderate number of dilated dynamic branches, which enhances performance by capturing more complex spectral patterns without causing excessive computational overhead.

In conclusion, the introduction of pre-convolution and the careful adjustment of dynamic branch configurations leads to better performance. Dilated dynamic branches help capture larger spectral dependencies, while non-dilated dynamic branches can improve fine-grained frequency adaptation. Thus, the key to enhancing the model’s performance lies in finding the right combination of dilated and non-dilated branches, along with the optimal number of dynamic branches to ensure the model remains efficient while still adapting to the spectral complexity of sound events.

\begin{figure}[t]
\centerline{\includegraphics[width=10cm]{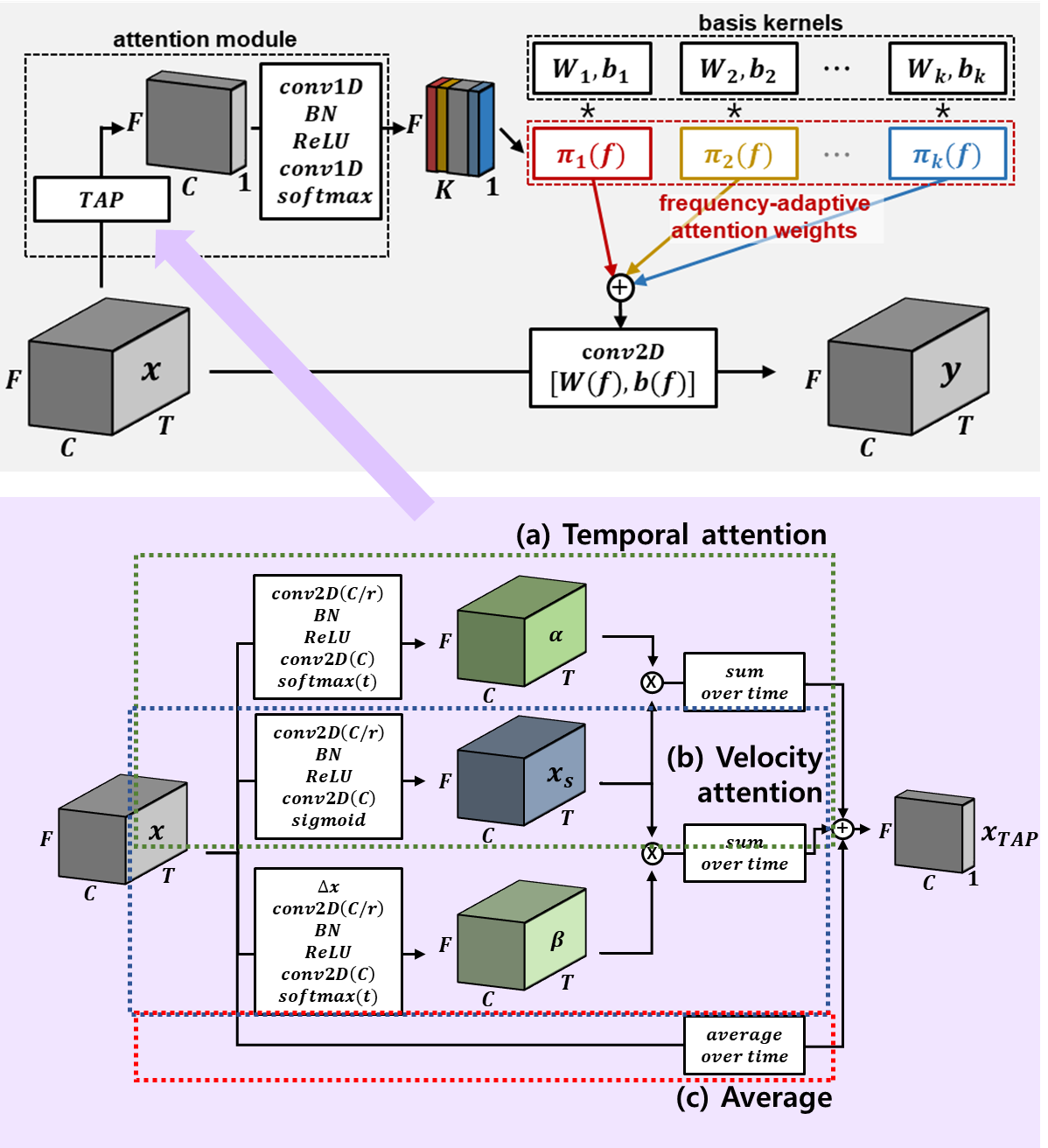}}
\caption{An illustration of TAP frequency dynamic convolution operation.}
\label{fig:TFDconv}
\end{figure}

\section{Temporal Attention Pooling for FDY conv}
While FDY conv has significantly improved SED by enabling frequency-adaptive feature extraction, it still relies on temporal average pooling for aggregating features along the time axis. Temporal average pooling assumes that all temporal frames contribute equally to the representation of a sound event, which can dilute the significance of transient sound events. Events such as alarm bells, door knocks, and speech plosives are often characterized by short bursts of sound that may not be effectively captured by this pooling method. This issue arises because temporal average pooling treats all time frames uniformly, without considering the varying importance of different frames in representing the event.

To address this limitation, we propose Temporal Attention Pooling (TAP) as a replacement for temporal average pooling. TAP enables adaptive weighting of temporal features, allowing the model to focus more on the transient parts of a sound event while still capturing the stationary aspects. Instead of averaging the features across all time frames, TAP uses learned attention weights to dynamically adjust the contribution of each time frame to the final representation. This ensures that the model places more emphasis on transient features that are critical for detecting events with brief, high-energy characteristics, such as speech bursts or sudden noises.

The core mechanism of TAP involves learning temporal attention weights that reflect the importance of each time frame in the context of the detected sound event. These attention weights are computed during training and allow the model to prioritize relevant frames based on the nature of the event. For example, for an alarm sound, TAP will assign higher attention weights to the time frames that contain the peak of the sound, improving the detection of such events. Similarly, for stationary events like hums or engine noise, the model will learn to assign more balanced attention across the time frames, ensuring that the event is represented consistently over time.

By replacing temporal average pooling with TAP, we enable FDY conv to better handle both transient and stationary sound events, making it more robust and flexible. TAP not only enhances the model’s ability to detect short-duration sound events but also ensures that it does not lose the important characteristics of longer, stationary events. This approach improves the overall performance of SED models, particularly in environments with diverse sound events that vary in both duration and frequency content.

\subsection{Temporal Attention Pooling (TAP)}

TAP is designed to dynamically adjust the importance of temporal features across time frames in order to improve the model's ability to handle both transient and stationary sound events. Unlike conventional temporal pooling methods, which treat all time frames equally, TAP allows the model to adaptively focus on the most relevant parts of the input signal, thus enhancing the detection of key sound event characteristics. TAP consists of three key pooling components, each targeting different aspects of temporal feature extraction:

\begin{itemize}
    \item Time Attention Pooling (TA): This component dynamically assigns attention weights to the salient temporal regions that are critical for detecting the onset and offset of sound events. For instance, in events like speech or alarms, the start and end times are important indicators. TA prioritizes these regions to improve the model’s ability to capture temporal boundaries effectively.
    
    \item Velocity Attention Pooling (VA): VA focuses on the transient characteristics of sound by applying attention based on the temporal differences between successive frames. This component is particularly useful for capturing events with rapid changes, such as plosives in speech or impulsive sounds like door knocks or glass breaking. By emphasizing temporal differences, VA helps the model focus on quick, transient changes that may be crucial for accurate detection.
    
    \item Average Pooling: Average pooling maintains robustness for stationary sound events by preserving the overall temporal structure of the input. For events that have a relatively stable frequency pattern over time, such as background hums or engine noise, average pooling ensures that the model does not lose important information across the entire duration of the event, while still allowing for dynamic attention in the other components.
\end{itemize}

These components are integrated into FDY conv, replacing the conventional temporal average pooling method. In FDY conv, temporal average pooling assumes that all time frames contribute equally to the representation of a sound event, which can dilute the importance of transient features. By introducing TAP, we enable the model to focus on the most relevant temporal features, enhancing its ability to detect transient events while still maintaining stability for stationary signals. 

The combination of these three pooling mechanisms allows TAP to balance the detection of both transient and stationary sound events, making it highly effective for a wide range of sound event detection tasks. By dynamically adjusting the attention given to different parts of the input, TAP helps the model capture critical features in complex acoustic environments, improving overall performance.

\subsection{Mathematical Formulation of TAP}
Let the input feature tensor be denoted as $\mathbf{x} \in \mathbb{R}^{C \times F \times T}$, where $C$ is the number of channels, $F$ is the number of frequency bins, and $T$ is the number of temporal frames. Temporal Attention Pooling (TAP) computes an output feature tensor $\mathbf{x}_{\text{TAP}}$ as:

\begin{equation}
    \mathbf{x}_{\text{TAP}} = \sum_{t=1}^{T} \alpha_t \odot \mathbf{x}_{s,t} + \sum_{t=1}^{T} \beta_t \odot \mathbf{x}_{s,t} + \frac{1}{T} \sum_{t=1}^{T} \mathbf{x}_t
\end{equation}

where:
\begin{itemize}
    \item $\alpha_t$ are the attention weights from Time Attention Pooling (TA), which modulate the importance of each time frame.
    \item $\beta_t$ are the attention weights from Velocity Attention Pooling (VA), which emphasize the temporal changes between consecutive time frames.
    \item The final term represents Average Pooling (AP), which computes the average of the input features over time, maintaining robustness for stationary sound events.
\end{itemize}

The attention weights $\alpha_t$ and $\beta_t$ are computed as follows:
\begin{equation}
    \alpha_t = \text{softmax}(W_{ta2} * \text{ReLU}(\text{BN}(W_{ta1} * \mathbf{x}_t + b_{ta1})) + b_{ta2})
\end{equation}

\begin{equation}
    \beta_t = \text{softmax}(W_{va2} * \text{ReLU}(\text{BN}(W_{va1} * (\mathbf{x}_t - \mathbf{x}_{t-1}) + b_{va1})) + b_{va2})
\end{equation}

where:
\begin{itemize}
    \item $W_{ta1}$, $b_{ta1}$, $W_{ta2}$, and $b_{ta2}$ are learnable parameters for Time Attention Pooling (TA).
    \item $W_{va1}$, $b_{va1}$, $W_{va2}$, and $b_{va2}$ are learnable parameters for Velocity Attention Pooling (VA).
    \item $\ast$ denotes a convolution operation, and BN represents Batch Normalization.
\end{itemize}

In the above equations:
- Time Attention Pooling (TA) applies softmax activation to temporal features for each time frame, adjusting the importance of each time frame based on its relevance to the sound event.
- Velocity Attention Pooling (VA) applies attention based on the temporal differences between consecutive frames, allowing the model to focus on rapid changes in the input signal that are typical in transient sound events.
- Average Pooling simply computes the time-averaged features to model stationary events, ensuring robustness to long-term, stable sound patterns.

These pooling mechanisms allow TAP to adaptively modulate the contribution of each time frame, giving more importance to transient sound events and ensuring consistent feature extraction for stationary sound events. This approach enhances the ability of FDY conv to detect dynamic sound events with varying temporal characteristics, while also maintaining the performance for more stable, background sounds.

\subsection{Experimental Results and Ablation Study}

To evaluate the effectiveness of Temporal Attention Pooling (TAP), we conducted an ablation study by incrementally adding the Time Attention Pooling (TA), Velocity Attention Pooling (VA), and Average Pooling (AP) components to FDY conv. The purpose of this study was to analyze how each pooling mechanism contributes to the overall performance of the model. The results are summarized in Table \ref{tab:tap_ablation}.

\begin{table}[t]
    \centering
    \caption{Ablation study on \textit{temporal attention pooling (TAP)} components.}
    \begin{tabular}{l|ccc|c}
        \hline
        \textbf{Model} & \textbf{Avg} & \textbf{TA} & \textbf{VA} & \textbf{PSDS1} \\
        \hline
        FDY (Baseline) & \checkmark &  &  & 0.441 \\
        FDY w/ TA & & \checkmark &  & 0.451 \\
        FDY w/ VA & &  & \checkmark & 0.444 \\
        FDY w/ avg+TA & \checkmark & \checkmark &  & 0.448 \\
        FDY w/ avg+VA & \checkmark &  & \checkmark & 0.448 \\
        FDY w/ TA+VA &  & \checkmark & \checkmark & 0.452 \\
        FDY w/ avg+TA+VA (TAP, best) & \checkmark & \checkmark & \checkmark & \textbf{0.455} \\
        \hline
    \end{tabular}
    \label{tab:tap_ablation}
\end{table}

As seen in the table, integrating both Time Attention Pooling (TA) and Velocity Attention Pooling (VA) significantly improves the PSDS1 score, increasing it from 0.441 (baseline FDY conv) to 0.455. This demonstrates the effectiveness of TAP in improving the model’s ability to capture transient sound features. While both TA and VA improve performance individually, combining both provides the best results, confirming that TAP enhances the detection of sound events by better emphasizing the most relevant temporal features.

TA focuses on the importance of event onset and offset, helping the model accurately detect the boundaries of transient events such as alarm bells, speech plosives, or other short-duration events. VA emphasizes rapid changes in the input signal, which is crucial for detecting impulsive sounds, such as knocks, claps, or sharp, transient sounds. By combining TA and VA, TAP ensures that the model can capture both the start and end points of sound events and the temporal dynamics of the events themselves, providing a comprehensive approach to handling both transient and stationary events.

The addition of Average Pooling further stabilizes the model, ensuring that stationary sound events, like background noise or mechanical hums, are still detected accurately. However, it is the combination of TA and VA that offers the most significant performance improvement. This combination allows the model to dynamically adjust attention to both transient and stationary features, improving its ability to capture the most relevant temporal characteristics.

In conclusion, TAP improves FDY conv’s performance by adapting to the temporal characteristics of sound events. The best performance is achieved when both TA and VA are used together. This study demonstrates TAP’s ability to focus attention on the most relevant time frames, effectively capturing both transient and stationary sound events, and ultimately providing more accurate sound event detection.

\subsection{Comparison with FDY conv Variants.}
We also compared TFD conv with various FDY conv variants (FDY conv, DFD conv, PFD conv, MDFD conv) on the DESED dataset to evaluate the effectiveness of TAP in enhancing sound event detection performance. The results are summarized in the following table:

\begin{table}[htbp]
    \centering
    \begin{tabular}{c|c|c}
    \hline
    Model & Parameters (M) & PSDS1 \\
    \hline
    FDY conv (baseline) & 11.061 & 0.444 \\
    DFD conv & 11.061 & 0.448 \\
    PFD conv & 5.041 & 0.441 \\
    MDFD conv & 18.157 & 0.455 \\
    TFD conv (proposed) & 12.703 & 0.455 \\
    \hline
    \end{tabular}
    \caption{Comparison of FDY conv variants with TFD conv.}
\end{table}

The table shows that TFD conv (referred to as TFD conv) outperforms FDY conv and PFD conv by improving the PSDS1 score, while achieving the same PSDS1 score as MDFD conv but with significantly fewer parameters (12.703M vs. 18.157M). This demonstrates that TFD conv is not only effective in improving the model’s performance in terms of sound event detection, but it also offers a computationally efficient solution by reducing the number of parameters compared to the more complex MDFD conv.

TAP effectively enhances temporal modeling by allowing the model to dynamically adjust attention to the most relevant time frames, which improves its ability to capture transient sound events. At the same time, TFD conv maintains a relatively low computational cost compared to MDFD conv, making it a strong candidate for real-time sound event detection in resource-constrained environments.

Moreover, PFD conv, which has fewer parameters (5.041M), performs slightly worse than FDY conv (0.441 vs. 0.444 in PSDS1). This suggests that while reducing the number of parameters can be beneficial for computational efficiency, it might come at the cost of a slight reduction in performance. On the other hand, MDFD conv, despite its higher parameter count (18.157M), achieves the best performance (0.455 in PSDS1), but the TFD conv model achieves the same performance with significantly fewer parameters.

In conclusion, TFD conv strikes a balanced trade-off between performance and computational cost, achieving state-of-the-art performance similar to MDFD conv while being more computationally efficient. This makes TFD conv a strong candidate for practical deployment in applications requiring efficient and accurate sound event detection.

%% file: figs/table_freqdepconvs.tex
\begin{table}[t]
\caption{Performance comparison between SED models with various frequency dependent convolution methods.}
\centering
\setlength{\tabcolsep}{4.25pt}
\begin{tabular}{l|lll}
\hline
\textbf{models} & \textbf{PSDS1$\uparrow$} & \textbf{PSDS2$\uparrow$} & Improvement \\
\hline
baseline        & 0.396                   & 0.598                   &             \\
 + FK conv      & 0.337                   & 0.512                   & -14.7\%    \\
 + FW Conv      & 0.408                   & 0.623                   & +3.72\%     \\
 + FDY Conv     & \textbf{0.434}          & \textbf{0.649}          & +8.95\%     \\
\hline
\end{tabular}
\label{tab:freqdepconvs}
\end{table}

%% file: figs/table_kernelsize.tex
\begin{table}[t]
\caption{Performance comparison between SED models with larger kernels and dilated kernels along frequency dimension.}
\centering
\setlength{\tabcolsep}{4.25pt}
\begin{tabular}{l|l|ll}
\hline
\textbf{models}                  & \textbf{params (M)} & \textbf{PSDS1$\uparrow$} & \textbf{PSDS2$\uparrow$} \\
\hline
CRNN                             & 4.428               & 0.410                    & 0.634                     \\
\hline
+ kernel size = 3,5              & 5.472               & 0.423                    & 0.645                     \\
+ kernel size = 3,7              & 6.517               & 0.418                    & 0.649                     \\
\hline
+ dilation size = 1,2            & 4.428               & 0.412                    & 0.643                     \\
+ dilation size = 1,3            & 4.428               & 0.404                    & 0.620                     \\
\hline
\end{tabular}
\label{tab:size_study}
\end{table}

%% file: figs/table_dfdtimefreq.tex
\begin{table*}[ht]
\caption{Performance comparison between SED models using frequency dynamic convolution with various dilation sizes on four and five basis kernels on DESED real validation dataset. $d_{i}$ implies dilation size of $i$th basis kernel in time and frequency dimensions.}
\centering
\setlength{\tabcolsep}{4.25pt}
\begin{tabular}{l|l|l|l|l|l|l|ll}
\hline
\textbf{models}        & $d_{1}$ & $d_{2}$ & $d_{3}$ & $d_{4}$ & $d_{5}$ & \textbf{params (M)} & \textbf{PSDS1$\uparrow$} & \textbf{PSDS2$\uparrow$} \\
\hline
FDY-CRNN (baseline)    & (1, 1)  & (1, 1)  & (1, 1)  & (1, 1)  & -       & 11.061             & 0.441                    & 0.668                     \\
\hline
DFD-CRNN, freq dilated & (1, 1)  & (1, 1)  & (1, 1)  & (1, 2)  & -       & 11.061             & \textbf{0.444}           & \textbf{0.676}            \\
DFD-CRNN, time dilated & (1, 1)  & (1, 1)  & (1, 1)  & (2, 1)  & -       & 11.061             & 0.442                    & 0.664                     \\
DFD-CRNN, both dilated & (1, 1)  & (1, 1)  & (1, 2)  & (2, 1)  & -       & 11.061             & 0.442                    & 0.659                     \\
\hline
DFD-CRNN, freq dilated & (1, 1)  & (1, 1)  & (1, 1)  & (1, 1)  & (1, 2)  & 12.628             & \textbf{0.445}           & \textbf{0.671}            \\
DFD-CRNN, time dilated & (1, 1)  & (1, 1)  & (1, 1)  & (1, 1)  & (2, 1)  & 12.628             & 0.442                    & 0.664                     \\

\hline
\end{tabular}
\label{tab:tf_results}
\end{table*}

%% file: figs/table_dfdsize.tex
\begin{table*}[ht]
\caption{Performance comparison between SED models using frequency dynamic convolution with various dilation sizes on DESED real validation dataset. $d_{i}$ implies dilation size of $i$th basis kernel in time and frequency dimensions.}
\centering
\setlength{\tabcolsep}{4.25pt}
\begin{tabular}{l|l|l|l|l|ll}
\hline
\textbf{models}                 & \textbf{$d_{1}$} & \textbf{$d_{2}$} & \textbf{$d_{3}$} & \textbf{$d_{4}$} & \textbf{PSDS1$\uparrow$} & \textbf{PSDS2$\uparrow$} \\
\hline
FDY-CRNN (baseline)             & (1, 1)           & (1, 1)           & (1, 1)           & (1, 1)           & 0.441                    & 0.668                     \\
\hline
DFD-CRNN, dilation size = 2     & (1, 1)           & (1, 1)           & (1, 1)           & (1, 2)           & 0.444                    & \textbf{0.676}            \\
                                & (1, 1)           & (1, 1)           & (1, 2)           & (1, 2)           & \textbf{0.447}           & 0.675                     \\
                                & (1, 1)           & (1, 2)           & (1, 2)           & (1, 2)           & 0.442                    & 0.666                     \\
                                & (1, 2)           & (1, 2)           & (1, 2)           & (1, 2)           & 0.438                    & 0.665                     \\
DFD-CRNN, dilation size = 3     & (1, 1)           & (1, 1)           & (1, 1)           & (1, 3)           & 0.444                    & 0.673                     \\
                                & (1, 1)           & (1, 1)           & (1, 3)           & (1, 3)           & \textbf{0.447}           & 0.673                     \\
DFD-CRNN, dilation size = 4     & (1, 1)           & (1, 1)           & (1, 1)           & (1, 4)           & 0.443                    & 0.672                     \\
                                & (1, 1)           & (1, 1)           & (1, 4)           & (1, 4)           & 0.441                    & 0.674                     \\
\hline
DFD-CRNN, varied dilation sizes & (1, 1)           & (1, 1)           & (1, 2)           & (1, 3)           & 0.442                    & 0.674                     \\
                                & (1, 1)           & (1, 2)           & (1, 2)           & (1, 3)           & \textbf{0.448}           & 0.672                     \\
DFD-CRNN (best)                 & (1, 1)           & (1, 2)           & (1, 3)           & (1, 3)           & \textbf{0.448}           & \textbf{0.688}            \\
                                & (1, 2)           & (1, 2)           & (1, 3)           & (1, 3)           & 0.447                    & 0.674                     \\
                                & (1, 1)           & (1, 2)           & (1, 3)           & (1, 4)           & 0.441                    & 0.672                     \\
\hline
\end{tabular}
\label{tab:size_results}
\end{table*}

%% file: figs/table_PFD.tex
\begin{table}[t]
\caption{Performance of partial frequency dynamic convolution models with varying proportion of dynamic branch.}
\centering
\begin{tabular}{c|c|c}
\hline
\textbf{models} & \textbf{Params(M)} & \textbf{PSDS1}\\ 
\hline
CRNN            &  4.428             & 0.410          \\ 
PFD-CRNN (1/32) &  4.794             & 0.436          \\
PFD-CRNN (1/16) &  4.996             & 0.434          \\ 
PFD-CRNN (1/8)  &  5.401             & \textbf{0.442} \\
PFD-CRNN (2/8)  &  6.209             & 0.439          \\ 
PFD-CRNN (3/8)  &  7.018             & \textbf{0.443} \\
PFD-CRNN (4/8)  &  7.827             & 0.439          \\
PFD-CRNN (5/8)  &  8.635             & 0.436          \\
PFD-CRNN (6/8)  &  9.444             & 0.441          \\
PFD-CRNN (7/8)  & 10.253             & \textbf{0.443} \\
FDY-CRNN        & 11.061             & 0.441          \\
\hline
\end{tabular}
\label{tab:PFD}
\end{table}

%% file: figs/table_MFD_one8th.tex
\begin{table}[t]
\caption{Performance of multi-frequency dynamic convolutions models.}
\centering
\begin{tabular}{c|cc|c}
\hline
\textbf{models} & \textbf{\# DYbr} & \textbf{Params(M)} & \textbf{PSDS1}  \\ 
\hline
FDY-CRNN        & 1                & 11.061             & 0.441        \\
PFD-CRNN (1/8)  & 1                & 5.401              & 0.442         \\ 
\hline
MFD-CRNN (1/32) & 4                & 5.896              & 0.430         \\
MFD-CRNN (1/16) & 2                & 5.566              & 0.439         \\
MFD-CRNN (1/8)  & 2                & 6.374              & 0.439         \\ 
MFD-CRNN (1/8)  & 3                & 7.348              & 0.444          \\ 
MFD-CRNN (1/8)  & 4                & 8.322              & 0.440         \\
MFD-CRNN (1/8)  & 5                & 9.296              & \textbf{0.449} \\ 
MFD-CRNN (1/8)  & 6                & 10.270             & \textbf{0.452} \\ 
MFD-CRNN (1/8)  & 7                & 11.243             & 0.445         \\
MFD-CRNN (1/8)  & 8                & 12.217             & 0.447          \\ 
\hline
\end{tabular}
\label{tab:MFD_one-eighth}
\end{table}

%% file: figs/table_MDFD_5.tex
\begin{table}[t]
\caption{Performance of multi-dilated frequency dynamic convolution models with varying dilation size sets. All partial dynamic branches have proportion of 1/8.}
\centering
\begin{tabular}{c|c|c}
\hline
\textbf{models} & \textbf{Dilation Sizes}             & \textbf{PSDS1} \\ 
\hline
FDY-CRNN  & (1)                               & 0.441        \\
PFD-CRNN  & (1)                               & 0.442          \\ 
MFD-CRNN  & (1)$\times$5                      & \textbf{0.449} \\ 
\hline
MDFD-CRNN &(1)$\times$4+(2)                   & 0.449          \\
MDFD-CRNN &(1)$\times$4+(3)                   & 0.448          \\
MDFD-CRNN &(1)$\times$4+(2,2)                 & 0.448          \\
MDFD-CRNN &(1)$\times$4+(2,3)                 & \textbf{0.451} \\
MDFD-CRNN &(1)$\times$4+(3,3)                 & 0.446          \\
MDFD-CRNN &(1)$\times$4+(2,2,3)               & 0.446          \\
MDFD-CRNN &(1)$\times$4+(2,3,3)               & \textbf{0.451} \\
\hline
MDFD-CRNN &(1)$\times$3+(2,3)$\times$2        & 0.448          \\
MDFD-CRNN &(1)$\times$3+(2,2,3)$\times$2      & 0.449          \\ 
MDFD-CRNN &(1)$\times$3+(2,3,3)$\times$2      & 0.450          \\
MDFD-CRNN &(1)$\times$3+(2,3)+(2,3,3)         & \textbf{0.451} \\
\hline
MDFD-CRNN &(1)$\times$2+(2,3)$\times$3        & 0.449          \\
MDFD-CRNN &(1)$\times$2+(2,2,3)$\times$3      & \textbf{0.452} \\
MDFD-CRNN &(1)$\times$2+(2,3,3)$\times$3      & 0.447          \\
MDFD-CRNN &(1)$\times$2+(2,3)+(2,2,3)+(2,3,3) & 0.447          \\ 
\hline
\end{tabular}
\label{tab:MDFD_5}
\end{table}

%% file: figs/table_preMDFD.tex
\begin{sidewaystable}[h]
\caption{Performance of multi-dilated frequency dynamic convolution models with pre-convolution and varying channel sizes.}
\centering
\begin{tabular}{c|ccc|c}
\hline
\textbf{models} & \textbf{\# Channels} & \textbf{Dilation Sizes}                                 & \textbf{Params(M)} & \textbf{PSDS1} \\
\hline
FDY-CRNN        & 8/8 (32,64,128,256)  & (1)                                                     & 11.061             & 0.441          \\
DFD-CRNN        & 8/8 (32,64,128,256)  & (2,3,3)                                                 & 11.061             & 0.448          \\
PFD-CRNN        & 8/8 (32,64,128,256)  & (1)                                                     & 5.401              & 0.442          \\
MFD-CRNN        & 8/8 (32,64,128,256)  & (1)$\times$5                                            & 9.296              & 0.449          \\ 
MDFD-CRNN      & 8/8 (32,64,128,256)  & (1)$\times$2+(2,3)+(2,2,3)+(2,3,3)                      & 9.296              & 0.447          \\ 
\hline
FDY-CRNN        & 11/8 (44,88,176,352) & (1)                                                     & 19.317             & 0.434          \\ 
MDFD-CRNN       & 11/8 (44,88,176,352) & (1)$\times$8                                            & 18.157             & 0.449          \\ 
MDFD-CRNN       & 11/8 (44,88,176,352) & (1)$\times$3+(2)+(3)+(2,3)+(2,2,3)+(2,3,3)              & 18.157             & 0.454          \\
MDFD-CRNN       & 11/8 (44,88,176,352) & (1)$\times$5+(2,3)+(2,2,3)+(2,3,3)                      & 18.157             & \textbf{0.455} \\
MDFD-CRNN       & 11/8 (44,88,176,352) & (1)$\times$6+(2,3)+(2,2,3)+(2,3,3)                      & 19.582             & 0.450          \\
MDFD-CRNN       & 13/8 (52,104,208,416)& (1)$\times$5+(2,2)+(3,3)+(2,3)+(2,2,3)+(2,3,3)          & 26.191             & 0.446          \\ 
MDFD-CRNN       & 14/8 (56,112,224,448)& (1)$\times$5+(2,2)+(3,3)+(2,2,3,3)+(2,3)+(2,2,3)+(2,3,3)& 30.894             & 0.433          \\ 
\hline
\end{tabular}
\label{tab:preMDFD}
\end{sidewaystable}

%% file: sections/4_experiments.tex
This section presents the experimental evaluation of our proposed methods for SED. We begin by detailing the experimental setup, including dataset composition, input features, model architectures, training procedures, and evaluation metrics. Subsequently, we provide comprehensive results and discussions covering each stage of our development: from the initial introduction of FDY conv to its advanced variants such as DFD conv, PFD conv, MDFD conv, and finally TFD conv.

Our experiments are conducted on the DESED dataset, a widely used benchmark in the DCASE (Detection and Classification of Acoustic Scenes and Events) Challenge \cite{DCASEtask4}. To evaluate system performance, we adopt the polyphonic sound detection score (PSDS), which is designed to assess polyphonic SED systems in a threshold-independent manner using intersection-based matching and receiver operating characteristic curves \cite{PSDS}. Among the two PSDS variants defined by the DCASE Challenge, we report only PSDS1, which emphasizes precise temporal localization of sound events. This is because our objective focuses on detecting temporally accurate event boundaries, whereas PSDS2 is more aligned with audio tagging tasks and is not considered in this study.

The following subsections present experimental results in a step-by-step manner, highlighting the individual contributions and limitations of each proposed module. We also include a final comparison with previous state-of-the-art models, as well as a class-wise performance analysis to examine the impact of each method on different sound event categories.

\section{Experimental Setup}
In this section, we describe the experimental setup, including the dataset, evaluation metrics, and the training framework used for the SED models. We also detail the configuration and application of the Temporal Attention Pooling Frequency Dynamic Convolution (TFD conv) method.

\subsection{Dataset}
The dataset used in this work is the Domestic Environment Sound Event Detection (DESED) dataset \cite{DCASEtask4}, which consists of 10-second audio recordings sampled at 16 kHz. DESED is designed to train and evaluate SED systems in realistic acoustic environments. The dataset includes recordings that reflect common domestic soundscapes such as kitchens, bathrooms, and living rooms, making it a suitable benchmark for evaluating models intended for real-world applications. DESED contains ten predefined sound event classes, which are listed in Table~\ref{table:sound_events}, along with their frequency statistics and average event durations in the real validation set.

The DESED dataset is composed of three distinct subsets that reflect varying levels of annotation quality and realism:
\begin{itemize}
    \item Strongly labeled synthetic dataset: This subset is generated by mixing isolated foreground sound events with realistic background noise recordings. Each resulting clip is annotated with precise onset and offset times for all sound events it contains. The use of synthetic mixtures allows for complete control over event composition and timing, providing high-quality ground truth for training with strong supervision. This subset consists of 10,000 clips for training and 2,500 clips for validation.
    
    \item Weakly labeled real dataset: This subset comprises actual domestic recordings captured in real acoustic settings. Unlike the synthetic dataset, it is only annotated at the clip level; that is, the presence or absence of each sound event class is provided, but without temporal information. This weakly labeled data reflects the common real-world challenge of having partially annotated data and is critical for training models under weak supervision. A total of 1,578 clips are included in this subset. For training purposes, it is randomly split into training and validation partitions in a 9:1 ratio.
    
    \item Unlabeled real dataset: This subset also consists of real domestic recordings, but it comes without any annotations. It is used exclusively in the semi-supervised learning framework to improve the generalization ability of the model. This portion is essential in enabling large-scale representation learning beyond the limited scope of labeled data. It contains 14,412 audio clips.

    \item Strongly labeled real dataset: Composed of 1,168 real-world audio clips, this subset provides precise temporal annotations (onset and offset) for sound events captured in actual environments. It is used solely for final model evaluation. As the only real dataset with strong labels, it serves as the most reliable benchmark for assessing how well the model performs in realistic SED scenarios.

\end{itemize}

These subsets are strategically integrated during training. At each training step, a mini-batch is constructed by sampling 12 audio clips from the strongly labeled synthetic dataset, 12 clips from the weakly labeled real dataset, and 24 clips from the unlabeled real dataset. This forms a total batch size of 48, balanced across supervision levels to maximize the benefit from strong, weak, and unlabeled data. Validation is performed using a batch size of 24, sampled from the union of the synthetic strongly labeled validation set and the weakly labeled validation set. Finally, testing is conducted on the real strongly labeled validation dataset, enabling accurate assessment of temporal localization performance.

This training strategy enables robust learning by combining the advantages of controlled synthetic data with the acoustic variability of real recordings. The inclusion of both labeled and unlabeled real-world data reflects a realistic training scenario and supports the use of semi-supervised learning strategies. Table~\ref{table:sound_events} summarizes the distribution and characteristics of sound events in the DESED real validation dataset.

\begin{table}[h]
\centering
\caption{Sound Event Classes in the DESED Dataset.}
\begin{tabular}{c|c|c}
\hline
\textbf{Sound Event Class} & \textbf{Number of Items} & \textbf{Average Length (s)} \\
\hline
\hline
Alarm/bell ringing & 420 & 1.96 \\
Blender & 96 & 5.13 \\
Cat & 341 & 1.38 \\
Dishes & 567 & 0.63 \\
Dog & 570 & 1.41 \\
Electric shaver/toothbrush & 65 & 7.73 \\
Frying & 94 & 8.26 \\
Running water & 237 & 5.25 \\
Speech & 1,754 & 1.50 \\
Vacuum cleaner & 92 & 8.48 \\
\hline
\end{tabular}
\label{table:sound_events}
\end{table}

\subsection{Training Framework}
To effectively utilize the large volume of unlabeled data provided in the DESED dataset, we adopt a semi-supervised learning framework based on the mean teacher method \cite{meanteacher}. Semi-supervised learning is particularly important in the context of SED, where precise annotations are costly to obtain and many real-world recordings lack time-aligned labels. By incorporating unlabeled data into the training process, the model can learn more generalized representations and become more robust to domain shifts and unseen acoustic conditions.

The mean teacher framework consists of two models: a student model, which is trained via backpropagation, and a teacher model, which acts as a slowly evolving target. The teacher model is not independently optimized but is instead updated as an exponential moving average (EMA) of the student model's parameters:
\begin{equation}
\theta_T \leftarrow \alpha \theta_T + (1 - \alpha) \theta_S,
\end{equation}
where $\theta_T$ and $\theta_S$ denote the teacher and student parameters respectively, and $\alpha$ is a smoothing hyperparameter (typically set to 0.99). This EMA update ensures that the teacher model provides a temporally stable reference, enabling consistency-based regularization during training.

The student model is trained to minimize a composite loss function that includes both a supervised classification loss and an unsupervised consistency loss:
\begin{equation}
\mathcal{L} = \mathcal{L}_{sup} + \lambda \mathcal{L}_{cons},
\end{equation}
where $\lambda$ controls the relative importance of the consistency loss. The supervised loss $\mathcal{L}_{sup}$ is computed using binary cross-entropy (BCE) on labeled data, reflecting the multi-label nature of SED. The consistency loss $\mathcal{L}_{cons}$ encourages the student model’s predictions to remain close to those of the teacher model for the same unlabeled input:
\begin{equation}
\mathcal{L}_{cons} = \frac{1}{N} \sum_{i=1}^{N} \| f_S(x_i) - \text{sg}(f_T(x_i)) \|^2,
\end{equation}
where $f_S(x)$ and $f_T(x)$ are the frame-wise predictions of the student and teacher models, respectively, for an unlabeled input $x$, and $N$ is the number of unlabeled samples in the batch. $\text{sg}(\cdot)$ represents the stop-gradient operation to prevent backpropagation through the teacher. 

This consistency regularization enables effective learning from unlabeled data. It acts as a form of implicit ensembling, where the student learns not only from ground truth labels but also from the smoothed predictions of the teacher. This helps prevent overfitting to the limited labeled data and stabilizes learning in the presence of noisy or ambiguous inputs.

Each training step is performed using mini-batches composed of diverse data sources: 12 audio clips are sampled from the strongly labeled synthetic dataset, 12 from the weakly labeled real dataset, and 24 from the unlabeled real dataset, yielding a total batch size of 48. The model is trained for 200 epochs using the Adam optimizer with an initial learning rate of 0.001. To further improve generalization, we apply cosine annealing to gradually decay the learning rate during training.

The mean teacher framework has proven particularly effective in SED settings, where temporal boundaries are often imprecise and event co-occurrence is common \cite{DCASEtask4}. By enforcing consistency over unlabeled data while learning from labeled examples, this framework allows the model to leverage large-scale real-world audio data without requiring additional annotation, ultimately improving both performance and robustness.

\begin{figure}[t]
\centerline{\includegraphics[width=16cm]{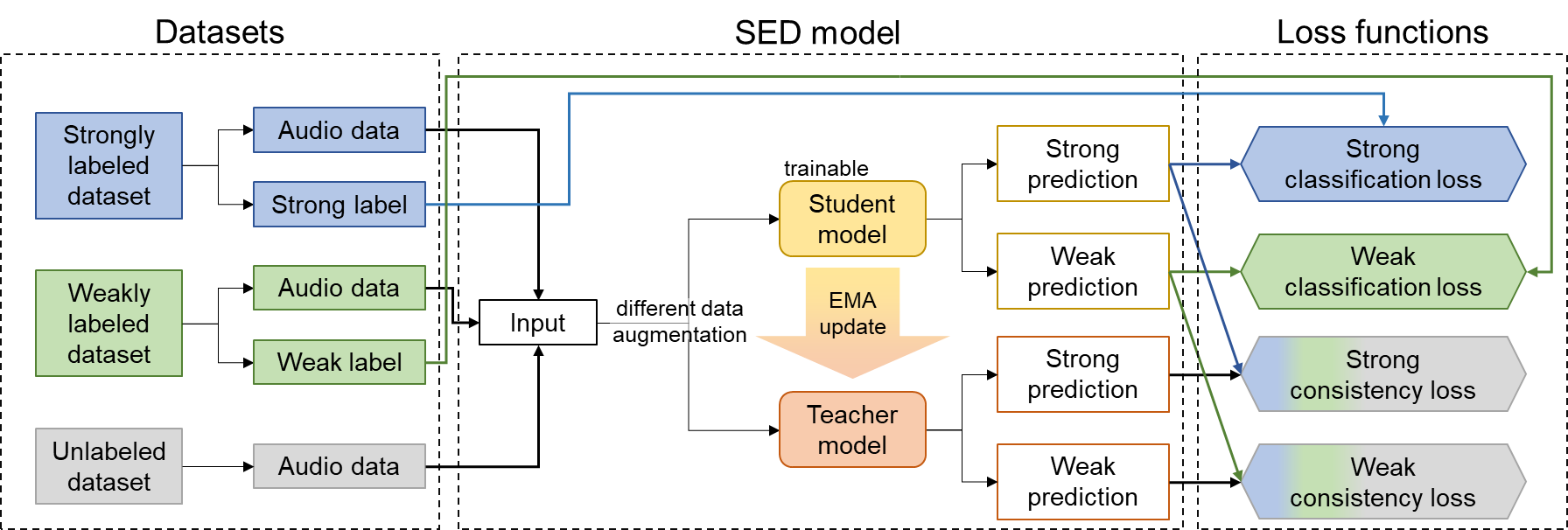}}
\caption{An illustration of the framework for training the SED models in this work. It applies mean teacher algorithm with strongly labeled dataset, weakly labeled dataset and unlabeled dataset to minimize four loss functions: strong classification loss, weak classification loss, strong consistency loss and weak consistency loss.}
\label{fig:meanteacher}
\end{figure}

\subsection{Input Feature}
The input feature used in this study is the log-mel spectrogram, a time-frequency representation that has become the de facto standard in SED. The log-mel spectrogram effectively captures the spectral energy distribution of audio signals while preserving temporal structure, enabling models to recognize diverse sound event patterns across time and frequency axes.

To extract this feature, raw audio signals are first resampled to 16 kHz to match the sampling rate of the DESED dataset. The audio is then converted into the time-frequency domain using the Short-Time Fourier Transform (STFT), computed with a window size of 2048 samples and a hop size of 256 samples. The resulting linear-frequency spectrogram is projected onto the mel scale using a filterbank with 128 mel bins, which mimics the frequency sensitivity of the human auditory system. This mel-scaled spectrogram is subsequently compressed using a logarithmic operation to suppress large magnitude differences and enhance the representation of lower-energy components.

The final log-mel spectrogram is a two-dimensional matrix of size $T \times 128$, where $T$ is the number of time frames. Each row corresponds to a mel frequency bin, and each column represents a short time segment of the input audio. This format allows CNNs to efficiently learn local patterns that are characteristic of different sound events.

Distinct sound events produce unique time-frequency patterns. For instance, quasi-stationary events such as vacuum cleaner or electric shaver sounds exhibit continuous and consistent spectral shapes over time. In contrast, non-stationary events like dog barking or human speech show rapid spectral fluctuations and require finer temporal modeling. The log-mel spectrogram effectively captures these contrasting characteristics, making it highly suitable for SED.

To improve training stability and generalization, per-frequency normalization is applied to the log-mel spectrogram. Specifically, the maximum and minimum values of the log-mel spectrogram are linearly scaled to be one and zero. This step reduces sensitivity to global amplitude variations and ensures that the model learns from relative spectral patterns rather than absolute signal energy.

The resulting normalized log-mel spectrograms are used as input to the convolutional layers of the SED model. The frequency-dependent convolution techniques proposed in this work—such as FDY conv and its variants—are particularly well-suited to this representation, as they dynamically adapt their kernels based on the spectral characteristics of each frequency band. This frequency-aware modeling improves the system’s ability to capture localized, frequency-specific structures, ultimately enhancing SED performance.

\subsection{Data Augmentation}
To improve the generalization ability and robustness of the SED models, various data augmentation techniques are applied during training. Data augmentation plays a vital role in preventing overfitting and enhancing the model’s ability to adapt to diverse acoustic environments, especially in real-world settings where recording conditions are highly variable. The augmentation methods employed in this work are designed to perturb both the temporal and spectral dimensions of the input features, thereby encouraging the model to learn invariant and discriminative representations.

The following augmentation techniques are used on log-mel spectrogram inputs:

1) Frame Shifting \cite{DCASEtask4}: To introduce small perturbations in the temporal alignment of sound events, the spectrogram is randomly shifted by a few frames along the time axis. This operation helps prevent the model from overfitting to the absolute timing of sound events and improves robustness to timing jitter in real recordings.

2) Mixup \cite{mixup}: Mixup generates virtual training examples by linearly interpolating between pairs of training samples. Given two spectrograms $X_1$ and $X_2$ with corresponding labels $y_1$ and $y_2$, the augmented sample is computed as:
\begin{equation}
X_{\text{new}} = \lambda X_1 + (1 - \lambda) X_2, \quad y_{\text{new}} = \lambda y_1 + (1 - \lambda) y_2,
\end{equation}
where $\lambda$ is sampled from a Beta distribution. This technique encourages the model to learn smoother decision boundaries and improves generalization by exposing it to interpolated samples that lie between existing training points.

3) Time Masking \cite{specaug}: A random temporal region of the spectrogram is zeroed out to simulate dropped or occluded time segments. This augmentation compels the model to rely on contextual information surrounding the masked region, making it more tolerant to temporal disruptions. For SED, label corresponding to the masked region is also zeroed out accordingly.

4) Frequency Masking \cite{specaug}: Similar to time masking, frequency masking randomly suppresses certain mel bands to simulate missing frequency components caused by environmental noise or microphone limitations. It enhances robustness by encouraging reliance on broad spectral context rather than narrow-band features.

5) FilterAugment \cite{specaug}: FilterAugment randomly emphasizes or suppresses specific frequency regions during training, simulating dynamic changes in spectral emphasis. Among all frequency perturbation techniques, this method has shown the most consistent performance improvements for SED, as it helps the model generalize across a wider range of spectral conditions.

6) Gaussian Noise Injection \cite{DCASEtask4}: Additive Gaussian noise is applied to the spectrogram to simulate sensor noise and real-world recording imperfections.

While multiple augmentation techniques were explored, FilterAugment was ultimately selected as the primary frequency-domain augmentation method due to its superior performance. Other techniques such as frequency masking and noise injection were evaluated but did not yield significant improvements in our experiments. All augmentations are applied stochastically with randomized parameters, ensuring a diverse set of variations is encountered in each training epoch. These augmentation strategies collectively enhance the model's ability to generalize across unseen environments, improve robustness to noisy and distorted inputs, and increase tolerance to minor timing and frequency deviations in real-world recordings.

\begin{figure}[t]
\centerline{\includegraphics[width=8cm]{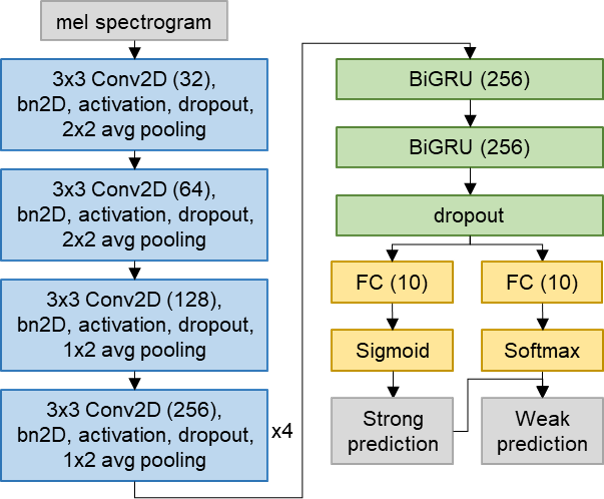}}
\caption{An illustration of SED baseline model architecture used in this work.}
\label{fig:architecture}
\end{figure}

\subsection{Baseline Model Architecture}
The baseline model employed in this study is a Convolutional Recurrent Neural Network (CRNN), a widely adopted architecture in SED due to its ability to jointly model spectral and temporal structures \cite{DCASEtask4, crnn}. The CRNN architecture comprises three main components: a convolutional front-end for feature extraction, a recurrent back-end for temporal modeling, and a fully connected output layer for event classification.

The convolutional front-end consists of four convolutional blocks, each composed of a $3 \times 3$ convolution, followed by batch normalization, rectified linear unit (ReLU) activation, dropout, and max-pooling. The number of output channels is increased progressively across layers: 32 filters in the first layer, 64 in the second, 128 in the third, and 256 in the rest of layers. Max-pooling is applied along both time and frequency axes to reduce the dimensionality of the feature maps while preserving the time-frequency structure relevant to sound events.

The extracted features are then passed to a bidirectional gated recurrent unit (biGRU) layer, which captures long-term temporal dependencies in the audio sequence. Two stacked biGRU layers are used, each with 256 hidden units in both forward and backward directions. This bidirectional processing enables the model to utilize both past and future context when detecting temporally overlapping or delayed sound events.

Finally, the model concludes with a fully connected output layer followed by a sigmoid activation function, producing frame-wise event probabilities for each class. The use of sigmoid activation enables multi-label classification, as multiple sound events may co-occur within a single audio clip.

To enhance the model’s ability to focus on informative segments within the sequence, an attention mechanism is applied after the recurrent layers \cite{DCASEtask4}. The attention module computes frame-wise importance weights and aggregates the temporal features into context-aware representations. This is especially effective for detecting sporadic or temporally ambiguous events embedded in complex acoustic scenes.

This CRNN architecture serves as the foundation upon which all frequency-dependent convolution variants are evaluated. In the proposed models, the standard convolutional layers are replaced with frequency-adaptive counterparts such as FDY conv, dilated DFD conv, and their multi-branch extensions. These methods enable the network to capture class-specific and frequency-localized patterns more effectively, thereby improving detection performance in complex acoustic environments.

In all frequency-adaptive variants, FDY conv and its extensions are applied from the second to the seventh convolutional layer. The first layer remains a conventional 2D convolution because its input has only a single channel, making it infeasible to extract meaningful frequency-adaptive attention weights. Furthermore, DFD conv and other dilated dynamic branches are applied only up to the sixth layer. This is because the input to the seventh convolutional layer is reduced to just two frequency bins due to pooling operations, and applying dilation in such a narrow frequency range is ineffective. These design choices ensure that frequency-dependent modeling is applied where it is most impactful while maintaining computational efficiency.

\subsection{Loss Function}
The loss function used in this study consists of four components: strong classification loss, weak classification loss, strong consistency loss, and weak consistency loss. This formulation enables the model to fully exploit all available supervision signals from the DESED dataset, which includes strongly labeled, weakly labeled, and unlabeled audio clips. The training objective follows the mean teacher framework and incorporates both supervised learning from labeled data and unsupervised consistency regularization from unlabeled data.

The overall loss is defined as:
\begin{equation}
\mathcal{L} = \mathcal{L}_{\text{strong-cls}} + w_W \cdot \mathcal{L}_{\text{weak-cls}} + w_C \cdot \mathcal{L}_{\text{cons}},
\end{equation}
where $w_W$ and $w_C$ are weighting coefficients for the weak classification loss and the consistency loss, respectively.

The strong classification loss $\mathcal{L}_{\text{strong-cls}}$ is computed using binary cross-entropy (BCE) between the model's frame-level predictions and the strong labels:
\begin{equation}
\mathcal{L}_{\text{strong-cls}} = BCE(SP_s, l_s),
\end{equation}
where $SP_s$ denotes the strong prediction from the student model and $l_s$ is the corresponding label.

The weak classification loss is similarly defined as:
\begin{equation}
\mathcal{L}_{\text{weak-cls}} = BCE(WP_s, l_w),
\end{equation}
where $WP_s$ is the weak prediction obtained by applying temporal pooling to the student model outputs, and $l_w$ is the weak label. Following previous work, $w_W$ is set to 0.5.

The binary cross-entropy loss is given by:
\begin{equation}
BCE(P, l) = - \frac{1}{N} \sum_{i=1}^{N} \sum_{c=1}^{C} \left[ l_{i,c} \log P_{i,c} + (1 - l_{i,c}) \log (1 - P_{i,c}) \right],
\end{equation}
where $N$ is the number of labeled samples, $C$ is the number of event classes, $l_{i,c}$ is the ground-truth label, and $P_{i,c}$ is the predicted probability.

The consistency loss $\mathcal{L}_{\text{cons}}$ encourages the student model to remain consistent with the teacher model's predictions under different augmentations. It is applied to both strong and weak outputs:
\begin{equation}
\mathcal{L}_{\text{cons}} = MSE(SP_s, \text{sg}(SP_t)) + MSE(WP_s, \text{sg}(WP_t)),
\end{equation}
where $\text{sg}(\cdot)$ denotes the stop-gradient operator, and $SP_t$, $WP_t$ are the strong and weak predictions from the teacher model.

The mean squared error (MSE) is defined as:
\begin{equation}
MSE(x, y) = \frac{1}{M} \sum_{j=1}^{M} \| x - y \|^2,
\end{equation}
where $M$ is the number of unlabeled samples. The consistency weight $w_C$ is linearly increased from zero to its target value during the first 50 epochs to stabilize training.

This four-part loss formulation enables the model to learn from all three data sources. Strongly labeled samples provide precise temporal supervision, weakly labeled data guide class-wise presence learning, and unlabeled data regularize the model through teacher-student consistency. This hybrid strategy allows for robust learning even under limited or partially labeled conditions.

\subsection{Post Processing}
Once the model generates frame-wise probability outputs for each sound event class, a series of post-processing steps is applied to convert these raw predictions into discrete, temporally localized event boundaries. These steps are essential for improving the reliability of the predictions by reducing noise, consolidating fragmented activations, and enforcing realistic duration and continuity constraints. The goal is to align the final predictions more closely with the perceptual and annotation standards used in real-world sound event detection.

The first stage of post-processing is weak prediction masking \cite{mytechreport}. While the model produces strong frame-level predictions for temporal localization, these outputs often include spurious activations caused by transient noise or overlapping acoustic events. To address this, weak clip-level predictions are used to suppress unreliable frame-level activations. For each event class, if the weak prediction probability falls below a predefined threshold, the corresponding strong predictions across all frames are set to zero. This gating mechanism filters out unlikely classes and acts as a soft prior, ensuring that strong predictions are only retained when the model is confident in the class-wise presence of an event.

Following masking, a median filtering step is applied to the remaining frame-wise predictions. The sigmoid activations from the model tend to exhibit sharp, frame-by-frame fluctuations, leading to fragmented detection results. Median filtering with a temporal window of seven frames (approximately 0.45 seconds) is used to smooth these predictions. This operation effectively removes short-duration false positives while preserving the temporal continuity of genuine events. Compared to averaging filters, median filtering is more robust to outliers, making it well-suited for eliminating isolated noise spikes.

Collectively, these post-processing steps improve the interpretability and usability of the model outputs. Weak prediction masking enforces semantic plausibility, and median filtering enhances temporal smoothness. The resulting event boundaries are more stable, less noisy, and better aligned with human perception, making the overall system more practical and reliable for deployment in real-world SED applications.

To further boost the detection performance, class-wise median filtering could be applied, allowing each event class to use a tailored smoothing window based on its temporal characteristics. This technique is known to improve boundary consistency for both short and long-duration events. However, to ensure a fair and invariant comparison across models, such class-dependent post-processing was not employed in this study. Instead, all models were evaluated under the same fixed post-processing configuration.

\subsection{Evaluation Metrics}
To quantitatively assess the performance of the sound event detection models, two complementary metrics are used: the Polyphonic Sound Detection Score (PSDS1) and the collar-based F1 score \cite{PSDS, sedmetrics, DCASEtask4}. These metrics jointly evaluate not only the detection accuracy but also the temporal precision of the predicted event boundaries.

PSDS1 serves as the primary evaluation metric in this study. It measures the trade-off between precision and recall across a range of decision thresholds, while explicitly incorporating constraints on onset and offset accuracy. Unlike traditional F1 scores computed at a fixed threshold, PSDS1 integrates over multiple operating points using a weighted intersection-based matching protocol. This allows for a threshold-independent and more robust evaluation, especially in polyphonic conditions where multiple events may overlap. In our experiments, PSDS1 is computed using the official DESED evaluation toolkit with default parameter settings, ensuring comparability with prior studies.

In addition to PSDS1, the collar-based F1 score is used for class-wise comparison. This score is calculated by matching predicted and reference events using a tolerance window, or “collar,” around the onset and offset boundaries. Specifically, a fixed collar of 200 milliseconds is applied to the onset time, and a dynamic collar equal to 20\% of the event’s duration is used for the offset. This tolerance accounts for small timing discrepancies that are often inevitable in both model predictions and human annotations, particularly for ambiguous or gradual-onset events.

Both metrics are computed on the real strongly labeled validation subset of the DESED dataset. Model outputs are evaluated across a set of decision thresholds to analyze the overall detection behavior and to select the optimal operating point. Together, PSDS1 and collar-based F1 provide a comprehensive evaluation framework that balances detection accuracy with temporal reliability. This allows for a nuanced comparison of different model variants, especially in terms of their ability to localize events with precision.

\begin{table}[h]
\centering
\begin{tabular}{c|c|c}
\hline
\textbf{Model} & \textbf{Params (M)} & \textbf{Max PSDS1} \\
\hline
Baseline (CRNN) & 4.428 & 0.410 \\
FDY conv & 11.061 & 0.441 \\
DFD conv & 11.061 & 0.448 \\
PFD conv & 5.041 & 0.442 \\
MDFD conv & 18.157 & \textbf{0.455} \\
TFD conv & 12.703 & \textbf{0.455} \\
\hline
\end{tabular}
\caption{Comparison of model size and PSDS1 score across baseline and frequency-adaptive convolution models.}
\label{table:baseline_comparison}
\end{table}

\section{Baseline Comparison}
To assess the effectiveness of the proposed frequency-dependent convolution methods, we compare their performance with a baseline CRNN architecture. The baseline model consists of standard two-dimensional convolutional layers followed by bidirectional gated recurrent units and a fully connected output layer. While this architecture has been widely used in SED, it applies the same convolutional filters across all frequency bands, implicitly assuming frequency shift-invariance. This assumption may be inappropriate for environmental sounds, which often exhibit class-specific and frequency-localized structures.

To address this limitation, we introduce a family of frequency-adaptive convolution models. The first variant, frequency dynamic convolution (FDY conv), replaces standard convolution with frequency-wise adaptive kernels that are modulated by attention weights specific to each mel frequency bin. This design enables the network to model frequency-dependent patterns more effectively and leads to a substantial performance improvement over the baseline.

Building on this foundation, we propose several enhanced variants. Dilated frequency dynamic convolution (DFD conv) extends FDY conv by introducing dilated kernels in the frequency dimension, which increases the effective receptive field and allows the model to capture broader spectral context. Partial frequency dynamic convolution (PFD conv), on the other hand, is designed to reduce computational cost by applying dynamic convolution only to a subset of the filters, while retaining static convolution in the remaining channels. This hybrid approach provides a favorable trade-off between model efficiency and detection accuracy.

Multi-dilated frequency dynamic convolution (MDFD conv) further generalizes DFD conv by combining multiple dynamic branches with different dilation rates. This structure enables the model to simultaneously capture short-range and long-range spectral dependencies, resulting in the highest PSDS1 performance among the frequency-only variants.

Finally, we propose temporal attention pooling frequency dynamic convolution (TFD conv), which augments FDY conv with a learnable temporal pooling mechanism. By replacing simple temporal average pooling with a mixture of attention-based pooling strategies, TFD conv allows the model to adaptively emphasize important frames during prediction. Although TFD conv does not explicitly modify frequency processing, it achieves performance on par with MDFD conv, highlighting the importance of both spectral and temporal adaptivity in SED.

Table~\ref{table:baseline_comparison} presents the number of model parameters and the maximum PSDS1 scores achieved by each model on the DESED real validation dataset. While frequency-adaptive models introduce additional parameters compared to the baseline, the gains in detection performance justify this increase. Notably, PFD conv achieves similar PSDS1 performance to FDY conv and DFD conv while using less than half the parameters, demonstrating its effectiveness as a compact and efficient alternative. Conversely, MDFD conv attains the highest PSDS1 score at the cost of a significantly larger model size, making it suitable for scenarios where computational resources are less constrained. TFD conv offers a competitive balance between performance and model complexity, combining high performance with moderate parameter growth.

These results collectively support the hypothesis that incorporating inductive biases about spectral and temporal variability into convolutional structures can substantially improve the performance and robustness of sound event detection systems. Moreover, they provide insights into the trade-off space between model complexity and detection performance, which is critical for real-world deployment where resource constraints may vary.

\section{TAP + other FDY variants}
\subsection{TAP + DFD Conv (Varying Dilation Sizes)}

To further explore the impact of TAP, we integrate it into DFD conv, which enhances FDY conv by applying dilation to the basis kernels on frequency dimension. Increasing dilation size expands the receptive field without increasing parameter count, thereby promoting kernel diversity and enabling better modeling of spectral variations. Since TAP focuses on temporal pooling, its combination with DFD conv is expected to improve overall time-frequency modeling if both mechanisms are harmonized effectively.

\begin{table}[t]
    \centering
    \caption{Ablation study on TAP integrated with DFD conv (various dilation sizes).}
    \begin{tabular}{l|c|c}
        \hline
        \textbf{Model} & \textbf{Dilation Sizes} & \textbf{PSDS1} \\
        \hline
        TFD conv   & (1) & 0.455 \\
        \hline
        TAP + DFD & (2) & 0.444 \\
        TAP + DFD & (3) & 0.445 \\
        TAP + DFD & (2,3) & 0.444 \\
        TAP + DFD & (2,2,3) & 0.441 \\
        TAP + DFD & (2,3,3) & 0.448 \\
        \hline
    \end{tabular}
    \label{tab:tap_dfd_ablation}
\end{table}

As shown in Table~\ref{tab:tap_dfd_ablation}, integrating TAP with DFD conv does not yield performance improvements over TFD conv. In fact, all tested dilation configurations result in lower PSDS1 scores, indicating that TAP and frequency dilation can interfere with each other when not carefully tuned. This is also consistent with prior findings~\cite{DFD} that suggest DFD conv contributes more to PSDS2—which emphasizes event classification—rather than PSDS1, which emphasizes precise temporal localization.

The baseline TFD conv without dilation achieves the highest PSDS1 of 0.448. When dilation is introduced, performance decreases across all configurations, with the lowest scores observed in (2,2,3). This suggests that excessive dilation may harm temporal granularity by over-smoothing, thus hindering TAP’s ability to emphasize transient details.

These findings imply that while DFD conv increases spectral flexibility, it may degrade temporal precision if combined naively with TAP. Careful co-design of temporal and spectral modeling components is necessary to fully leverage their complementary strengths. Future research could investigate adaptive dilation strategies or attention-conditioned dilation that responds dynamically to the temporal variability of the input signal.

\subsection{TAP + PFD Conv (Varying Channel Proportion)}

To further investigate the adaptability of TAP, we integrate it into PFD conv, which reduces model complexity by balancing static and dynamic convolution branches. PFD conv achieves this by introducing a static convolutional branch alongside the frequency-adaptive dynamic branch, significantly reducing the number of parameters while maintaining competitive performance. However, reducing the proportion of dynamic channels may lead to a loss of adaptive frequency modeling capability. In this context, TAP is expected to mitigate this drawback by enhancing the temporal feature representation, compensating for the reduced dynamic processing capacity.

\begin{table}[t]
    \centering
    \caption{Ablation study on TAP integrated with PFD conv (various proportions).}
    \begin{tabular}{l|cc|c}
        \hline
        \textbf{Model} & \textbf{Channel Proportion} & \textbf{Params(M)} & \textbf{PSDS1} \\
        \hline
        Baseline (CRNN)  & 0 & 4.428 & 0.410 \\
        PFD  & 1/8 & 5.041 & 0.442 \\
        TAP + PFD & 1/32 & 6.435 & 0.439 \\
        TAP + PFD & 1/16 & 6.637 & 0.432 \\
        TAP + PFD & 1/8 & 7.042 & 0.438 \\
        TAP + PFD & 2/8 & 7.850 & 0.449 \\
        TAP + PFD & 3/8 & 8.659 & 0.445 \\
        TAP + PFD & 4/8 & 9.468 & 0.447 \\
        TAP + PFD & 5/8 & 10.276 & \textbf{0.455} \\
        TAP + PFD & 6/8 & 11.085 & 0.446 \\
        TAP + PFD & 7/8 & 11.894 & 0.449 \\
        TFD conv & 8/8 & 12.703 & \textbf{0.455} \\
        \hline
    \end{tabular}
    \label{tab:tap_pfd_ablation}
\end{table}

Table~\ref{tab:tap_pfd_ablation} presents the results of integrating TAP into PFD conv while varying the proportion of dynamic channels. The performance exhibits a non-linear trend depending on the dynamic-to-static channel ratio. The best performance (PSDS1 = 0.455) is observed when 5/8 of the channels are dynamic, suggesting that an optimal balance between dynamic and static branches is essential. A lower proportion of dynamic channels (e.g., 1/32 or 1/16) leads to noticeable performance degradation, likely due to insufficient frequency adaptivity in static branches. As the dynamic ratio increases, the model’s ability to handle complex frequency variations improves, culminating in a PSDS1 of 0.455 at the 8/8 configuration—equivalent to TFD conv.

These findings indicate that TAP can successfully complement PFD conv by enhancing temporal modeling, allowing reduced-parameter models to achieve performance levels close to fully dynamic configurations. While PFD conv helps reduce model complexity, excessively static configurations are suboptimal as they fail to capture intricate spectral patterns. The results underscore the synergy between frequency-adaptive convolution and temporal attention pooling, emphasizing that both frequency and temporal adaptivity are critical to achieving robust and efficient SED systems.

\subsection{TAP + MFD and MDFD Conv (Various Configurations)}

To further evaluate the robustness of TAP, we integrate it into two frequency-adaptive convolution variants: MFD conv, which employs multiple non-dilated dynamic branches, and MDFD conv, which extends MFD by incorporating a variety of dilation sizes across branches. These architectures aim to capture diverse spectral patterns by increasing kernel diversity. However, their reliance on temporal average pooling may limit performance on transient-rich sound events. Integrating TAP into both structures is intended to refine temporal sensitivity while preserving their spectral adaptability.

\begin{table*}[t]
    \centering
    \caption{Ablation study on TAP integrated with MFD and MDFD conv (various configurations).}
    \begin{tabular}{l|c|c|c|c}
        \hline
        \textbf{Model} & \textbf{\# Channels} & \textbf{Dilation Sizes} & \textbf{Params (M)} & \textbf{PSDS1} \\
        \hline
        TFD conv & 1/1 (32, 64, 128, 256) & (1) & 12.703 & 0.455 \\
        MDFD(1/8) & 11/8 (44, 88, 176, 352) & (1)×5+(2,3)+(2,2,3)+(2,3,3) & 18.157 & 0.455 \\
        \hline
        TAP + MFD(1/4) & 4/4 (32, 64, 128, 256) & (1)×2 & 11.274 & 0.441 \\
        TAP + MFD(1/4) & 4/4 (32, 64, 128, 256) & (1)×3 & 14.697 & 0.452 \\
        TAP + MFD(1/8) & 8/8 (32, 64, 128, 256) & (1)×5 & 17.501 & 0.451 \\
        TAP + MFD(1/8) & 8/8 (32, 64, 128, 256) & (1)×6 & 20.116 & 0.452 \\
        TAP + MFD(1/16) & 16/16 (32, 64, 128, 256) & (1)×10 & 26.532 & 0.448 \\
        TAP + MFD(1/16) & 16/16 (32, 64, 128, 256) & (1)×11 & 28.742 & 0.450 \\
        TAP + MFD(1/16) & 16/16 (32, 64, 128, 256) & (1)×12 & 30.953 & 0.447 \\
        \hline
        \hline
        TAP + MFD(1/4) & 5/4 (40, 80, 160, 320) & (1)×4 & 25.266 & \textbf{0.459} \\
        TAP + MFD(1/4) & 6/4 (48, 96, 192, 384) & (1)×5 & 40.100 & 0.446 \\
        \hline
        TAP + MDFD(1/4) & 5/4 (40, 80, 160, 320) & (1)×3+(2,2,3) & 25.266 & 0.451 \\
        TAP + MDFD(1/4) & 5/4 (40, 80, 160, 320) & (1)×3+(2,3,3) & 25.266 & 0.446 \\
        TAP + MDFD(1/4) & 5/4 (40, 80, 160, 320) & (1)×2+(2,2,3)+(2,3,3) & 25.266 & 0.446 \\
        TAP + MDFD(1/4) & 5/4 (40, 80, 160, 320) & (1)+(2,3)+(2,2,3)+(2,3,3) & 25.266 & 0.445 \\
        \hline
    \end{tabular}
    \label{tab:tap_mdfd_ablation}
\end{table*}

As shown in Table~\ref{tab:tap_mdfd_ablation}, combining TAP with MFD yields mixed results. While shallow or moderately deep configurations (e.g., TAP + MFD(1/4) with (1)×3) perform reasonably (PSDS1 = 0.452), deeper MFD structures with up to 12 branches show performance degradation, likely due to redundancy and optimization instability. This trend is particularly evident in TAP + MFD(1/16), where over-branched structures introduce diminishing returns and disrupt TAP’s transient sensitivity.

On the other hand, integrating TAP with MDFD shows the negative effect of combining temporal and multi-scale spectral modeling. Especially, overly diverse dilation configurations, such as (1)+(2,3)+(2,2,3)+(2,3,3), lead to oversmoothing and suboptimal detection performance. The best performing MDFD combination (PSDS1 = 0.451) was achieved with a moderate multi-dilation configuration (1)×3+(2,2,3), indicating the need for restrained dilation diversity.

The highest PSDS1 of 0.459 is achieved with TAP + MFD(1/4) using 5/4 widened channels and 4 non-dilated dynamic branches, suggesting that richer channel capacity and moderately deep non-dilated structures are most compatible with TAP. This implies that while TAP greatly improves transient modeling, its synergy with spectral diversity is maximized under balanced architectural design—not through brute-force depth or excessive dilation.

\begin{table}[t]
\caption{Performance comparison of MDFD-CRNN with state-of-the-art models without external dataset. \textit{pp} = post-processing, \textit{mf cw} = class-wise median filter, \textit{ssl} = semi-supervised learning.}
\centering
\begin{tabular}{c|cc|cc}
\hline
\textbf{Model}       & \textbf{PP}  & \textbf{SSL} & \textbf{PSDS1} & \textbf{tPSDS1} \\ 
\hline
FDY-CRNN~\cite{FDY}   & cw-mf        & MT           & 0.451          & -              \\
DFD-CRNN~\cite{DFD}   & cw-mf        & MT           & 0.455          & 0.465          \\
MFD-CRNN~\cite{mdfdy} & cw-mf        & MT           & 0.461          & -              \\
MFD-CRNN~\cite{mdfdy} & cw-mf        & CMT          & \textbf{0.470} & -              \\
\hline
TFD-CRNN~\cite{TFD}   & mf           & MT           & 0.455          & -              \\ 
MDFD-CRNN~\cite{PFD}  & mf           & MT           & 0.455          & 0.468          \\ 
MDFD-CRNN~\cite{PFD}  & cw-mf        & MT           & 0.461          & 0.474          \\ 
MDFD-CRNN~\cite{PFD}  & cSEBBs       & MT           & -              & \textbf{0.485} \\ 
\hline
\end{tabular}
\label{tab:sotawdext}
\end{table}

\begin{table}[t]
\caption{Performance comparison of PFD/MDFD-CRNN with state-of-the-art models using pretrained audio models, without AudioSet or ensemble. \textit{pp} = post-processing, \textit{mf cw} = class-wise median filter, \textit{ft} = fine-tuning. ABC = ATST-frame + BEATs + CRNN.}
\centering
\begin{tabular}{c|c|c}
\hline
\textbf{Model}                                 & \textbf{PP} & \textbf{tPSDS1} \\ 
\hline
FDY-LKA-CRNN + BEATs~\cite{dcase2023a_1st}     & cw-mf       & 0.525          \\
CRNN + ATST-frame~\cite{atstsed}               & cw-mf       & 0.492          \\
CRNN + ATST-frame + ft~\cite{atstsed}          & cw-mf       & \textbf{0.583} \\
CRNN + BEATs + ft~\cite{dcase2024_1st}         & cw-mf       & 0.539          \\
\hline
ABC                                            & mf          & 0.507          \\ 
ABC w/ PFD-CRNN                                & mf          & 0.517          \\
ABC w/ MDFD-CRNN                               & mf          & 0.524          \\ 
ABC w/ MDFD-CRNN + ft                          & cw-mf       & 0.550          \\ 
\hline
ABC                                            & cSEBBs      & 0.546          \\ 
ABC w/ PFD-CRNN                                & cSEBBs      & 0.558          \\
ABC w/ MDFD-CRNN                               & cSEBBs      & \textbf{0.577} \\ 
\hline
\end{tabular}
\label{tab:dcase}
\end{table}

\section{Comparison with State-of-the-art Models}
To contextualize the performance of our proposed MDFD-CRNN model, we compare it with recent state-of-the-art (SOTA) models both with and without external datasets. Table~\ref{tab:sotawdext} presents a comparison of models evaluated under the constraint of not using external datasets such as AudioSet \cite{audioset}. Under this setting, MDFD-CRNN achieves competitive PSDS1 and true PSDS1 (tPSDS1) scores while maintaining a simpler architecture and moderate parameter count \cite{PSDS, truePSDS}.

With the use of a class-wise median filter during post-processing, MDFD-CRNN achieves a PSDS1 score of 0.461, matching the performance of the MFD-CRNN model~\cite{mdfdy}. While the integration of more advanced training strategies such as confident mean teacher (CMT) or MFD convolution modules could potentially improve performance further, we leave such extensions for future work. Notably, when change-detection-based sound event bounding boxes (cSEBBs)~\cite{SEBB} are applied to MDFD-CRNN, the model achieves a tPSDS1 of 0.485, which represents the current state-of-the-art on the DESED dataset without using external datasets~\cite{truePSDS}.

It is important to clarify that the true PSDS (tPSDS1) metric used in this study is computed in a threshold-independent manner, as proposed in~\cite{truePSDS}. This contrasts with the standard PSDS used in earlier DCASE challenges (e.g., DCASE 2021 and 2022), which average over a fixed set of 50 thresholds~\cite{DCASEtask4}. The tPSDS1 metric provides a more reliable estimate of model robustness across threshold settings, making it particularly suitable for fair model comparison.

To further evaluate the competitiveness of MDFD convolution in a broader setting, we implemented our architecture within the DCASE 2024 Task 4 challenge framework and compared it with leading models that utilize pretrained audio encoders. Table~\ref{tab:dcase} summarizes this comparison. For fairness, we exclude models that leverage ensembling or large-scale self-training with AudioSet, as those methods are primarily aimed at performance maximization rather than comparative model analysis.

Among the pretrained baselines, models such as ATST-frame and BEATs with fine-tuning achieve strong results. However, our hybrid ABC model (consisting of ATST-frame, BEATs, and CRNN) combined with MDFD-CRNN achieves a tPSDS1 of 0.550 with class-wise median filtering, and further improves to 0.577 when combined with cSEBBs. These results outperform several prior DCASE challenge winners~\cite{dcase2023a_1st, dcase2024_1st} under comparable settings, without relying on additional training data or model ensembles.

Although our proposed model does not directly target outperforming large transformer-based architectures, it demonstrates highly competitive results with a relatively simple and efficient design. Given the increasing complexity of SED benchmarks, we emphasize that model efficiency, parameter count, and transparency of architecture are critical factors in evaluating the practical viability of modern sound event detection systems.

\section{Comparison with Waveform-inspired 1D Convolutional Architectures}
In recent years, 1D convolutional encoders have gained popularity in speech processing tasks, particularly in end-to-end pretrained models such as Wav2Vec2.0 and HuBERT \cite{wav2vec2.0, hubert}. These models adopt deep 1D CNN stacks over raw waveform inputs or spectrograms, followed by transformer encoders. Motivated by their success, we investigate whether replacing the 2D convolutional encoder in our CRNN-based SED architecture with a 1D convolutional encoder can improve performance.

To this end, we construct a variant of the baseline CRNN model by replacing the 2D convolutional layers with a stack of 1D convolutional blocks. The architecture of this 1D CRNN is inspired by the encoder configurations of Wav2Vec2.0 and HuBERT, but the transformer modules are substituted with BiGRU layers to isolate the effect of convolutional front-end architecture. Both models—2D CRNN and 1D CRNN—share the same recurrent and prediction heads, allowing a fair comparison of convolutional design choices.

\begin{figure}[t]
    \centering
    \includegraphics[width=0.95\linewidth]{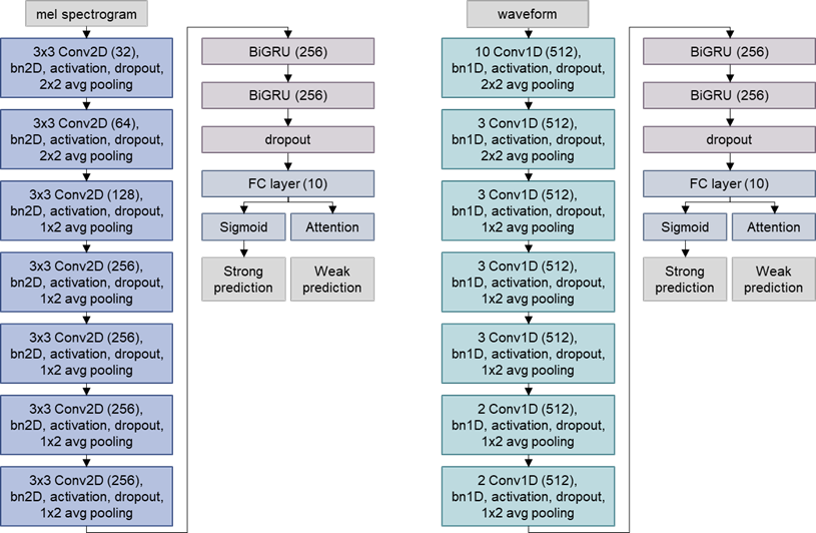}
    \caption{Comparison between 2D CRNN and 1D CRNN architectures. The 1D convolutional stack is adapted from waveform-based models such as Wav2Vec2.0 and HuBERT.}
    \label{fig:crnn_architectures}
\end{figure}

Figure~\ref{fig:crnn_architectures} illustrates the architectures of the 2D and 1D CRNN models compared in this study. Table~\ref{tab:1d_vs_2d} reports their parameter counts and PSDS1 performance on the DESED validation set.

\begin{table}[t]
    \centering
    \caption{Performance comparison between 2D CRNN and 1D CRNN architectures on DESED validation set.}
    \begin{tabular}{|c|c|c|}
        \hline
        \textbf{Model} & \textbf{Params (M)} & \textbf{PSDS1} \\
        \hline
        2D-CRNN & 4.428 & 0.410 \\
        1D-CRNN & 6.586 & 0.192 \\
        \hline
    \end{tabular}
    \label{tab:1d_vs_2d}
\end{table}

Despite having fewer parameters, the 2D CRNN model significantly outperforms the 1D CRNN counterpart. The 1D architecture, while deeper and nominally more expressive, shows a substantial drop in PSDS1 score. This result highlights a critical observation: models that perform well in large-scale speech recognition tasks may not generalize effectively to sound event detection, particularly in semi-supervised and limited-data settings.

There are several possible explanations for this outcome. First, 2D convolution preserves both time and frequency resolution in early layers, which is essential for capturing the localized spectral-temporal patterns that characterize sound events. In contrast, 1D convolution only processes along the time axis and collapses frequency information early, potentially discarding useful structure. Second, sound events in domestic environments often exhibit broad-band frequency characteristics that benefit from 2D spatial filtering. Finally, the success of waveform-based 1D models in other domains may depend heavily on large-scale pretraining and transformer encoders, which are absent in this controlled comparison.

These findings emphasize the importance of preserving both spectral and temporal resolution in SED architectures. While waveform-inspired designs may be promising in other audio domains, their effectiveness in real-world SED remains limited when not coupled with extensive pretraining or complex back-ends. We conclude that 2D convolution remains more suitable as the front-end feature extractor in CRNN-based models for sound event detection under realistic training conditions.

\begin{figure}[h]
    \centering
    \includegraphics[width=\linewidth]{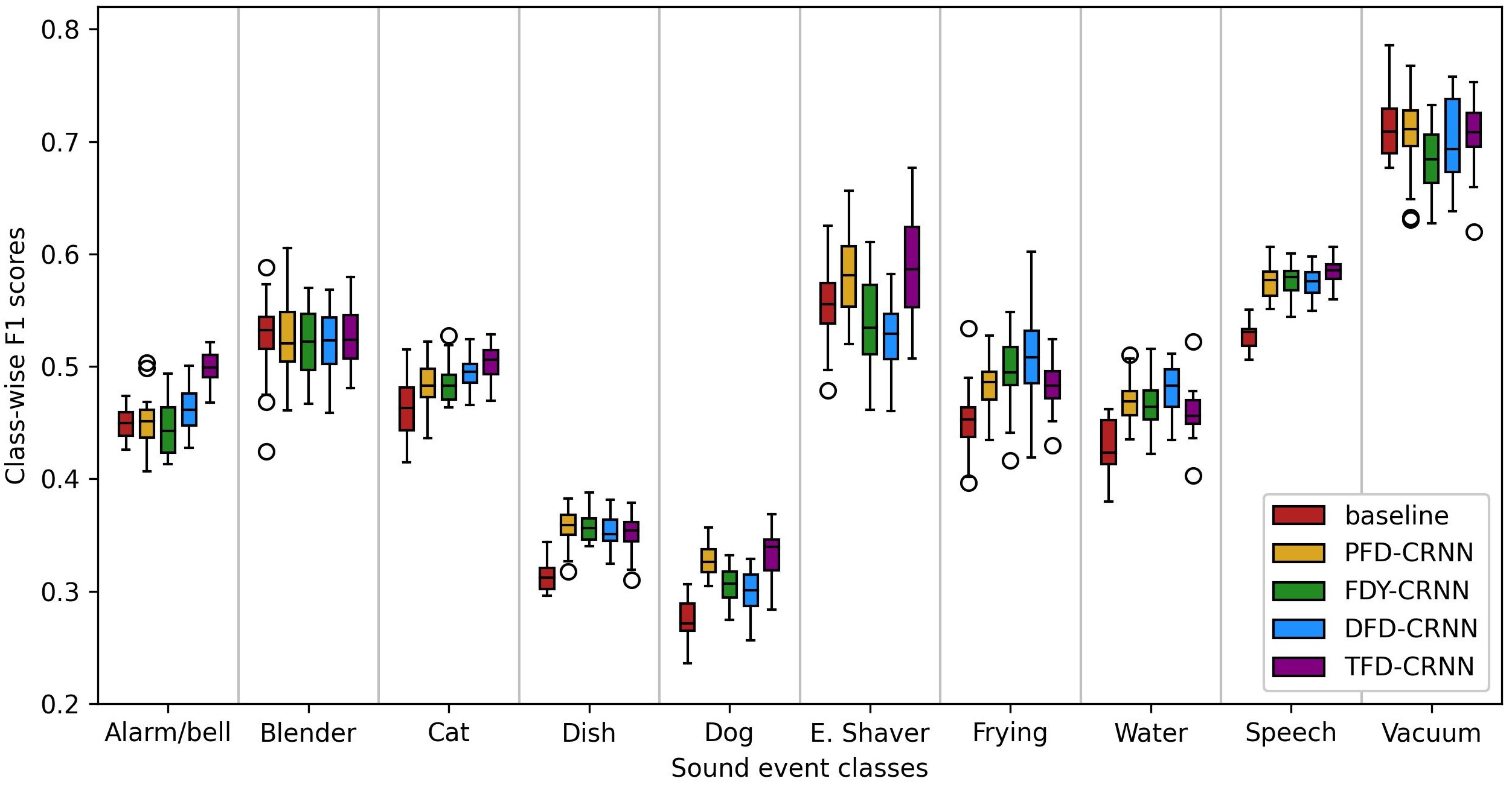}
    \caption{Class-wise PSDS1 scores for different models. Frequency-adaptive convolution methods show notable improvements, especially for non-stationary events.}
    \label{fig:classwise_psds}
\end{figure}

\begin{table}[t]
    \centering
    \caption{ANOVA + Tukey HSD post-hoc analysis results for class-wise F1 scores.}
    \begin{tabular}{l|c}
        \hline
        \textbf{Class} & \textbf{ANOVA + Tukey HSD post-hoc analysis} \\
        \hline
        Alarm/bell & FDY = PFD = Baseline $\leq$ DFD $<$ TFD \\
        Blender & Baseline = PFD = FDY = DFD = TFD \\
        Cat & Baseline $<$ PFD = FDY = DFD $\leq$ TFD \\
        Dish & Baseline $<$ PFD = FDY = DFD = TFD \\
        Dog & Baseline $<$ DFD $\leq$ FDY $\approx$ PFD $\leq$ TFD \\
        Electric Shaver & FDY, DFD $\leq$ Baseline $\approx$ PFD $\leq$ TFD \\
        Frying & Baseline $<$ PFD = FDY = DFD = TFD \\
        Running Water & Baseline $<$ PFD = FDY = DFD = TFD \\
        Speech & Baseline $<$ PFD = FDY = DFD = TFD \\
        Vacuum Cleaner & FDY $\leq$ PFD = DFD = TFD $\leq$ Baseline \\
        \hline
    \end{tabular}
    \label{tab:anova_results}
\end{table}

\section{Class-wise Performance Comparison with Statistical Analysis}
To gain deeper insights into how frequency-dependent convolution methods affect the detection of individual sound events, we conduct a class-wise performance analysis. Given that different sound event classes exhibit varying degrees of spectral stationarity, duration, and temporal sharpness, the effect of model architecture is expected to vary significantly by class. In particular, we hypothesize that transient or non-stationary events will benefit more from frequency-adaptive feature extraction.

Figure~\ref{fig:classwise_psds} illustrates the distribution of class-wise PSDS1 scores across five models: the baseline CRNN and four frequency-adaptive variants (PFD, FDY, DFD, and TFD conv). To statistically validate these observations, we perform a one-way analysis of variance (ANOVA) followed by Tukey's honest significant difference (HSD) post-hoc test. The ANOVA results confirm that model type has a statistically significant impact on F1 scores across most sound event classes ($p < 0.05$), and the detailed post-hoc comparisons in Table~\ref{tab:anova_results} further clarify the relative ranking of models for each class.

The baseline model exhibits strong performance on quasi-stationary events such as \textit{vacuum cleaner} and \textit{electric shaver}, which maintain relatively consistent spectral content over time. However, its performance drops significantly on more dynamic events such as \textit{cat}, \textit{dishes}, and \textit{dog barking}, likely due to its use of frequency-invariant convolutional kernels.

FDY-CRNN improves over the baseline in a broad range of non-stationary classes, including \textit{cat}, \textit{dish}, \textit{dog}, \textit{frying}, \textit{running water}, and \textit{speech}. This validates the effectiveness of frequency-dynamic convolution in capturing localized spectral variations. PFD-CRNN, while being more lightweight, shows competitive performance with FDY and even outperforms it in certain categories such as \textit{dog} and \textit{electric shaver}, indicating that partial dynamic convolution can strike a useful balance between model complexity and performance.

DFD-CRNN demonstrates improved detection for events like \textit{alarm/bell}, which may benefit from its expanded frequency receptive field enabled by dilation. However, for most other classes, its performance is statistically similar to that of FDY. In contrast, TFD-CRNN achieves the most consistent improvements across all event types, significantly outperforming the other models in both non-stationary classes (e.g., \textit{cat}, \textit{dog}, \textit{speech}) and even some stationary ones such as \textit{alarm/bell} and \textit{electric shaver}. These results highlight the strength of combining temporal attention mechanisms with frequency-dynamic kernels.

In summary, the statistical analysis confirms that frequency-adaptive convolution techniques offer clear advantages for detecting complex, rapidly changing sound events. While FDY, PFD, and DFD conv provide incremental improvements, TFD conv emerges as the most robust architecture across diverse sound types. These findings support the hypothesis that jointly modeling spectral dynamics and temporal saliency is key to building high-performing SED systems.

%% file: sections/5_casestudy.tex
\section{Overview}
SED is an essential area of research aimed at accurately identifying sound events and providing precise temporal information about their occurrences. This research is particularly relevant to numerous applications, such as robotics, surveillance systems, and automated monitoring in various environments. Traditional SED models, particularly those based on 2D convolutional layers, are often limited by the assumption of translational equivariance, which is more suited to image data than audio data. Specifically, 2D convolution imposes shift-invariance along both the time and frequency axes, which is not appropriate for audio, where frequency-dependent features often define the characteristics of sound events.

In response to this issue, frequency-dependent convolution methods such as FDY conv, DFD conv, PFD conv, and MDFD conv have been proposed. These methods aim to enhance the performance of SED systems by allowing convolution kernels to adapt to the frequency content of the audio input, addressing the translational equivariance problem along the frequency axis. This allows the models to better capture the varying spectral features that are typical in audio signals, particularly in sound events that exhibit rich, dynamic, and frequency-dependent patterns.

This study builds upon these approaches by introducing TFD conv, which integrates temporal attention pooling into the frequency-adaptive convolution framework. The goal of this new model is to improve the system’s sensitivity to transient sound events, such as alarms or speech, which often contain critical information in narrow time windows. TFD conv focuses on learning the temporal aspects of transient events, thereby improving detection performance for events that are both brief and time-sensitive.

The chapter procee with a detailed application of FDY conv, DFD conv, PFD conv, and TFD conv to real-world case studies. These case studies are designed to evaluate the performance of these models under practical conditions, examining their strengths and limitations in various environments. Specifically, the effectiveness of these models in dynamic and quasi-stationardsy sound event recognition will be explored. The case studies highlight the advantages of frequency-dependent methods in handling diverse acoustic patterns encountered in real-world applications, from sound event detection in dynamic environments to motor fault recognition in industrial settings.

\begin{figure}[h!]
\centering
\includegraphics[width=0.95\textwidth]{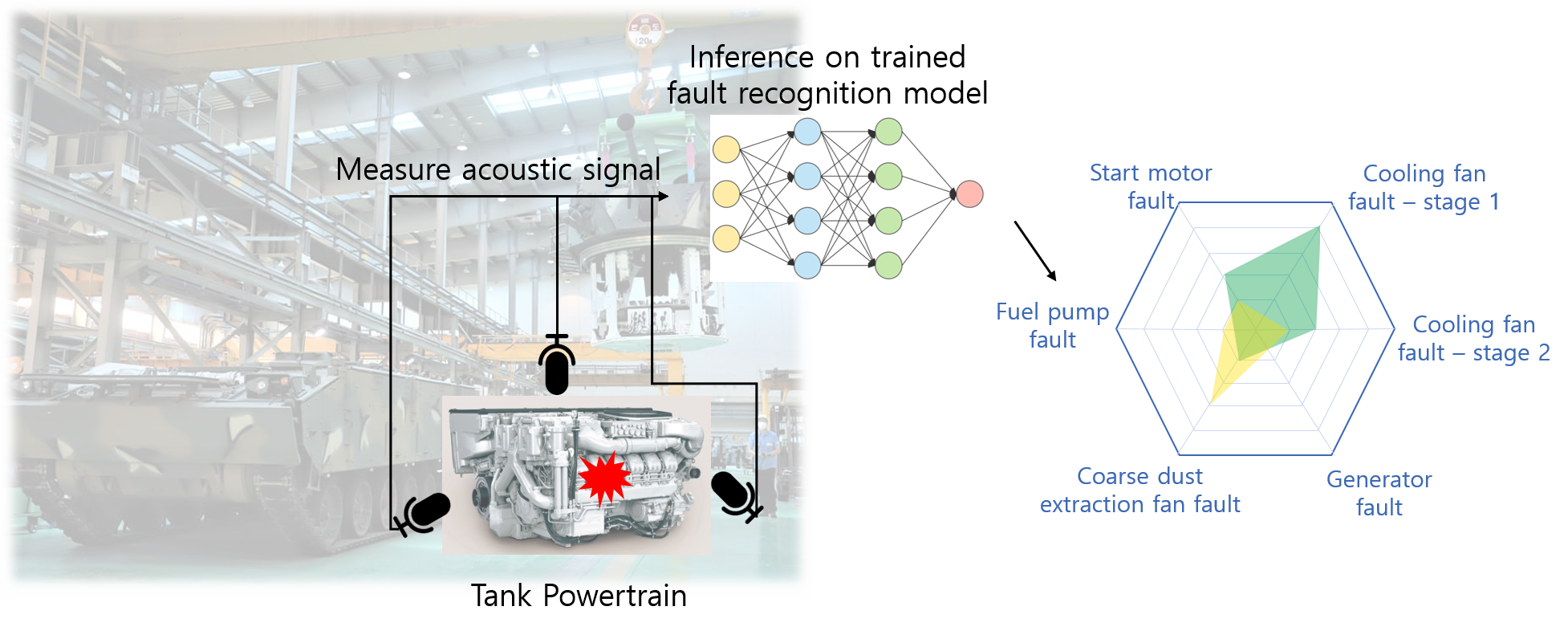}
\caption{The overall task workflow of the proposed fault diagnosis system for the K1 tank powertrain. The process includes data collection, AI model training, real-time fault detection, and visual representation.}
\label{fig:tankpowertrain}
\end{figure}

\section{Case Study 1: Tank Powertrain Fault Recognition}

\subsection{Problem Statement} 
In military operations, ensuring the operational readiness of tanks and other heavy military equipment is paramount for mission success. Given the strategic importance of these vehicles in combat situations, their downtime due to malfunctions can lead to severe consequences, both tactically and operationally. Traditional maintenance methods, which often rely on the expertise of mechanics combined with basic diagnostic tools, tend to be time-consuming and susceptible to human error. These methods are also typically reactive rather than predictive, often addressing faults only after they have led to significant degradation in performance. This reactive approach is not ideal in high-stakes environments where the timely detection and resolution of faults is crucial for maintaining combat readiness.

Powertrain failures are particularly critical in tanks, as these components are responsible for the vehicle’s movement, energy generation, and overall functioning. Malfunctions in the powertrain, if not identified early, can lead to catastrophic system failures that leave the vehicle inoperable, potentially jeopardizing the safety and effectiveness of military operations. The need for a more efficient, accurate, and real-time fault detection system is therefore increasingly urgent. By implementing advanced diagnostic technologies, military vehicles can achieve higher levels of operational reliability, and maintenance personnel can respond more proactively to faults, minimizing downtime and maximizing mission success.

This study aims to address these challenges by developing a state-of-the-art fault diagnosis system for the K1 tank powertrain, leveraging sound signals collected during its operation. The proposed system utilizes cutting-edge artificial intelligence algorithms to continuously monitor the health of the powertrain, enabling real-time fault detection. By analyzing sound patterns associated with specific failure modes, the system can identify issues before they escalate into major problems, thus improving the efficiency of maintenance operations, reducing downtime, and significantly enhancing the overall operational readiness of the K1 tank. This proactive approach to fault detection promises not only to streamline maintenance workflows but also to enhance the safety and tactical effectiveness of military forces by ensuring their equipment remains fully operational when it matters most.

\begin{figure}[htbp]
    \centering
    \includegraphics[width=\linewidth]{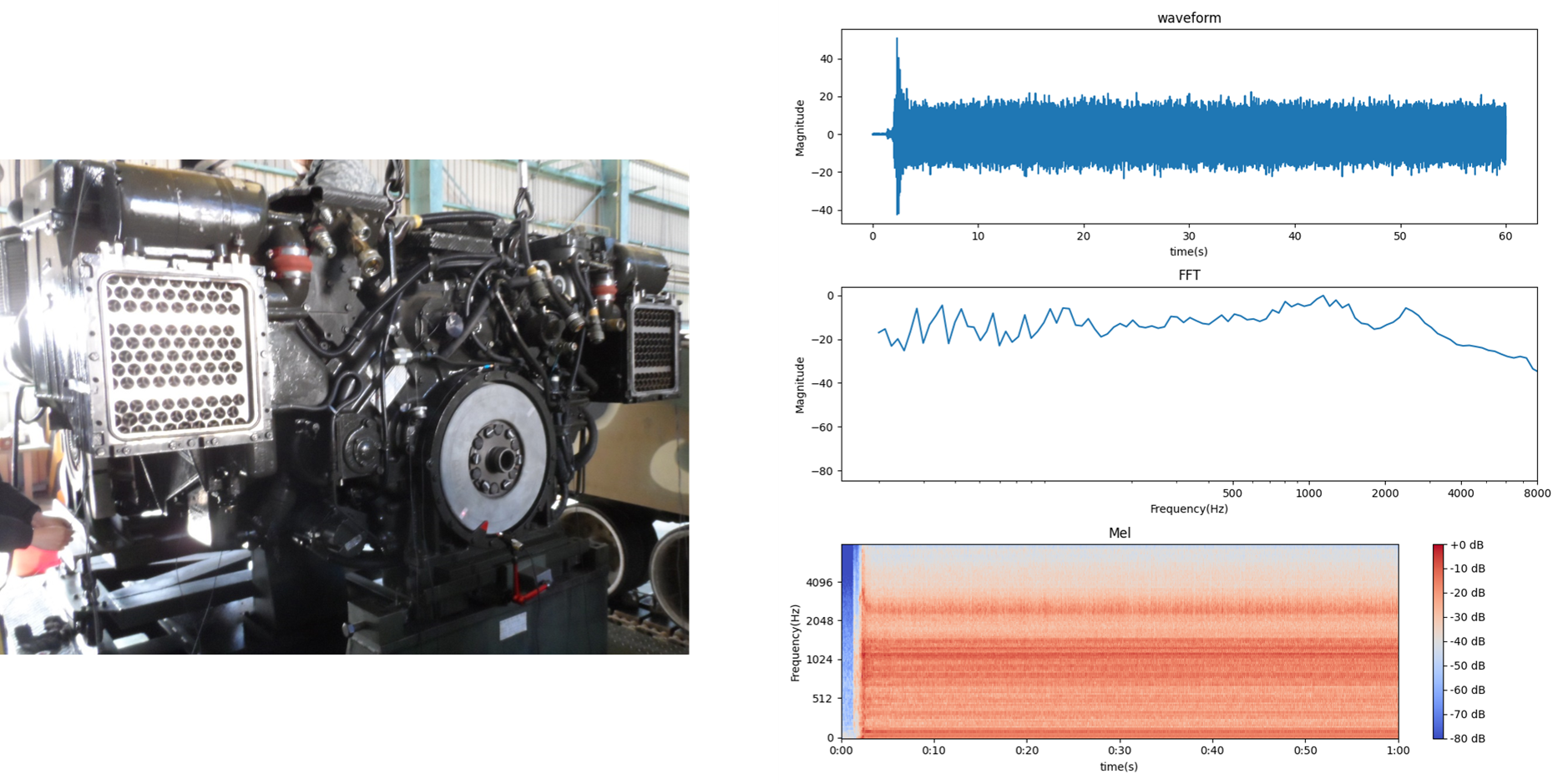}
    \caption{Visual representation of data collected from the K1 tank powertrain system. The left panel shows the powertrain setup with sensor placements. The right panels present the corresponding acoustic signal in three forms: raw waveform, frequency-domain representation (FFT), and Mel-spectrogram. These recordings were acquired using high-fidelity microphones and accelerometers placed at critical locations on the powertrain. Data was collected from 16 K1 tanks, including both normal and faulty states, with sampling rates up to 102.4 kHz to enable precise analysis of mechanical faults such as fan drive, starter motor, dust extractor, and generator failures.}
    \label{fig:k1_powertrain_data}
\end{figure}

\subsection{Data Description}
The dataset used in this study comprises sound signals collected from various components of the K1 tank's powertrain. A total of 16 K1 tanks were used to gather data, providing a diverse range of conditions including both normal operational states and various fault types. The data was collected through a robust sensor system that was designed to capture high-fidelity acoustic signals under real-world operating conditions. Specifically, five high-sensitivity microphones were used to capture sound signals, which were strategically placed around the powertrain to capture a wide range of acoustic signals.

The microphones were placed around the powertrain, including front, side, back, bottom and top, to capture a wide range of acoustic signals. The microphones were specifically chosen to ensure they captured both stationary and transient sound events that are indicative of mechanical faults. 

The data was recorded at a high sampling rate of 102.4 kHz, ensuring that even brief and transient sound events could be captured with precision. Each recording lasted for at least 60 seconds, allowing the system to monitor the powertrain under various operational conditions. In total, over 3,761 recordings were collected, stored in the TDMS (Technical Data Management Streaming) file format. This format allows for efficient data storage and retrieval, making it suitable for large-scale analysis and model training.

The dataset was meticulously labeled to differentiate between normal operational states and the following fault types:
\begin{itemize}
    \item \textbf{Fan Drive Motor Fault} – This fault occurs when the fan fails to rotate at the correct speed due to a malfunction, which can impact cooling and air flow.
    \item \textbf{Starter Motor Fault} – A malfunction in the starter motor's pinion gear that prevents the engine from starting properly.
    \item \textbf{Dust Extractor Fan Fault} – A failure in the dust extractor fan motor that reduces its efficiency, potentially affecting air quality and operational performance.
    \item \textbf{Generator Fault} – A failure in the generator's gear mechanism that affects power generation, which is critical for the operation of the tank's electrical systems.
\end{itemize}

These fault types were selected based on their impact on the overall performance of the powertrain. The signals collected from the microphones were carefully synchronized to ensure that all data from the microphones were aligned temporally, enabling an accurate comparison between normal and faulty states. This synchronization allows for a comprehensive analysis of how the different fault types manifest in the acoustic data, which is essential for training the AI-based fault diagnosis system.

The dataset serves as the foundation for developing a machine learning model that can automatically detect these faults in real-time, improving the efficiency of maintenance operations and reducing downtime. Through the analysis of these sound signals, the study aims to demonstrate how advanced AI algorithms can be leveraged to improve the operational readiness of military vehicles such as the K1 tank.

\subsection{Experimental Setup}
The experimental setup for this study was designed to collect high-quality sound data from the K1 tank powertrains under both normal and faulty conditions. The primary goal was to simulate real-world operational scenarios where the tank’s powertrain could exhibit both typical and malfunctioning behavior. A versatile and adaptable data acquisition system was implemented to ensure that it could be easily deployed across various environments and capture a wide range of acoustic signals, crucial for fault detection and system performance analysis.

To capture these acoustic signals, five high-fidelity microphones were strategically positioned at critical locations around the powertrain. These sensors were specifically chosen to monitor the sound emitted by various components of the powertrain, including the engine, transmission, and other vital mechanical parts. The microphones were placed at multiple angles around the powertrain, such as the front, sides, and back, to ensure that all relevant sound events, including transient and stationary faults, were effectively captured. The sensors were placed at a distance of approximately 1 meter from the surface of the powertrain, allowing for a balance between signal capture and interference from other mechanical noises.

The sensors were configured to capture sound signals at a high temporal resolution of 102.4 kHz, ensuring that even subtle acoustic anomalies, such as transient fault signatures, were detected with high accuracy. By capturing fine temporal details, the system was able to provide a comprehensive picture of the powertrain's sound profile, making it possible to discern between normal operational noise and early-stage faults that could lead to major system failures if left undetected.

The recorded data was stored in the TDMS (Technical Data Management Streaming) file format, which allowed for efficient data management and preservation of the high-resolution sound signals. TDMS format, with its hierarchical structure, ensured that large volumes of data could be handled and processed effectively, while maintaining the integrity of the raw sound data. This allowed for seamless retrieval and analysis of the collected data, ensuring that no important acoustic features were lost during the storage process.

\begin{table}[htbp]
\centering
\begin{tabular}{l|c|c|c|c|c}
\hline
\textbf{Models} & \textbf{Overall} & \textbf{Cooling fan} & \textbf{Generator} & \textbf{Starter} & \textbf{Dust extractor} \\
\hline
CNN & \textbf{93.1$\pm$9} & \textbf{98.2$\pm$2.6} & \textbf{96.9$\pm$4.9} & 97.2$\pm$8.6 & \textbf{96.7$\pm$3.0} \\
FDY-CNN & 89.8$\pm$10.7 & 96.2$\pm$5.5 & 95.2$\pm$5.1 & 96.8$\pm$8.4 & 95.3$\pm$4.8 \\
PFD-CNN & 92.0$\pm$8.8 & 97.3$\pm$4.5 & 96.1$\pm$5.3 & 98.1$\pm$6.0 & 95.8$\pm$4.5 \\
DFD-CNN & 88.4$\pm$10.5 & 96.1$\pm$4.8 & 94.4$\pm$5.9 & 96.7$\pm$8.0 & 94.6$\pm$4.6 \\
TFD-CNN & 90.6$\pm$8.8 & 96.9$\pm$4.6 & 95.6$\pm$5.1 & \textbf{98.6$\pm$3.0} & 94.8$\pm$4.5 \\
\hline
\end{tabular}
\caption{Performance comparison (\%) across different machine types and overall.}
\label{tab:comparison}
\end{table}

\subsection{Results and Discussion}
The collected sound data was processed using advanced signal processing techniques to extract meaningful features that could indicate faults in the powertrain. The primary tool used for feature extraction was Mel Spectrogram analysis. This allowed the study to capture key frequency bands and temporal patterns that are typically associated with specific fault types. By analyzing the time-frequency domain representation of the data, the study could differentiate the acoustic signatures of faulty powertrain components from those of healthy ones.

A critical step in the analysis was comparing the sound signals from both normal and faulty states. The system identified distinct differences in the frequency and amplitude of the signals, which were then used to classify the fault types accurately. For example, faults in components like the cooling fan or the generator exhibited specific frequency patterns that could be isolated using the Mel-spectrogram representation. This method enabled a high degree of accuracy in distinguishing between operational sounds and fault signals, which is key for real-time fault detection in military vehicles.

The results of the study, as shown in Table \ref{tab:comparison}, highlight the effectiveness of the AI-based fault diagnosis system. The system achieved an overall accuracy rate of over 85\% in detecting and classifying the four primary fault types: fan drive motor faults, starter motor faults, dust extractor fan faults, and generator faults. The system was also able to distinguish between normal and faulty conditions with high precision, even in the presence of background noise and other operational variations. CNN performs the best in overall, since the tank powertrain signals are mostly quasi-stationary sounds thus results in simple horizontal time-frequency patterns which could be easily captured by simple CNN. Along FDY conv variants, PFD-CNN with static branch performed the best, followed by TFD-CNN with temporal attention feature. Notably, the performance of the TFD-CNN model was superior to the other models, achieving the highest accuracy in detecting starter motor faults, which are typically harder to identify due to their transient nature. This result is coherent to the results and analysis in the previous chapter.

Another key feature of the system is its ability to process data in real-time. The AI model can provide instant feedback to maintenance personnel, allowing for timely intervention before a small issue escalates into a larger, more costly failure. This real-time capability is essential for maintaining the operational readiness of military vehicles, particularly in combat scenarios where downtime can have critical consequences.

\subsection{Conclusion}
This case study demonstrates the successful application of advanced AI techniques in improving the fault diagnosis process for the K1 tank powertrain. By leveraging sound data collected from high-fidelity microphones placed strategically around the powertrain, the proposed AI-based system was able to achieve high accuracy in real-time fault detection and classification. The system's ability to distinguish between normal operational states and various fault conditions, such as fan drive motor faults, starter motor faults, dust extractor fan faults, and generator faults, showcases its potential to significantly enhance the operational efficiency of military vehicles.

The results highlight the power of AI in revolutionizing maintenance practices, particularly in military applications where operational readiness is critical. By automating fault detection and providing real-time feedback to maintenance personnel, the system reduces downtime, increases vehicle availability, and minimizes the risk of catastrophic failures. These advantages are especially important in combat or other mission-critical environments, where equipment failure can severely impact mission success.

The study also underscores the importance of continuous refinement and optimization in AI-based diagnostic systems. While the system showed promising results, several challenges remain, including the need for more robust noise filtering techniques and improvements in the model's ability to generalize across different operational conditions. Future work will focus on addressing these challenges, with particular emphasis on enhancing the system's ability to handle more complex fault scenarios, such as those involving subtle or intermittent faults. Additionally, further optimization of the feature extraction and classification models will improve the system’s robustness and reliability in diverse environments.

Ultimately, this case study paves the way for smarter, more efficient maintenance practices in military vehicles. By integrating advanced AI techniques, the system not only improves fault detection accuracy but also enables more proactive and predictive maintenance strategies, ensuring that critical military equipment remains operational and ready for deployment at all times.

\begin{figure}[h!]
\centering
\includegraphics[width=0.95\textwidth]{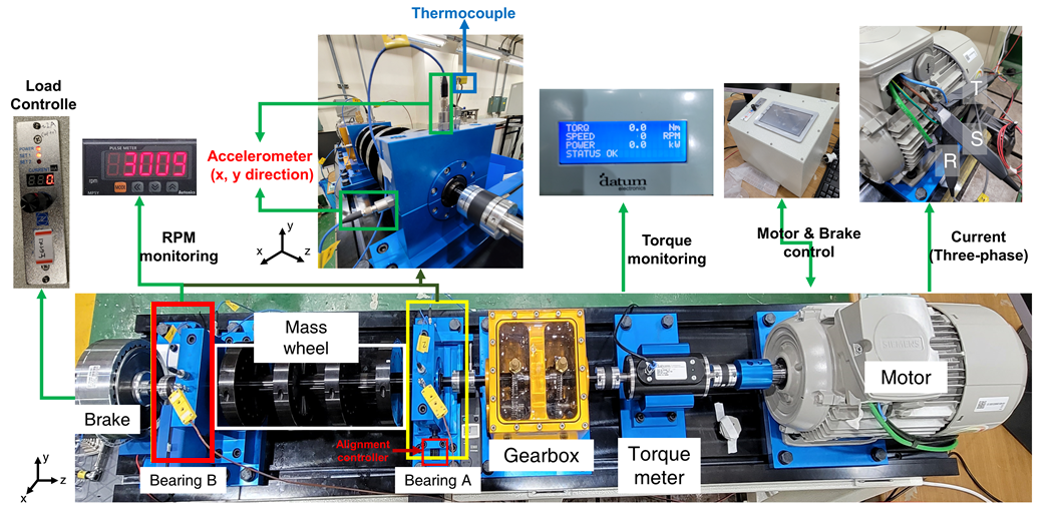}
\caption{Detailed configuration of the testbed including RPM monitoring, torque sensing, thermocouple, and multi-fault injection units. The system was designed to emulate realistic compound motor faults under variable operational conditions.}
\label{fig:motorfault}
\end{figure}

\section{Case Study 2: Speed-Varying Motor Fault Recognition}
\subsection{Problem Statement}
In this study, we aim to explore the generalizability of frequency-adaptive convolution methods beyond their traditional use in SED. Specifically, we conduct a second case study focusing on compound motor fault recognition under variable-speed conditions. Motor faults, especially in industrial and automotive applications, are often complex and multifaceted, requiring advanced methods for accurate detection. Unlike stationary sound events, motor faults are influenced by a variety of dynamic factors such as changes in rotational speed, load variations, and operational wear, which can significantly alter the spectral and temporal characteristics of vibration and acoustic signals \cite{motor1, motor2}.

The main challenge in recognizing motor faults under variable-speed conditions lies in the spectral and temporal variability introduced by these dynamic operational conditions. As the motor speed fluctuates, the frequency content of the emitted signals also changes, creating a moving target for conventional detection methods. The standard convolutional approaches often struggle to adapt to these shifts, as they tend to assume stationary patterns over time. Therefore, this study explores how frequency-adaptive convolution methods, such as FDY-CNN, DFD-CNN, and PFD-CNN, can be applied to capture these varying patterns effectively, offering a potential solution for more robust motor fault detection.

Additionally, motor faults can manifest differently depending on the type of fault and its severity. For example, faults such as inner race faults (IRF), outer race faults (ORF), shaft misalignment, and unbalanced mass each produce distinct vibration patterns. These patterns can be significantly affected by changes in motor speed, which may further complicate fault detection. This dynamic nature introduces a layer of complexity, as the system needs to accurately identify these fault types despite the continuous variation in signal characteristics due to changing operational conditions. Addressing this requires a model that is not only sensitive to the presence of faults but also robust enough to handle the variability introduced by the dynamic system.

Moreover, the traditional approach of analyzing stationary signals might miss transient fault events or fail to generalize across varying speeds. To tackle this, we incorporate frequency-adaptive convolutional models that can dynamically adjust to frequency shifts, enabling better detection of motor faults under these variable conditions. The ultimate goal of this case study is to assess the ability of these advanced convolution models to handle the complexity and variability inherent in motor systems, demonstrating their potential for more reliable and accurate fault detection in real-world industrial applications where motor speeds and operating conditions fluctuate continuously.

\begin{figure}[h!]
\centering
\includegraphics[width=0.95\textwidth]{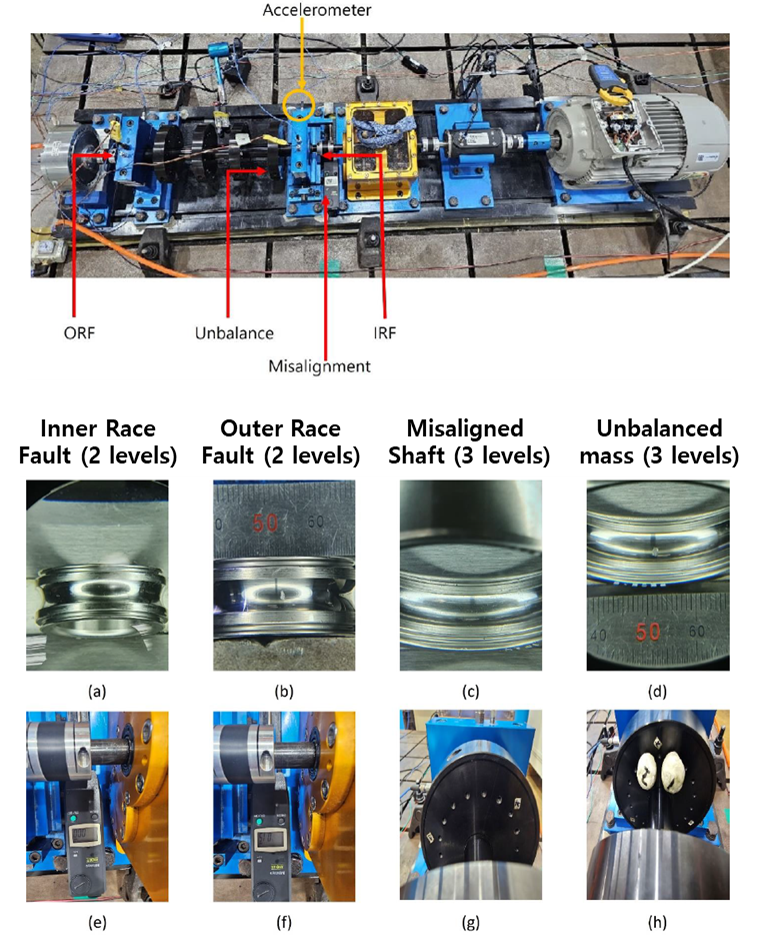}
\caption{Experimental setup for compound motor fault diagnosis under variable speed conditions. Four fault types (IRF, ORF, misalignment, unbalance) were implemented with multiple severity levels. Vibration signals were collected using accelerometers mounted on the bearing housing.}
\label{fig:motorfaults}
\end{figure}

\begin{figure}[h!]
\centering
\includegraphics[width=0.95\textwidth]{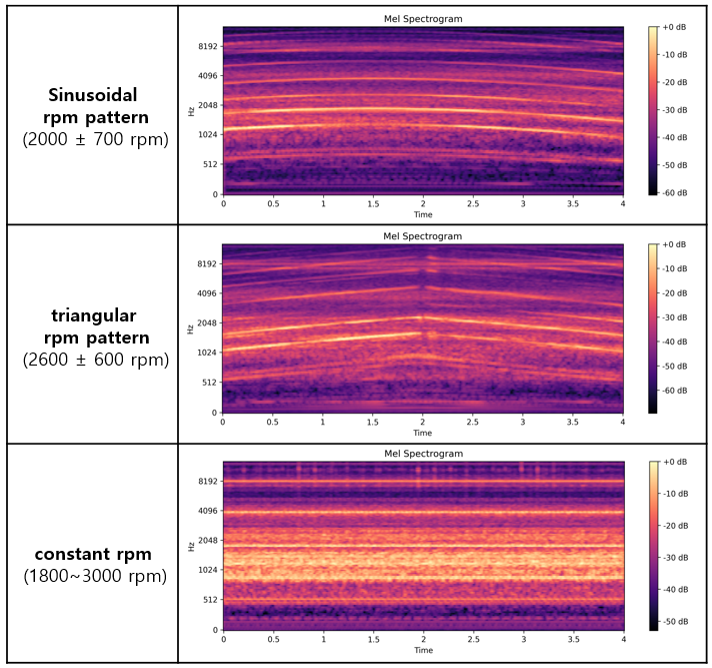}
\caption{Mel spectrograms of vibration signals under three speed profiles: (top) sinusoidal (2000 ± 700 rpm), (middle) triangular (2600 ± 600 rpm), and (bottom) constant (1800–3000 rpm). Each rpm pattern produces distinct time-frequency characteristics.}
\label{fig:motorspeed}
\end{figure}

\subsection{Data Description}
The testbed used for this study consisted of a motor-driven mechanical system designed to simulate real-world conditions for fault detection. The system included controllable fault injection at four critical locations: inner race fault (IRF), outer race fault (ORF), shaft misalignment, and unbalanced mass. Each fault type was implemented with varying severity levels to simulate different operational scenarios. Specifically, the IRF and ORF each had two severity levels, while shaft misalignment and unbalanced mass faults were implemented with three different severity levels. This resulted in a total of 36 unique compound fault combinations, providing a diverse set of data to train and test the fault detection models.

The fault injection system was coupled with vibration data collection using a high-sensitivity accelerometer mounted on the bearing housing. The accelerometer was positioned to capture vibrations generated by the motor during operation, which are indicative of faults in the system. The collected vibration signals were then processed and transformed into spectrograms using Short-Time Fourier Transform (STFT). Spectrograms were chosen for their ability to capture both time and frequency information, which is crucial for analyzing the spectral features of motor vibrations and detecting faults.

To ensure that the testbed accurately represented real-world operating conditions, the system was run under three different rpm (revolutions per minute) profiles: Sinusoidal (2000 ± 700 rpm), which simulates periodic speed variations typical of many mechanical systems, Triangular (2600 ± 600 rpm), which mimics more abrupt and cyclical speed fluctuations, often seen in systems with irregular load conditions, Constant (1800–3000 rpm), which represents steady operation where the motor runs at a relatively constant speed, typical for systems operating under constant load conditions.

Each of these rpm profiles induced distinct time-frequency patterns in the motor's vibration signals, which were captured by the accelerometer and transformed into spectrograms for analysis. The variability in rpm profiles is critical for emulating the dynamic operating conditions found in real-world scenarios, where motor speeds can fluctuate due to load changes, temperature variations, and other factors. The different rpm profiles allowed the study to explore how fault detection models could handle a variety of motor operating states and better generalize across real-world conditions.

The resulting dataset, which consists of vibration signals under varying fault and operational conditions, serves as the foundation for training and evaluating the performance of the proposed AI-based fault detection models. By utilizing these diverse operational scenarios, the dataset enables the models to learn how motor faults manifest under dynamic conditions, improving their robustness and accuracy in real-world applications.

\subsection{Experimental Setup}
For the motor fault recognition task, we formulated the problem as a multi-output classification task. The goal was to predict the severity levels of four different fault types: inner race fault (IRF), outer race fault (ORF), shaft misalignment, and unbalanced mass. Each fault type had a different number of severity levels, with IRF and ORF having 2 severity levels each, while shaft misalignment and unbalanced mass had 3 severity levels each. This resulted in a total of 36 unique fault combinations, each corresponding to a specific severity pattern, which allowed us to frame the task as a 36-class compound classification problem.

To solve this, we employed a 2D convolutional Neural Network (CNN) architecture, a model widely used for tasks involving time-frequency representations such as spectrograms. The input to the network was a Mel spectrogram generated from vibration signals. The 2D CNN model was designed with one output branch per fault type, ensuring that each branch could independently predict the severity level of its corresponding fault. This architecture allows for the simultaneous prediction of multiple faults and their severity levels, making the model highly suitable for compound fault detection scenarios.

During training, the Cross Entropy loss function was utilized, which is commonly used for classification tasks. The 36-class compound labels were used as the target output during training, where each label represents one of the possible fault combinations, with its severity levels encoded in the labels. The dataset used for training was carefully labeled, ensuring that each fault type and its associated severity level were clearly identified in the recordings. This allowed the model to learn the correlation between different severity levels and the acoustic patterns that represent each fault.

The model was trained using a combination of batch normalization and dropout techniques to improve generalization and prevent overfitting. Data augmentation techniques, including time-shifting and mixup, were also applied to make the model more robust to small variations in the input data, which is common in real-world industrial applications. The training process was performed on a large dataset containing recordings from various motor states under different fault conditions, with the goal of ensuring the model could generalize across various operational environments and fault severities.

By designing the network in this way, the system was able to simultaneously handle multiple fault types and their severity levels, making it a comprehensive solution for real-time motor fault recognition under variable-speed conditions.

\subsection{Evaluation Results}
To evaluate the performance of the proposed models, we compared five different convolutional architectures: the standard 2D CNN, FDY-CNN, PFD-CNN, DFD-CNN, and TFD-CNN. These models were trained on the dataset described earlier, and their performance was measured using the accuracy of fault detection across different motor fault types.

As shown in Table~\ref{tab:motor_fault_results}, DFD-CNN achieved the highest overall accuracy of 86.0\%, outperforming all other models. This result suggests that the integration of frequency dilation in DFD-CNN plays a crucial role in capturing the spectrally rich characteristics of motor vibration signals. By expanding the receptive field of the convolutional kernels in the frequency dimension, DFD-CNN is better able to adapt to the varying frequency patterns that characterize motor faults, especially in systems where the fault-induced vibrations span a wide range of frequencies. The dilation technique helps the model learn and generalize more effectively across these varying spectral patterns, thus achieving superior performance in detecting motor faults.

On the other hand, FDY-CNN and PFD-CNN also showed competitive performance, achieving 84.2\% and 83.9\% accuracy, respectively. These models also leverage frequency-adaptive kernels, which enable them to capture frequency-dependent features of the motor vibration signals. While they performed well, their results were slightly behind DFD-CNN, suggesting that the frequency dilation mechanism in DFD-CNN provides an additional advantage in handling the complex spectral variations of motor fault signals.

In contrast, TFD-CNN showed the weakest performance among all models, with an accuracy of 82.0\%. This lower performance highlights the limitations of temporal pooling when dealing with relatively stationary signals such as those produced by motor faults. Temporal pooling, which is effective in tasks with transient-rich events (such as sound event detection), does not capture the subtleties of stationary motor fault signals as effectively. Unlike transient-rich tasks where rapid changes in the signal are crucial, stationary faults tend to exhibit more consistent patterns over time, and temporal attention pooling alone may not be sufficient to capture the important spectral features that differentiate the fault types.

Overall, the results demonstrate that while TFD-CNN shows promise in transient-heavy domains like sound event detection (SED), models like DFD-CNN that incorporate frequency dilation are more effective for tasks involving stationary or quasi-stationary motor fault signals, where frequency-dependent features are more critical for fault recognition.

\begin{table}[t]
\centering
\caption{Classification accuracy (\%) for each fault type and overall.}
\label{tab:motor_fault_results}
\begin{tabular}{l|c|cccc}
\hline
\textbf{Models} & \textbf{Overall} & \textbf{IRF} & \textbf{ORF} & \textbf{Misalignment} & \textbf{Unbalance} \\
\hline
CNN        & 83.7$\pm$2.4 & \textbf{99.8$\pm$0.4} & 98.5$\pm$0.9 & 89.2$\pm$1.6 & 92.9$\pm$1.7 \\
FDY-CNN    & 84.2$\pm$2.4 & 99.6$\pm$0.5 & 98.6$\pm$1.0 & 89.3$\pm$2.0 & 93.9$\pm$1.6 \\
PFD-CNN    & 83.9$\pm$2.1 & \textbf{99.8$\pm$0.5} & 98.5$\pm$0.9 & 89.0$\pm$1.8 & 93.4$\pm$1.5 \\
DFD-CNN    & \textbf{86.0$\pm$2.3} & 99.7$\pm$0.5 & \textbf{99.0$\pm$0.9} & \textbf{90.7$\pm$1.7} & \textbf{94.3$\pm$1.7} \\
TFD-CNN    & 82.0$\pm$3.2 & 99.4$\pm$0.7 & 97.1$\pm$1.5 & 86.8$\pm$2.6 & \textbf{94.3$\pm$1.7} \\
\hline
\end{tabular}
\end{table}

\subsection{Conclusion}
This case study highlights the strength of dilated frequency dynamic convolution (DFD-CNN) in effectively handling motor signals that are stationary but spectrally diverse. The DFD-CNN model, with its ability to capture frequency-dependent features through dilation in the frequency dimension, demonstrated superior performance in recognizing faults in motor systems. The results suggest that frequency dilation is particularly beneficial when dealing with motor fault signals that exhibit rich, varying spectral patterns across different fault types and severities. By expanding the receptive field of the convolutional kernels, DFD-CNN was able to better capture the subtle differences in frequency characteristics, making it highly effective for motor fault recognition in variable-speed conditions.

One of the key takeaways from this study is that the design of a model should closely align with the characteristics of the signal being analyzed. In this case, TFD-CNN, which incorporates temporal attention pooling, performed well in transient-heavy tasks like sound event detection (SED) but did not perform as effectively for the relatively stationary motor fault signals. While TFD-CNN excels in domains where rapid temporal changes are the primary distinguishing factor, such as in SED, it falls short in tasks where frequency-dependent patterns play a more critical role. This highlights the importance of selecting the appropriate model architecture based on the nature of the signal and the specific task at hand.

Overall, this study emphasizes the importance of frequency modeling in fault detection tasks that involve stationary or quasi-stationary signals. The findings suggest that DFD-CNN, by incorporating frequency-dilated convolutions, offers a more robust solution for motor fault detection, while TFD-CNN and similar temporal models are better suited for tasks involving more transient, time-varying sound events. These insights can guide future research and application in both industrial motor fault detection and other domains where frequency-dependent patterns are essential.

\section{Case Study 3: Arc Detection in Offshore Environments}
\subsection{Problem Statement}
Fires in maritime environments, particularly on ships, present unique and highly critical challenges due to the confined nature of the spaces, the limited external firefighting resources, and the difficulty of evacuating personnel in emergency situations. Fires can spread rapidly in the absence of immediate detection and response, putting lives, equipment, and cargo at risk. Electrical arc discharges are one of the primary causes of fires on ships, as they can lead to significant damage to electrical systems, which are crucial for the operation of modern vessels. These arc discharges often occur without prior warning, making early detection of such events a vital component in preventing catastrophic fires and ensuring the safety of the crew and vessel.

In offshore environments, where ships operate in isolated conditions far from immediate help, the ability to quickly detect and respond to potential fire hazards is paramount. The detection of electrical arcs, both Alternating Current (AC) and Direct Current (DC), is particularly challenging because these events are transient and can vary significantly in frequency and amplitude depending on the type of fault and environmental conditions. Traditional methods of arc detection, such as visual inspection and manual monitoring of electrical systems, are inadequate for providing timely alerts, particularly in large or complex marine environments.

To address these challenges, this study focuses on the development of an AI-based sound event recognition (SER) system for the detection of electrical arc discharges in offshore environments. Unlike traditional systems that focus on temporal localization of events, this system performs clipwise prediction, classifying audio clips based on the presence or absence of arc discharges. The system leverages advanced machine learning techniques to analyze acoustic signals captured by strategically placed microphones on the ship. By detecting the distinctive sound patterns associated with electrical arcs, the proposed system enables rapid response to potential fire hazards, improving overall safety.

This clipwise prediction system provides a reliable, real-time solution for detecting AC and DC arc discharges, enabling proactive measures to mitigate the risk of fire. The AI-based system offers an efficient alternative to manual inspection methods, helping to ensure that potential fire hazards are identified and addressed before they lead to catastrophic consequences.

\begin{figure}[htbp]
    \centering
    \includegraphics[width=\linewidth]{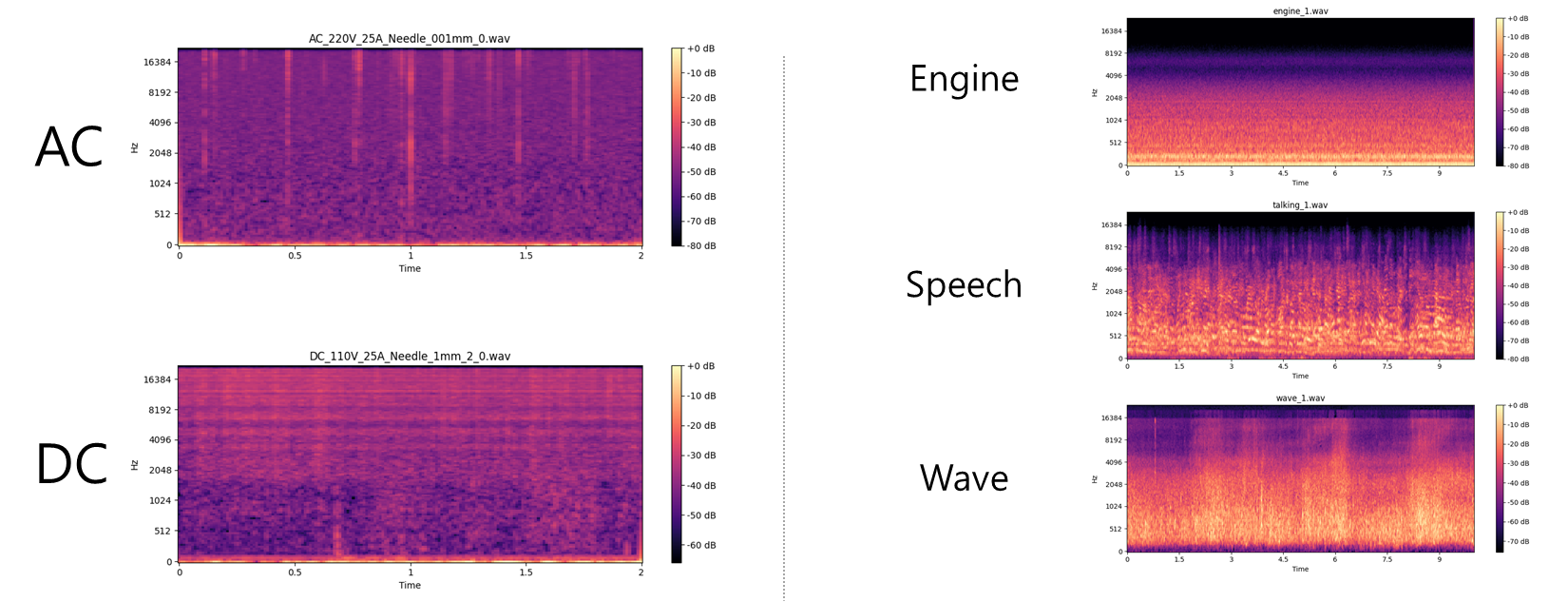}
    \caption{Spectrogram comparison of different acoustic signals recorded in offshore environments. The left column shows spectrograms of arc discharge sounds: (top) AC arc discharge with irregular transient spikes, and (bottom) DC arc discharge with more stationary harmonic patterns. The right column displays examples of background noise: (top) engine noise with low-frequency components, (middle) speech signals characterized by dense modulations, and (bottom) wave sounds with smooth, periodic textures. This dataset enables the AI-based model to learn distinctions between arc discharges and various ambient shipboard sounds.}
    \label{fig:arc_vs_background_spectrogram}
\end{figure}

\subsection{Data Description}
The dataset used in this study consists of acoustic signals collected from various electrical equipment and operational environments aboard offshore vessels. The primary objective of this dataset is to facilitate the detection of electrical arc discharges, which are a common source of fire hazards on ships, while distinguishing them from common ambient noises found in maritime settings.

The dataset includes both arc discharge sounds (AC and DC) and various background noise types. The arc discharge sounds are essential for identifying electrical faults, while the background noises—such as engine sounds, wave noises, and speeches—are crucial for training the system to differentiate between normal operational sounds and fault signals. These signals were collected using a sophisticated sensor setup, which includes five high-sensitivity microphones placed strategically around the ship's electrical equipment and power systems. The microphones are specifically tuned to capture high-frequency transient events, which are characteristic of arc discharges.

The data was recorded at 32kHz to ensure high temporal resolution, allowing for the detection of even brief and subtle arc discharges that might otherwise go unnoticed. This high sampling rate was necessary to capture the intricate details of the arc discharge sounds, which can include very rapid transient spikes and harmonic patterns. Each recording lasted at least 60 seconds to ensure that the system could monitor the equipment under different operational conditions.

In total, 307 arc discharge and 135 background recordings were collected, encompassing a wide range of conditions. The dataset was manually labeled to identify the type of event present in each recording. The main categories of events are as follows:

\begin{itemize}
    \item \textbf{Arc Discharge} – AC arc characterized by irregular high-frequency bursts of sound, with sharp transient spikes typically observed in alternating current faults, and DC arc characterized by a more stationary, harmonic frequency pattern, with weaker and more consistent transient components compared to AC arcs.
    \item \textbf{Background Noise} – This category includes sounds from stationary engine operations, somewhat stationary ambient wave noise, and non-stationary speech.
\end{itemize}

The data labeling process focused on accurately identifying the onset and offset times for each arc discharge event to allow for precise detection. Each label indicates the presence of an arc discharge within a specific clip, allowing the AI-based model to perform clipwise prediction, identifying the type of event in each audio clip without temporal localization. This data was used for training and evaluating the performance of the AI model, which aims to detect and differentiate electrical arc discharges from background noise in real-time, offering significant improvements in fire hazard detection on offshore vessels.

This dataset serves as a crucial resource for training the machine learning models, ensuring that they can effectively identify and classify arc discharge events amidst the complex soundscape of an offshore environment. By learning to distinguish between different types of electrical faults and ambient sounds, the system can provide rapid alerts and improve the safety of the vessel.

\subsection{Experimental Setup}
For the arc discharge detection task, the collected acoustic data was processed to extract time-frequency features that could best represent the characteristics of arc discharges in comparison to background noise. Key techniques such as the Short-Time Fourier Transform (STFT) and Mel-Spectrogram were employed to transform the raw audio signals into time-frequency representations. These representations allow the system to capture both the spectral content and temporal evolution of the sound events, which is crucial for differentiating between the transient nature of arc discharges and the more stationary characteristics of background noises.

Once the features were extracted, they were fed into a Convolutional Neural Network (CNN)-based model for learning. CNNs are particularly effective at identifying spatial patterns in time-frequency representations, which makes them ideal for classifying sound events like arc discharges. The experimental setup included training multiple models with different variants of frequency-dependent convolutions, namely FDY-CNN, DFD-CNN, and PFD-CNN, to explore how well these models could capture the frequency-adaptive characteristics of arc discharge sounds.

The goal of this experiment was to compare the performance of these models in detecting the transient, short-duration nature of arc discharges. Additionally, the Temporal Attention Pooling-based model, TFD-CNN, was also tested. TFD-CNN introduces an attention mechanism that helps the model focus on the most salient temporal features, which is particularly beneficial for detecting transient events like arc discharges that contain critical information within short time windows.

The models were trained on the labeled dataset, with a training-validation-test split of 70-15-15\%. The training process utilized binary cross-entropy as the loss function, which is commonly used for classification tasks where each sample belongs to one of two classes (arc discharge or background noise). The Adam optimizer was employed to minimize the loss function, ensuring efficient training of the models.

To improve the models' ability to generalize across different noise conditions and ensure robustness in diverse environments, data augmentation techniques were applied. These techniques included frequency masking and time masking, which help the models become more invariant to small variations in time and frequency, as they simulate real-world distortions like noise and signal interruptions. These augmentations also helped prevent overfitting, allowing the models to better handle previously unseen data during evaluation.

Overall, the experimental setup was designed to evaluate and compare the performance of different convolutional models in detecting electrical arc discharges in offshore environments, with a focus on improving the detection of transient, high-frequency sound events.
\begin{table}[htbp]
\centering
\begin{tabular}{l|c}
\hline
\textbf{Models} & \textbf{Accuracy (\%)} \\
\hline
CNN & 79.44$\pm$7.91 \\
FDY-CNN & 79.56$\pm$8.28 \\
PFD-CNN & 80.46$\pm$8.39 \\
DFD-CNN & 80.57$\pm$7.80 \\
TFD-CNN & \textbf{81.53$\pm$6.06} \\
\hline
\end{tabular}
\caption{Accuracy comparison (\%) across different models.}
\label{tab:accuracy}
\end{table}

\subsection{Results and Discussion}
The results of the arc discharge detection task revealed that the AI-based models were highly effective in distinguishing arc discharges from background noise. Among the models tested, the TFD-CNN model demonstrated the most superior performance in detecting transient events such as AC and DC arc discharges. The key advantage of TFD-CNN lies in its integration of Temporal Attention Pooling (TAP), which allows the model to focus on the most salient features in the acoustic signal. By giving higher weight to the transient, high-energy portions of the signal, TAP improved the model’s ability to detect rapid, brief events like arc discharges, which are typically difficult to capture using traditional methods.

In terms of overall performance, the TFD-CNN model achieved the highest accuracy of 81.53\%, surpassing all other models, including the baseline CNN model (79.44\%) and other variants such as FDY-CNN (79.56\%), PFD-CNN (80.46\%), and DFD-CNN (80.57\%). This improvement underscores the effectiveness of Temporal Attention Pooling in enhancing the detection of transient faults, making TFD-CNN the most suitable model for this task.

The model’s performance was especially dependent on its ability to capture the transient features of arc discharge events, which are characterized by rapid fluctuations in sound intensity and frequency. As demonstrated in the results, TFD-CNN outperformed all other models in this regard, particularly in detecting rapid arc events, which are typically harder to identify due to their fleeting nature. The model’s ability to focus on these critical transient features allowed it to achieve a significant accuracy improvement over the other models.

Overall, the results demonstrate the robustness and efficiency of the TFD-CNN model for detecting arc discharges in offshore environments. The integration of Temporal Attention Pooling significantly improved the model’s ability to detect transient sound events, which is crucial for early fire detection and preventing damage to electrical systems in maritime environments.

\subsection{Conclusion}
This case study demonstrates the effectiveness of frequency-dependent convolutional models, particularly TFD-CNN, in detecting electrical arc discharges in offshore environments. The proposed AI-based detection system successfully distinguishes between arc discharges and other ambient background noises, achieving high detection accuracy and robustness even in challenging real-world offshore settings. The integration of Temporal Attention Pooling in the TFD-CNN model allowed it to focus on the most salient temporal features of the sound signals, enhancing the detection of rapid, transient events such as electrical arc discharges. This attention mechanism significantly improved the system's ability to detect both AC and DC arc discharges, even in the presence of other complex, interfering sounds typical in offshore environments.

The model demonstrated strong generalization performance, achieving an accuracy of 81.53\% in classifying arc discharges and background noise. This makes the proposed system a promising solution for enhancing fire detection systems on ships, where early identification of electrical arc discharges can prevent catastrophic fires. 

Overall, this study highlights the potential of AI-based fault detection systems in improving operational safety and reducing the risks associated with electrical faults in maritime environments, providing a foundation for the development of smarter, more efficient maintenance and safety systems for offshore vessels.

\section{Conclusion}
This chapter presented two case studies that demonstrate the effectiveness of advanced AI-based systems in improving the detection and diagnosis of faults in complex and dynamic environments. The first case study focused on the development of a fault diagnosis system for the K1 tank powertrain, while the second case study explored the application of AI for the early detection of electrical arc discharges in offshore environments. Both case studies leveraged frequency-dependent convolutional models, which proved highly effective in enhancing the detection of transient and frequency-dependent sound events, highlighting the potential of AI for real-time fault detection, preventive maintenance, and safety enhancement in both military and maritime contexts.

In the first case study, the proposed system for the K1 tank powertrain utilized sound signals to achieve high accuracy in detecting various faults such as fan drive motor failures, starter motor faults, and dust extractor fan motor issues. The integration of frequency-dependent convolutional networks—specifically, FDY-CNN, DFD-CNN, PFD-CNN, and TFD-CNN—proved effective in distinguishing between normal and faulty conditions. While these models outperformed traditional methods like CNN in some cases, the study highlighted that TFD-CNN showed the best performance, achieving the highest accuracy of 81.53\%. This result underscores the model's ability to effectively capture transient fault events and enhance real-time maintenance capabilities. The success of this system emphasizes the potential of AI-based models in optimizing military equipment maintenance, improving operational readiness, and significantly reducing downtime in real-world environments.

The second case study demonstrated the application of an AI-based detection system for electrical arc discharges in offshore environments, an area with critical safety implications. The proposed system successfully distinguished between AC and DC arc discharges and various background noises, providing essential real-time detection that is crucial for preventing fires caused by electrical faults on ships. Here, the TFD-CNN model once again outperformed other models, achieving high detection accuracy by focusing on the most salient temporal features in the acoustic signals. The success of this case study highlights how AI-driven detection systems can greatly improve safety in maritime operations, offering early warnings for critical electrical faults, reducing the risk of fire, and enhancing the overall safety and reliability of offshore vessels.

Together, these two case studies highlight the transformative potential of AI in fault detection and diagnosis. They showcase how advanced frequency-dependent convolutional models—like FDY-CNN, DFD-CNN, PFD-CNN, and TFD-CNN—can be successfully applied to a wide range of fields, from military maintenance to offshore safety. Both systems achieved significant improvements in detection accuracy, and the integration of Temporal Attention Pooling in TFD-CNN further enhanced the models' ability to detect transient events, which are often critical for early fault detection. These results provide a strong foundation for future developments in AI-based systems that can continuously improve performance, reliability, and safety in diverse real-world applications.

%% file: sections/6_Conclusionandfutureworks.tex
\section{Contributions}

This dissertation presents a series of novel contributions to the field of SED, focusing on improving the physical consistency, computational efficiency, and detection accuracy of deep learning-based models through frequency-dependent convolution techniques. The key advancements proposed in this work are summarized as follows.

First, we introduce frequency dynamic convolution (FDY conv), a method designed to overcome the limitation of translation equivariance along the frequency axis inherent in conventional 2D convolution. By allowing convolution kernels to adapt to the frequency content of input features, FDY conv aligns the model’s inductive bias with the physical nature of sound events, which often exhibit frequency-localized and non-stationary characteristics. This architectural shift enables significantly improved frequency-dependent pattern recognition, leading to strong gains in SED performance over conventional convolutional baselines.

As a further enhancement of kernel expressiveness, we propose dilated frequency dynamic convolution (DFD conv), which applies different dilation sizes to each basis kernel within FDY conv. This enables each kernel to attend to distinct spectral resolutions, thereby expanding the receptive field along the frequency axis without increasing the number of parameters. DFD conv encourages functional diversity among basis kernels by explicitly differentiating their roles, which leads to more robust frequency-adaptive feature extraction. Experimental results show that frequency dilation is more effective than temporal dilation and that mixing various dilation sizes improves SED performance beyond standard FDY conv.

To address the increased computational cost introduced by FDY conv’s dynamic kernel mechanism, we propose partial frequency dynamic convolution (PFD conv). This architecture incorporates a static 2D convolutional branch alongside the dynamic FDY conv branch and concatenates their outputs. By carefully balancing the proportion between static and dynamic paths, PFD conv maintains most of the performance benefits of FDY conv while substantially reducing the model size—achieving over 50\% parameter reduction in practice. This allows for deployment in resource-constrained environments, broadening the practical applicability of frequency-adaptive methods.

Building upon these approaches, we further propose multi-dilated frequency dynamic convolution (MDFD conv), which integrates multiple FDY or DFD (dilated FDY) branches within a single convolutional layer. By assigning different dilation sizes to each dynamic branch, MDFD conv expands the spectral receptive field and enriches kernel diversity. This structural augmentation enables the model to simultaneously capture narrow-band and wide-band frequency patterns, resulting in improved generalization to diverse acoustic scenes and superior overall detection performance.

In parallel, we identify a major limitation of FDY conv’s use of temporal average pooling, which treats all time frames equally and fails to highlight transient audio cues. To address this, we develop temporal attention pooling (TAP), a novel pooling mechanism that adaptively emphasizes salient or rapidly changing time segments. TAP integrates time-based attention, velocity-based attention, and residual average pooling to balance sensitivity to transients with robustness to stationary events. When combined with FDY conv, the resulting TFD conv model consistently outperforms existing variants, particularly for transient-heavy classes such as alarms, bells, and speech segments.

To quantitatively verify the effectiveness of the proposed architectures, we conduct detailed class-wise performance analyses, along with statistical tests such as ANOVA and Tukey HSD. These evaluations demonstrate that the proposed models not only achieve overall improvements in polyphonic detection metrics, but also exhibit superior detection capability for transient events, which are often more challenging for traditional SED systems.

In summary, this dissertation contributes a unified and extensible framework for frequency-dependent and temporally-aware modeling in SED. The proposed methods—FDY conv, PFD conv, MDFD conv, and TAP—collectively advance the state of the art by enhancing physical consistency with the acoustic domain, improving efficiency for real-world applications, and boosting detection robustness across both stationary and transient sound events. These contributions lay a solid foundation for future research, which may explore integration with pre-trained audio-language models, adaptation to new domains (e.g., bioacoustics or egocentric audio), or deployment on low-power edge devices.

\section{Key Findings}
This dissertation yields several key findings that advance the understanding and effectiveness of SED systems through the use of frequency-dependent convolution and temporal attention mechanisms. These findings clarify how each proposed method contributes to different aspects of SED—especially in detecting non-stationary, transient, and quasi-stationary sound events.

Frequency-Adaptive kernels significantly improve detection of non-stationary sound events. One of the most notable findings is the effectiveness of FDY conv in capturing frequency-dependent patterns inherent in non-stationary events such as alarms, speech, and animal sounds. By relaxing the translation equivariance constraint along the frequency axis and adapting kernels dynamically to the spectral characteristics of each input, FDY conv enables the model to better recognize localized spectral variations. This results in a marked improvement in detecting sound events with rapid and complex changes in frequency over time.

Dilated frequency dynamic convolution enhances receptive field diversity and generalization. This work also demonstrates the effectiveness of DFD conv, which introduces distinct dilation sizes to each basis kernel. DFD conv expands the model’s frequency receptive field and encourages functional diversity among kernels. Furthermore, MDFD conv, which combines multiple FDY/DFD branches with different dilation patterns, achieves even greater performance by capturing multi-scale spectral patterns. The ablation studies confirm that varying dilation sizes across branches plays a crucial role in detecting events that span wide or disjoint frequency ranges.

Partial frequency dynamic convolution balances efficiency and performance. To address the computational burden of dynamic convolutions, PFD conv was introduced. By combining static 2D convolution with FDY conv in a hybrid architecture, PFD conv reduces parameter count by up to 51.9\% while maintaining comparable detection performance. This finding underscores the feasibility of deploying frequency-adaptive SED models in real-time or resource-constrained environments without significant loss of accuracy.

Temporal attention pooling boosts transient event detection. Another major finding is the superiority of temporal attention pooling (TAP) over conventional average pooling for transient-heavy sound events. When integrated into FDY conv to form TFD conv, TAP enables the model to emphasize salient temporal segments by incorporating both saliency-based and velocity-based attention. This leads to a significant improvement in detecting short-duration events such as door knocks, alarms, and plosive speech sounds, which are typically diluted by uniform temporal pooling mechanisms.

Class-Wise analyses reveal model-specific strengths and limitations. Detailed class-wise performance evaluation reveals that TFD conv is particularly effective for transient-heavy classes, outperforming all other variants for events like speech, alarms, and cat/dog sounds. Conversely, models like PFD conv and MDFD conv showed competitive or superior performance on quasi-stationary classes such as vacuum cleaners and electric shavers. This contrast indicates that while TAP improves temporal resolution, a mixture of dynamic and static modeling is better suited for stationary events. These insights suggest a potential for model ensembles or adaptive hybrid architectures depending on sound class characteristics.

In summary, this work demonstrates that the combination of frequency-adaptive convolutional kernels and temporal attention pooling mechanisms offers substantial improvements in sound event detection. The key findings highlight the complementary roles of FDY, DFD, MDFD, PFD, and TAP in capturing the spectral and temporal nuances of diverse sound events. These results emphasize the importance of tailoring convolutional processing to both frequency structure and temporal dynamics in order to robustly detect a broad spectrum of real-world acoustic events.

\section{Limitations}
While the proposed frequency-dependent convolution methods have demonstrated significant advancements in SED, several limitations remain that should be addressed in future research to further enhance their robustness, generalizability, and practical applicability.

A notable limitation of FDY conv and its variants is their relatively lower performance in detecting quasi-stationary sound events, such as vacuum cleaners, electric shavers, and other mechanical sounds that exhibit consistent frequency patterns over time. These models are inherently designed to capture dynamic and localized spectral changes, which makes them highly effective for non-stationary events but less suited for sounds with prolonged steady-state characteristics. This suggests a need for complementary mechanisms that explicitly model spectral stability to balance the strengths of frequency-adaptive methods.

Although MDFD conv achieves superior performance by leveraging diverse receptive fields, its architecture introduces considerable computational cost due to the use of multiple parallel dynamic branches. This limits its feasibility in real-time or embedded systems. Future research should explore architectural compression techniques, such as weight sharing, pruning, or knowledge distillation, to maintain performance while reducing computational demands.

While the introduction of TAP significantly improves the detection of short-duration events, certain ultra-transient or overlapping events remain challenging, especially when they occur under high noise or with frequency masking. TAP-FDY conv improves temporal resolution, but its reliance on frame-wise attention may still miss extremely brief or ambiguous cues. More expressive attention mechanisms, such as hierarchical or event-centric attention, may be required for finer-grained temporal discrimination.

PFD conv was proposed to reduce the parameter footprint of FDY conv by combining it with static convolution branches. While this design succeeded in improving efficiency, it also restricted the full flexibility of dynamic modeling, potentially limiting its performance on complex acoustic scenes. Finding the optimal static–dynamic balance, possibly through adaptive gating or conditional execution, remains an open design challenge.

All evaluations in this work were conducted on the DESED dataset, which, although widely adopted, represents a curated domestic soundscape with a specific event taxonomy. Consequently, the generalization of the proposed methods to other domains (e.g., urban environments, industrial monitoring, wildlife acoustics) remains unverified. Factors such as diverse background noise, channel mismatch, and reverberation could affect real-world performance. Future studies should validate these methods across broader datasets and conduct domain adaptation or robustness tests.

In summary, the limitations of this research highlight important areas for future work: improving detection of quasi-stationary events, optimizing high-performing architectures for low-resource settings, refining attention mechanisms for fine-grained temporal cues, achieving a better balance between model expressiveness and efficiency, and ensuring robustness across diverse acoustic environments. Addressing these challenges will further increase the practical impact and deployment potential of frequency-dependent SED systems.

\section{Implications for Future Work}
Building on the advances achieved in this dissertation, several promising directions emerge for future research in frequency-dependent convolutional approaches to SED. These directions aim to address current limitations, enhance model robustness, and broaden applicability to diverse acoustic environments and resource-constrained settings.

A key challenge identified in this work is the relatively weak performance of frequency-adaptive models on quasi-stationary sound events (e.g., vacuum cleaners, electric shavers). Future research could explore hybrid architectures that combine dynamic frequency modeling with mechanisms optimized for stable spectral patterns. For instance, frequency-dependent feature masking, low-frequency temporal modeling, or explicit stationary pattern modeling could complement the dynamic behavior of current models. Integrating domain knowledge specific to machine-generated or constant-harmonic sound sources may also provide tailored improvements.

While MDFD conv demonstrates strong performance, its computational complexity may hinder deployment in latency-sensitive or embedded systems. Future efforts should explore model compression techniques such as structured pruning, knowledge distillation, and parameter sharing across branches to reduce computational load. Additionally, the design of lightweight yet expressive architectures, potentially using mobile-friendly convolution modules or attention-based approximations, could make real-time deployment feasible. Benchmarking on edge AI hardware (e.g., Jetson, Coral) would further ground these developments in practical contexts.

Although TAP enhances transient detection, future work could explore more expressive attention modules that go beyond frame-wise weighting. For example, hierarchical attention, event-conditioned attention, or context-aware dynamic pooling may better capture rapid changes, overlaps, or subtle transitions in time-frequency patterns. Another important avenue is extending TAP to support multi-label and overlapping event detection with greater temporal precision.

Balancing efficiency and flexibility remains a critical task. Future models may benefit from input-adaptive processing, where the model dynamically adjusts the proportion of dynamic vs. static computation based on the characteristics of incoming audio. This could be framed as a conditional computation strategy, enabling resource-aware inference depending on the complexity of the input scene. Integration with neural architecture search or runtime optimization frameworks could further improve adaptability.

Current evaluations have been centered on the DESED dataset, which, while useful, may not fully capture the variability encountered in real-world environments. Future work should prioritize cross-dataset validation, including urban soundscapes, industrial monitoring, wildlife recordings, or egocentric audio. Moreover, leveraging self-supervised or semi-supervised learning could enable models to adapt to new domains with minimal labeled data, increasing robustness and transferability. Investigating domain generalization and unsupervised adaptation techniques would also strengthen model applicability in unseen conditions.

In summary, future research should address key areas including detection of quasi-stationary events, efficient real-time implementation, advanced temporal modeling for transients, input-aware adaptivity, and broader generalization to complex environments. By tackling these challenges, the models developed in this dissertation can evolve into more versatile and deployable SED systems for applications ranging from smart homes to industrial safety and autonomous agents.

\section{Conclusion}
This dissertation proposed a series of novel frequency-dependent convolution methods to advance the state of SED. Motivated by the need to improve frequency modeling, computational efficiency, and sensitivity to transient sound events, we introduced and thoroughly evaluated a family of architectures: frequency dynamic convolution (FDY conv), partial frequency dynamic convolution (PFD conv), dilated and multi-dilated frequency dynamic convolution (DFD conv, MDFD conv), and temporal attention pooling frequency dynamic convolution (TFD conv).

The first major contribution, FDY conv, addressed the physical inconsistency of standard 2D convolution by releasing translation equivariance along the frequency axis. By allowing convolution kernels to adapt to local spectral content, FDY conv significantly improved detection accuracy for non-stationary sound events with complex time-frequency patterns, such as alarms, speech, and animal sounds. However, FDY conv showed weaker performance on quasi-stationary events like vacuum cleaners or electric shavers, which led to the development of complementary architectures.

To address computational overhead and improve coverage of stationary patterns, we introduced PFD conv, which combines static and dynamic convolution branches to retain the benefits of FDY conv while reducing parameter count by over 50\%. Additionally, DFD conv extended FDY conv by incorporating dilation into the basis kernels, diversifying the receptive fields and enabling better modeling of frequency-wide patterns. This idea was further expanded in MDFD conv, which integrates multiple dynamic branches with varying dilation sizes to capture multi-scale spectral features. MDFD conv yielded state-of-the-art performance across a range of event types.

Recognizing the limitations of temporal average pooling, we proposed TAP to replace it within the FDY framework. The resulting TAP-FDY conv model adaptively emphasized salient and rapidly changing temporal segments, significantly enhancing detection of short-duration events such as plosive speech sounds and alarms. TAP-FDY demonstrated superior performance on transient-heavy sound classes and highlighted the importance of integrating temporal context awareness into frequency-adaptive models.

While these contributions represent meaningful progress, several limitations were identified. The reduced performance on quasi-stationary events, the high computational cost of multi-branch models, and the reliance on a single dataset (DESED) suggest opportunities for future improvements. Further research should explore hybrid architectures for stationary events, efficient dynamic modeling for deployment on resource-constrained devices, and the use of self- or semi-supervised learning for adaptation to diverse environments and novel sound classes.

In summary, this dissertation lays a strong foundation for building more physically consistent, efficient, and transient-aware SED systems. By rethinking convolutional design in light of the spectral and temporal structure of sound, the proposed methods contribute not only to the field of SED but also to broader directions in auditory intelligence, with applications ranging from robotics and smart devices to environmental sensing and human-computer interaction.

%% file: sections/7_acknowledgement.tex
카이스트에 학부 신입생으로써 제일 처음 교문을 지나던 날이 생각납니다. 저는 부모님의 차 안에서 "이 곳이 내가 20대를 다 바칠 곳이구나"하고 생각하며 여러 복잡한 감정을 동시에 느꼈습니다. 그 날로부터 12년이 지났습니다. 20대는 물론 30대의 초반까지 모두 이 곳에서 보내버렸습니다. 그 날과는 다른, 하지만 마찬가지로 여러 복잡한 감정을 동시에 느낍니다. 저는 이 곳에서 많은 것을 배우고, 아내를 비롯한 많은 소중한 친구들과 동료들을 만났으며, 많이 즐거웠고 힘들었고 아팠고 성장했습니다. 카이스트에서 보낸 12년은 저에게 큰 행운이자 축복이었던 것 같습니다. 저라는 사람이 한 사람의 구실을 할 수 있도록 배움을 이어나가는 기간이 참 길었습니다. 그 과정에서 제가 만난 모든 분들은 저를 더욱 나은 사람이 되도록 도움주었던 것 같습니다. 모두 감사의 말씀 드리고 싶습니다.

먼저, 부족하고 고집 센 제자를 이끌어주시느라 가장 고생하신 박용화 교수님께 감사의 말씀 드립니다. 돌이켜보면 저는 참 피곤한 제자였습니다. 박용화 교수님께서는 항상 바쁜 시간을 쪼개어 제자들을 지도하고 과제를 관리하며 강의 등 다양한 교내 업무 또한 보았습니다. 이 외에 제가 알지 못하는 많은 일들을 신경 쓰시며 얼마나 힘드셨을지 감히 상상하기도 어렵습니다. 제가 교수님께서 제자들을 위해 최선을 다하고 있음을 깨닫는데 너무 오랜 시간이 걸렸습니다. 교수님께서 베풀어주신 많은 가르침을 닦고 정진하여 사회에 기여하고 선한 영향을 끼치는 사람이 되도록 하겠습니다. 항상 감사드립니다.

또한 부족한 제 박사학위연구를 심사하고 코멘트 주셔서 더욱 훌륭한 연구로 만드는데 도움을 주신 윤용진 교수님, 이승철 교수님, 최정우 교수님, 그리고 김성영 교수님께 감사의 말씀 드립니다. 제가 생각치 못한 부분을 짚어주시고 따뜻하면서도 날카롭게 지적해주신 덕분에 제 연구를 다양한 관점에서 돌아보고 발전시키는 경험을 할 수 있었습니다. 제 연구를 심사해주시고 승인해주셨기 때문에 제가 박사학위를 받을 수 있었다는 점 잊지 않고, 사회에 나가서도 교수님들 이름에 먹칠하지 않는 모자람 없는 박사로서의 삶을 살도록 하겠습니다.

제가 박사를 받는 모습은 커녕 석사를 마치는 모습도 보지 못하고 떠나신 어머니께도 감사의 말씀 드립니다. 누구보다 기뻐하고 자랑스러워하셨을 당신이 이 순간을 함께하지 못함이 너무나도 안타깝습니다. 차마 갚지 못한 커다란 사랑으로 키워주셔서, 그리고 잘못했을때는 따끔히 혼내주셔서 제가 모자라나마 한사람의 몫을 할 수 있게 되었습니다. 어릴 땐 너무 공부하라 다그치던 모습이 원망스러울 때도 있었지만, 지금 와서 돌아보면 그 또한 어머니께서 제게 주실 수 있는 최고의 사랑이었음이라 짐작합니다. 저는 아직도 가끔씩 초등학생 시절 전국수학경시대회에서 수상을 해서 다른 어머니들의 부러움을 받았다며 즐거워하던 어머니의 얼굴이 생각납니다. 제가 작지나마 효도를 한 것 같아 뿌듯했습니다. 제가 열심히 공부하고, 좋은 연구를 하고, 박사를 딴 이후에도 계속 사회에 기여하는 훌륭한 사람이 되는 모든 과정을 지켜봐주시고 자랑스러워하시리라 생각합니다. 앞으로도 자랑스러운 아들로써 살아가도록 하겠습니다.

제 삶의 방향이 묘연해질때마다 아낌없이 격려와 조언을 주신 아버지, 먼저 박사과정을 밟아온 탓에 당신과 같은 인고의 길을 가는 제 모습을 다시 아파하셨을 아버지께도 감사의 말씀 드립니다. 항상 아버지께선 제게서 당신의 모습이 보인다고 하셨습니다. 저로썬 아버지처럼 살 수 있을까 항상 막막하고 어렵지만, 당신의 아들이기에 힘 내 볼 수 있었고, 앞으로도 힘내보려 합니다. 아버지는 훌륭한 가장이자 남편이며 아버지셨습니다. 이 모든 역할을 한번에 수행하면서 한번도 모자람 없으셨기에, 제가 나름대로의 유복하고 화목한 가정에서 무탈하게 자라, 남들 쉽게 못 겪어보는 해외생활경험도 하고 석사박사과정도 밟을 수 있었습니다. 금전적으로는 물론이고, 정서적으로도 아버지께서 끌어주시고 어깨토닥여주셨기 때문에 여기까지 올 수 있었습니다. 못난 아들이지만 항상 존경하고 우러러보는 마음 알아주셨으면 합니다.

세상에 하나뿐인 제 누나에게도 감사의 말씀 드립니다. 항상 어릴땐 말도 안듣고, 뭣도 모르고 누나에게 상처주는 말만 했던 것 같습니다. 제가 어릴땐 주변에 누나에 대해 불평을 늘어놓고 그랬지만, 제가 누나가 있어서 든든하고 의지가 많이 되었습니다. 제가 어렸을때 좀 모나기도 하고 별나기도 했던 구석이 참 많았는데, 누나가 또래이자 선배로써 잘 타일러줘서 정상인의 범주에 잘 안착하지 않았나 싶습니다. 항상 절 싫어하는것처럼 굴면서도 이것저것 잘 챙겨준 덕분에 저는 다른 친구들처럼 맞고 자라지도 않고 오히려 간식도 많이 얻어먹고 썩 즐거웠습니다. 어렸을 적 서로 컴퓨터하고 싸울때도, 서로 왕래없이 각자 싱가폴에 살때도, 한국에 돌아와서 각자 대학원과 회사를 다닐 때도, 어디선가 누나가 착실하게 자기 삶을 살고 있다는 사실 자체가 제게 위로가 되기도 했습니다. 동생으로써 미안한 마음이 더 많을 뿐이지만, 그리고 아직도 좀 별난 동생이지만 그래도 항상 응원합니다.

모자란 사위를 따뜻하게 환영해주시고 가족으로 맞아주신 장인어른과 장모님께 감사의 말씀 드립니다. 아직 이룬 것 없고 아직도 공부중인 대학원생에게 귀한 의사 따님을 시집보내는건 쉽지 않은 결정이였을 것이라 생각됩니다. 무뚝뚝하고 말수도 적은 저를, 모자라고 부족한 저를 좋게 봐주시고 믿어주시고 항상 뵐때마다 아들처럼 대해주셔서 언제나 감사한 마음뿐입니다. 너무 반겨주셔서 처가에 갈때마다 제 친가에 가는 기분입니다. 제가 언젠가 대성해서 성영이랑 장인어른 장모님 모두 호강시켜드리도록 하겠습니다. 든든한 아들이 되도록 노력하겠습니다.

또 저를 따뜻하게 가족으로 맞아주신 처남 준현이와 처제들 수민, 수빈이에게도 감사의 말씀 드립니다. 형 오빠로써 든든한 모습을 보여야 하는데, 부족한 모습을 많이 보여준 것 같아 마음이 쓰입니다. 그래도 저를 가족으로써 환영하고 맞아줘서 너무 기뻤습니다. 앞으로 좋은 형 오빠의 모습을 보여줘서 모범이 되도록 노력하겠습니다.

싱가폴에서 제게 멘토가 되어주셨던 최재혁 박사님께도 감사의 말씀 드립니다. 돌이켜보면 선생님께서도 당시 포닥생활이 만만치 않으셨을텐데, 말도 안듣고 숙제도 거의 안해가다시피하는 제자를 위해 기꺼이 시간을 내주셨습니다. 학부시절 대학원시절 저도 모르게 제게 베어있던 공부습관들에 놀랐다가도, 선생님을 많이 떠올렸습니다. 제게 단순히 공부를 가르쳐주신게 아니라, 공부를 대하는 태도와 자세를 스스로 익히게 해주셨습니다. 덕분에 제가 여기까지 올라올 수 있었다고 생각됩니다. 한국에 오신 이후로 많이 찾아뵙진 못했지만, 항상 감사하는 마음입니다. 감사드립니다.

지금까지 함께해온 휴먼랩 연구실동료들에게도 감사의 말씀 드립니다. 연구에서도 인생에서도 제게 많은 조언을 주고 제게 많은 깨달음을 주신 경태형 감사드립니다. 제가 힘들때마다 많이 시간 뺏었는데, 덕분에 저는 많은 지혜를 얻어간 것 같습니다. 가장 연구와 토의 많이했던 성후도 고맙습니다. 성후의 연구가 없었다면 이 논문의 제목 자체가 아예 달랐을 것이기 때문에, 아마 이 사사에서 반복되는 '당신 없이는 이 논문이 나오지 못했을것입니다'라는 표현의 가장 말그대로의 주인공일겁니다. 연구실에서도 졸업해서도 함께 서로 연구 코멘트해주고 디스커션해준 음향팀 동기 준혁에게도 감사합니다. 항상 많이 배우고 또 덕분에 열심히 해야겠단 생각을 많이 합니다. 남은 박사학위도 잘 마치고 미국에 잘 자리잡을 것이라고 믿습니다. 같이 석사동기로 시작해, 카이스트에서 각자의 박사과정을 마친 대희와 상현이도 감사합니다. 가끔씩 얼굴보고 삶얘기, 연구얘기하며 힘든 연구실 생활 버틸 수 있었습니다. 제일 늦었지만 저도 결국 박사가 됩니다. 동갑내기 석사동기 인규도 석사시절 같이 수영도 하고 방도 쓰면서 많이 든든하고 의지했었습니다. 수학공부도 정말 좋아하고 끈기있어서 박사를 안하는게 아쉬웠는데, 그래도 미국 유학을 가게 되어 다행입니다. 남은 기간도 잘 마쳐서 좋은 박사가 되리라고 생각합니다. 그리고 한학기지만 헬스센서 연구 함께했던 석사동기 혜원이도 감사합니다. 같이 고생하고 열심히하고 연구실에서 힘든 일도 많이 공감했는데, 박사때도 있엇으면 좋았을텐데 안타깝기도 하고, 먼저 사회에 나가서 열심히 살아나가는 모습을 보면 멋지기도 합니다. 앞으로도 잘 해낼것이라 믿습니다. 음향팀 후배 병윤이에게도 감사합니다. 까칠한 선배 밑에서 고생이 많았습니다. 말은 하지 않았지만 항상 긍정적인 에너지를 나눠주었고, 또 일을 맡길때면 매번 든든했습니다. 또 제 첫 연구실 후배인 보미, 정민이, 현준이에게도 감사의 말씀 드립니다. 선배로써 멋진 모습보다 못난 모습을 훨씬 많이 보인 것 같아 부끄럽지만, 그래도 덕분에 더 즐거운 연구실생활이였습니다. 모두 훌륭한 박사가 되리라 믿습니다. 제일 많이 도움받은 후배인 덕기에게도 감사의 말씀 드립니다. 연구도 과제도 항상 군말없이 다 따라와주고, 실력도 있지만 인성도 훌륭해서 세상에 이런 후배가 있기도하네 싶었습니다. 제가 누군가의 후배가 된다면 이렇게 될 수 있을까 생각해보다가 발끝에도 미치지 못할 것 같아서 고개를 내젓곤 합니다. 이 외에도 동갑내기 친구이자 든든한 후배였던 상민이, 학부시절부터 오랫동안 잘 따라와준 승덕씨, 든든히 밑고 맡겼던 유정씨, 인한이, 성민씨, 제현씨 등 음향팀 후배님들께도 감사의 말씀 드립니다. 제가 선배로써 더 잘챙겨드리고싶었는데 제 컨디션이 왔다갔다해서 봐주다가 못봐주다가 했습니다. 그래도 잘 따라와주신 덕분에 저도 더 다양한 경험과 연구를 하고 일할 수 있어습니다. 이 외에도 준오형, 재덕이, 강재, 길용이형, 성현이형, 재학이 등 앞길을 닦아주신 선배들 덕분에 좋은 연구실에서 연구할 수 있었습니다. 외에도, 앞서 언급된 분들과 윤섭이, 준호씨 등 랩장하셨던 분들께도 랩원들 위해 고생해주셔서 감사드립니다. 그리고 자주 괴롭히는 원호에게도 감사의 말씀 드리며 크게 될 사람이라 미리 잘 보이고싶어서 그랬으니 양해를 부탁드립니다. 그리고 과제제안서도 같이 쓰고 졸업도 같이하는 대근이형께도 감사의 말씀 드립니다. 함께 해서 더욱 든든했습니다.

학부 새내기때부터 지금까지 함께 해온 승모, 학기에게도 감사의 말씀 드립니다. 하필 기계과를 골라서 학부때도 고생했는데, 대학원와서도 어김없이 다 같이 고생하고 힘들어하는 모습을 보면 웃기면서도 위로가 됐습니다. 덕분에 힘든 길 더 많이 웃으며 버틸 수 있었습니다. 다들 많이 힘들었고 많은 책임을 졌고 많은 고민이 있었습니다. 여기까지 함께 고생 많았던 만큼 모두 미래에는 웃을일이 더 많았으면 좋겠습니다. 박사까진 함께하지 않았지만, 석사까지 함께하고 학부때도 룸메이트로써 신세 많이졌던 성진이에게도 감사의 말씀 드립니다.

연구하느라 지친 몸과 마음을 함께 신나게 털어버리는 것을 도와줬던 밴드 오얀을 함께했던 동훈이, 혜상이, 정훈이형, 석중이, 윤성이, 효진이에게 감사의 말씀 드립니다. 돌이켜보면 오얀을 쉬던 시절이 제 대학원 시절에 가장 힘들던 시절이였습니다. 알게 모르게 많은 치유와 위로 받아왔던 것 같습니다. 제가 까탈스러워서 자꾸 편곡하자고 자작곡하자고 졸라대서 힘들었을 것도 같습니다. 그래도 함께 즐거운 무대 많이 했으니 잘 되지 않았나도 싶습니다. 즐거운 음악 함께 했고, 몇명은 각자 몇명은 계속 같이 하게 되었지만 모두가 제겐 소중한 오얀 멤버들입니다. 언제든 웃으며 다시 만나면 좋겠습니다.

대학생 새내기부터 함께 해온 13후기들, 특히 호구톡방 친구들인 규철이, 근민이, 근우, 의곤이, 호형이, 호준이, 현석이, 수민이, 진혁이, 경제, 기석이에게도 감사의 말씀 드립니다. 이렇게 오랫동안 활발한 톡방에 속해있다는 소속감에 항상 든든하고, 매번 유쾌하고 어이없는 말들로 즐거웠습니다. 물론 제가 잠깐 나가있던 반년을 지적하고 싶겠지만, 그땐 제가 석사 최종발표가 코앞이라 예민했었으니 너른 이해 부탁드립니다. 언제까지나 12년 전처럼, 혹은 지금처럼, 변함없이 철부지처럼 같이 어울리며 술 마시면 더 바랄게 없겠습니다. 특히 제 아내를 소개시켜준 강호형 박사에게는 항상 감사하는 마음뿐인 점 알아주세요.

학부시절 함께 즐겁게 동아리 활동했던 뮤즈 친구들에게도 감사의 말씀 드립니다. 해일이, 재윤이, 상명이, 홍창이, 영준이, 지수, 현상이, 태경이, 희윤이, 수한이, 은진이, 유진이, 태희, 은설이, 상은이, 찬희, 종원이, 진우(2명), 우석이, 우진이(2명), 현우, 선일이, 민수, 창균이, 강욱이, 유경이, 소원이, 성민이, 윤헌이 등등 덕분에 제 대학생활은 다채롭고 즐거웠습니다. 추천해준 덕분에 뮤즈에 들어올 수 있었어서 혜인이에게 특별히 감사의 말씀 드립니다.

싱가폴에서부터 카이스트, 그리고 부부친구로 17년간의 세월을 함께해온 지윤이에게도 감사의 말씀드립니다. 어려서부터 같이 음악도 이것저것 도전해보고, 다투기도 했었고, 축가도 불러주고, 추억이 정말 많습니다. 언제나처럼 반갑고 따뜻한 친구라서 고맙습니다.

고등학교때 싱가폴에서 같이 밴드했던 성현이, 영주, 현경이, 시진이, 창윤이에게도 감사의 말씀 드립니다. 같이 밴드하면서 진짜 많이 싸우고 화해도 하고 많은 추억을 남기면서, 저도 인격적인 성장이 많았던 것 같습니다. 그때의 모자란 제가 많이 상처줘서 죄송합니다. 그래도 여러분과 함께한 시간이 제게 음악이라는 큰 선물을 주었습니다. 여러분께도 그랬으면 합니다. 음악 덕분에 제가 음향을 전공하였고 대학원 시간을 버텼으니 여러분께도 큰 감사를 빚졌습니다. 이 외에도 한국에서 같이 박사과정 하던 지현이를 비롯한 친하게 지내던 NUS High School 선후배들께도 좋은 추억 함께해줘서 감사합니다.

중학교부터 함께 놀던 지균이, 승욱이, 호태에게도 감사의 말씀 드립니다. 싱가폴에서 살면서 한국 놀러올때 놀아주고, 대전에 살면서 서울 올라갈때 놀아줘서 항상 즐거웠습니다. 중학생때로 돌아간 것처럼 실없이 웃고 떠들 수 있어 항상 만날때마다 즐겁습니다. 더 자주 놀러가지 못해 미안합니다. 유부남인 관계로 앞으로도 계속 그럴 가능성이 농후하지만 양해 부탁드립니다.

그간 제게 가르침을 주셨던 카이스트의 많은 교수님들과, 제 논문들을 읽고 코멘트주셨던 많은 익명의 리뷰어분들께도 감사의 말씀 드립니다. 덕분에 저의 지식과 지혜가 한층 더 깊어지고, 제가 더 나은 학자로 거듭날 수 있었습니다. 저도 학계에 남아 제가 받은 은혜를 후대에 갚을 수 있으면 더할나위 없겠습니다.

마지막으로, 제게 주어진 가장 큰 행복이자 행운이자 선물인 제 아내 박성영에게 감사의 말씀 드립니다. 항상 부족한 남편의 쓰잘데기없는 잔소리와 예민함을 견디느라 고생이 많습니다. 더 잘해주고 싶으면서도 몸이 마음을 따라주지 못하는 걸 보면 전 아직도 멀었나봅니다. 그래도 덕분에 웃고, 행복하고, 더 나은사람이 되고싶단 욕심을 냅니다. 박사과정 남편 뒷바라지 해준다고 고생 많았습니다. 앞으로도 좀 더 고생할 지 모르겠지만, 언젠간 꼭 호강시켜드리겠습니다. 석사 2년차 중간쯤에 처음 만나서, 덕분에 힘을 내어 석사를 졸업할 수 있었습니다. 그리고 박사과정에 결혼하여 덕분에 이 험난한 과정을 함께 헤쳐나가고 끝내 이 자리에 다다랐습니다. 종종 아무 생각없이 저랑 결혼했다고 장난삼아 말하지만, 아직 박사과정뿐일 저를, 제 연구분야에 대한 지식이 없어서 저에 대해 확신을 가지기도 힘들었을텐데 믿고 따라주고 버팀목이 되어주어 너무 감사합니다. 앞으로도 계속 노력하여 더욱 든든하고 지혜롭고 다정한 남편으로써 당신의 가장 행복할때와 가장 힘들때를 모두 함께해주도록 하겠습니다. 사랑합니다.

This dissertation marks my very first step into the world of academia. Though still immature in many respects, it is a work I hold with quiet pride. I sincerely hope that it may serve, however humbly, as a stepping stone for those who follow. May this work become a small foundation for bridging machines and the world through sound.

2025년 6월

남현욱